\documentclass[aps,prd,superscriptsize,superscriptaddress,preprint,longbibliography,
nofootinbib,notitlepage]{revtex4-1}
\usepackage[utf8]{inputenc} 
\usepackage{graphicx,xcolor,overpic,mathtools}
\usepackage[colorlinks]{hyperref}
\hypersetup{ 
	colorlinks=true, 
	linkcolor=black, 
	filecolor=black, 
	citecolor = blue,       
	urlcolor=blue, 
} 
\newcommand{\beq}{\begin{equation}}
\newcommand{\eeq}{\end{equation}}

\usepackage{upgreek}
\usepackage{csquotes}
\numberwithin{equation}{section}

\usepackage[title]{appendix}
\usepackage{verbatim}
\usepackage{tensor}
\usepackage{tikz}
\usetikzlibrary{patterns,positioning}
\usepackage[compat=1.1.0]{tikz-feynman}

\def\beq{\begin{equation}} 
\def\eeq{\end{equation}}
\def\beqa{\begin{eqnarray}}
\def\eeqa{\end{eqnarray}}

\DeclareMathOperator*{\sumint}{\int\hspace{-1.55em}\sum}

\usepackage{amsthm,amsmath,amssymb}

\PassOptionsToPackage{normalem}{ulem}
\usepackage{ulem} 
\usepackage{sidenotes}
\usepackage{bm}
\usepackage{url}
\usepackage{physics}

\usepackage{xfrac}
\usepackage[makeroom]{cancel}
\usepackage{tensor}
\usepackage{mathrsfs}
\usepackage{slashed}
\usepackage{tikz}
\usepackage[mode=buildnew]{standalone}
\usepackage{bbm}
\usepackage{booktabs}
\usepackage{cleveref}
\usepackage{placeins}

\crefname{section}{Sec.}{sec.}
\crefname{figure}{Fig.}{Fig.}
\Crefname{section}{Section}{Sections}
\usepackage{caption,subcaption}
\newcommand{\lag}{\mathcal{L}}

\NewDocumentCommand{\evat}{sO{\bigg}mm}{%
  \IfBooleanTF{#1}
   {\mleft. #3 \mright|_{#4}}
   {#3#2|_{#4}}%
}

\begin{document}

\count\footins = 1000 
\title[
]{Quantum fermion emission from excited kinks}
\author{Sergio Alameda-Calvo}
\email{sergio.alameda@ehu.eus}
\affiliation{Department of Physics, University of the Basque Country UPV/EHU, Bilbao, Spain}
\author{Jose J. Blanco-Pillado}
\email{josejuan.blanco@ehu.eus}
\affiliation{Department of Physics, University of the Basque Country UPV/EHU, Bilbao, Spain}
\affiliation{EHU Quantum Center, University of the Basque Country UPV/EHU, Bilbao, Spain}
\affiliation{IKERBASQUE, Basque Foundation for Science, 48011, Bilbao, Spain}
\author{Alberto García Martín-Caro}
\email{alberto.garcia.martin-caro@uvigo.es}
\affiliation{Department of Physics, University of the Basque Country UPV/EHU, Bilbao, Spain}
\affiliation{EHU Quantum Center, University of the Basque Country UPV/EHU, Bilbao, Spain}
\affiliation{Instituto de Física e Ciencias Aeroespaciais (IFCAE), University of Vigo, 32004 Ourense, Spain}

\date[ Date: ]{\today}
\begin{abstract} 
The amplitude of an excited shape mode in a kink is expected to decay with a well-known power law via scalar radiation emission due to the nonlinear self-coupling of the scalar field. In this work we propose an alternative decay mechanism via pair production of fermions in a simple extension of the $\phi^4$ model in which the scalar field is coupled to a (quantum) fermionic field through a Yukawa-like interaction term. We study the power emitted through fermions as a function of the coupling constant in the semi-classical limit (without backreaction) and compare it to the case of purely scalar radiation emission.

\end{abstract}
\maketitle
\newpage
\begin{minipage}{\textwidth}
\end{minipage}
\section{Introduction}

Fermionic fields play a fundamental role in our current understanding of the particle content of the universe, and, in particular, they have been extensively studied in the context of particle creation processes in the cosmological evolution of the universe, where they could have been produced during inflation  due to cosmological inhomogeneities \cite{Campos:1991ff,Greene:1998nh}. As opposed to their bosonic analogues, fermion fields are inherently quantum in their nature, a defining property that must be taken into account when studying their dynamics. Hence, the mechanism of cosmological fermion production is usually analyzed from a semiclassical approach with techniques of quantum field theory in curved spacetimes \cite{Parker:1971pt,Baacke:1998di,Fuentes:2010dt}.    

On the other hand, topological (and non-topological) solitons appear quite generically in the spectrum of many different, higher-energy extensions of the Standard Model, and may have been formed during the evolution of the Early Universe \cite{Vilenkin_2000}. 
In most of these extensions, such as in supersymmetric
theories, couplings between bosonic and fermionic particles appear rather naturally, which justifies the study of fermionic field dynamics in the presence of defects.

Indeed, it is well known that the presence of a topological soliton generically modifies the spectrum of an otherwise free quantum fermionic field, in a way that may generate the existence of bound states, spatially localized around the soliton \cite{Rubakov:2002fi,Weinberg:2012pjx}. It is natural then to assume that a time-dependent solitonic background would also produce an effect of fermion particle production in a similar fashion as a non-trivial and time-dependent spacetime does. Such effect has been previously addressed (both in the case of bosonic and fermionic radiation) in Q-balls \cite{Cohen:1986ct,Clark:2005zc}, oscillons \cite{Hertzberg:2010yz,Saffin:2016kof,Evslin:2023qbv}, breathers \cite{Olle:2019skb,Mukhopadhyay:2021wmu,Rout:2024kbz} as well as other solitonic solutions \cite{Contri:2025eod}. In this paper, we will analyze a similar process of fermionic radiation but for the case of an excited
topological kink configuration within a 1+1 dimensional toy model.

The model we are going to focus on was firstly proposed by R. Jackiw and C. Rebbi \cite{jackiw-1976}.  In this model, a Dirac fermion  interacts with a background scalar field with a non-trivial topology that takes the form of a kink in (1+1) dimensions. Such model has also attracted some interest in the physics of condensed matter, where the kink models the interface at the boundaries of a topological insulator, with the localized zero modes interpreted as topologically-protected boundary states \cite{Angelakis:2013dia,2024Kumar,jana2019jackiw} . 

The structure of the paper is the following. Firstly, a brief review of the basics of the $\lambda\phi^4$ theory is given in \cref{sec: lambda_phi_4}, with particular emphasis on the kink solution, its spectrum of perturbations and the bosonic decay of its shape mode. In \cref{sec:classicaldyn} we present a model in which a scalar field that admits a kink solution (we focus on the $\phi^4$ model) is coupled to a fermionic field via a Yukawa interaction. We firstly discuss the classical field theory of the fermion in the static kink background (leading to a time-independent Dirac equation) and afterwards, we consider the kink to be excited with its shape mode and characterize the solutions of the resulting time-dependent Dirac equation. Next, in \cref{sec: quant_dyn} we proceed to canonically quantize the fermionic sector of the model, employing the established formalism of quantum fields in non-trivial backgrounds, which allows us to characterize the phenomenon of particle production, indicating the possibility of fermion emission by the excited soliton. In \cref{sec: numerical_results}, numerical simulations are conducted to determine the viability of this new decay channel and identify the conditions or regimes under which it can be taken into consideration. The work ends with some conclusions and some appendices are included in which technical details related to some of the analytical developments are provided.

In this paper we will restrict ourselves to (1+1)-dimensional Minkowski spacetime, and the metric signature is taken to be $g_{\mu\nu}=\text{diag}(+1,-1)$. Furthermore, natural units $(c=\hbar=1)$ will be used, so that all dimensionful quantities have 
dimensions of mass (energy) to some power.

\section{Review of the $\lambda\phi^4$ model}
\label{sec: lambda_phi_4}
The 1+1 dimensional model for a real scalar field we are interested in, the so-called $\lambda\phi^4$ model, is given by the following action:

\begin{equation}
    S=\int d^2x\left(\frac{1}{2}\partial_\mu\phi\partial^\mu\phi-V(\phi)\right)=\int d^2x\left(\frac{1}{2}\partial_\mu\phi\partial^\mu\phi-\frac{\lambda}{4}(\phi^2-\eta^2)^2\right)\,.
    \label{S phi4}
\end{equation}

The positive constants $\lambda$ and $\eta$ represent the quartic self-coupling and the vacuum expectation value of the field, respectively. The Euler-Lagrange equation is
\begin{equation}
    \ddot{\phi}-\phi''+\lambda(\phi^2-\eta^2)\phi=0\,,
    \label{EOM phi4}
\end{equation}
where dots and primes denote differentiation with respect to time and space, respectively. The trivial vacuum solutions of this model are $\phi=\pm\eta$, i.e., the minima of the double-well potential $V(\phi)$. The mass of the (scalar) excitations around one of these vacua is given by $m_s=\sqrt{2\lambda}\eta$. 

Most importantly, this model presents a family of static and non-perturbative solutions, commonly known as \emph{kinks}, that interpolate between the asymptotic vacua of the theory as $\phi(\pm\infty)=\pm\eta$. Their expression is given by
\begin{equation}
    \phi_k(x)=\eta \tanh\left(\sqrt{\frac{\lambda}{2}}\eta(x-x_0)\right)\,,
\end{equation}
where $x_0$ is the free parameter of this set of solutions and marks the position of the kink. By inverting the boundary conditions, that is, imposing $\phi(\pm\infty)=\mp\eta$, one can find a complementary family of solutions, known as \emph{antikinks}.

The energy density of both kinks and antikinks is
\begin{equation}
    \mathcal{E}(x)=\frac{\lambda\eta^4}{2}\sech^4\left(\sqrt{\frac{\lambda}{2}}\eta(x-x_0)\right)\,.
\end{equation}
which is localized around $x_0$, in a region whose width is of the order of $w\sim\left(\sqrt{\lambda}\eta\right)^{-1}$. Its total energy, the classical mass of the kink, is
\begin{equation}
    E=\frac{2\sqrt{2\lambda}}{3}\eta^3\,.
    \label{energy kink}
\end{equation}

In order to find the spectrum of excitations around the kink, let us parametrize the field configuration as the kink solution plus some perturbations, namely
\begin{equation}
    \phi(x,t)=\phi_k(x)+\psi(x,t)\,,
\end{equation}
where we will assume that $|\psi|\ll\langle\phi\rangle=\eta$. Plugging this definition back into (\ref{EOM phi4}), the equation of motion for these perturbations up to a linear order is given by,
\begin{equation}
    \Ddot{\psi}-\psi''+\lambda\left(3\phi_k^2-\eta^2\right)\psi=0\,.
    \label{EOM linear}
\end{equation}

Assuming the fluctuations oscillate in time with some frequency $\omega$, we can use the ansatz $\psi(x,t)\propto e^{-i\omega t}f(x)$ and find that the EOM for the spatial part of the perturbations is
\begin{equation}
    -f''(x)+U(x)f(x)=\omega^2f(x)\,,\qquad U(x)=\lambda(3\phi_k^2-\eta^2)\,.
    \label{EOM f(x)}
\end{equation}

This Schrödinger-like equation is analytically solvable \cite{morse-1953}. Its spectrum is composed by a zero-energy mode and a bound state, followed by an infinite continuum of scattering states.

The zero mode
\begin{equation}
    \quad f_0(x)=N_0\sech^2\left(\frac{m_sx}{2}\right),\quad\omega_0=0\\
\end{equation}
is directly related to small rigid displacements of the position of the kink and reflects the translational invariance of the model. The first excited state
\begin{equation}   
    \quad f_1(x)=N_1\sinh\left(\frac{m_sx}{2}\right)\sech^2\left(\frac{m_sx}{2}\right),\quad\omega_1=\frac{\sqrt{3}m_s}{2}
    \label{shape mode}
\end{equation}
deforms the profile of the kink at its origin. Unlike the zero mode, the position of the kink is not affected, but its width. Because of this, this bound state is generically known in the literature as \emph{shape mode}.

As far as scattering states go, they have the following form
\begin{equation}
f_k(x)=N_ke^{ikx}\left[3\tanh^2\left(\frac{m_sx}{2}\right)-1-w^2k^2-3iwk\tanh\left(\frac{m_sx}{2}\right)\right]\,.
\end{equation}

Their eigenmodes are given by the dispersion relation
\begin{equation}
    \omega_k=\sqrt{k^2+m_s^2}\,,
\end{equation}
where $k>0$ and, thus, the frequency ranges from $m_s$ to infinity. These states will asymptotically tend to plane waves and can be identified as radiative modes. Furthermore
the $N_0$, $N_1$ and $N_k$ constants can also be found from the normalization conditions of each mode.

\subsection{Discussion of the decay of the shape mode}

From the following section onward we will be dealing with the fermionic extension of the $\lambda\phi^4$ theory. Thus, as a closing remark of the review of the scalar case, we will present the decay of the shape mode in the purely scalar model. For our purposes, a qualitative explanation of the decay suffices to get the general ideas, some of which we will bring back during the final part of the work. A more detailed, mathematically rigorous treatment covering all the relevant aspects would constitute a digression from our main objectives. Hence, we refer the interested reader to the original work by Manton and Merabet \cite{manton-1997}, as well as more recent work in \cite{Blanco-Pillado:2020smt,Navarro-Obregon:2023}.

In the final part of this section we introduce a dimensionless formulation of our scalar field theory by implementing the following re-scalings,
\begin{equation}
\phi=\eta\tilde{\phi}\,,\quad~~~ x=\frac{1}{\eta}\sqrt{\frac{2}{\lambda}}\tilde{x}\,.
\end{equation}

With the previous redefinitions, the expression of the kink solution (centered around the origin) is simplified to 
\begin{equation}
    \phi_k(\tilde{x})=\tanh(\tilde{x})\,,
    \label{kink}
\end{equation}
and the (normalized) shape mode becomes,
\begin{equation}
    f_s(\tilde{x})=\sqrt{\frac{3}{2}}\sech{\tilde{x}}\tanh{\tilde{x}}\,.
    \label{shape mode_sp}
\end{equation}

The nonlinear coupling of the shape mode to the radiation modes beyond the linearized approximation is what causes the decay of the shape mode. To see this, one can use the following parametrization of the scalar field
\begin{equation}
    \phi(x,t)=\phi_k(x)+A_s(t)f_s(x)+f(x,t)\,,
    \label{general_config}
\end{equation}
where $f_s(x)$ denotes the profile of the first excited state and $f(x,t)$ accounts for the radiative modes around the kink. If we substitute this field configuration into the dimensionless version of the equation of motion (\ref{EOM phi4}) we find that, at $\mathcal{O}(A_s)$ order, the shape mode has a purely oscillatory behavior with frequency $\omega_s=\sqrt{3}$ and there is no source for radiation. At a quadratic order in $A_s$, the system becomes
\begin{equation}
    (\ddot{A}_s+3A_s)f_s+\ddot{f}-f''+2(\phi_k^2-1)f=-6\phi_kA_s^2f_s^2\,.
    \label{EOM_O(A^2)}
\end{equation}

After projecting both sides onto $f_s$,  and knowing that the eigenstates of this spectral problem are orthogonal to each other, we arrive at
\begin{equation}
    \ddot{A_s}+3A_s=-6\alpha A_s^2,\quad\alpha=\int dx\phi_kf_s^3=\frac{3}{32}\sqrt{\frac{3}{2}}\pi\,.
\end{equation}

By substituting this result back into \cref{EOM_O(A^2)}, one gets the following differential equation for $f$:
\begin{equation}
   \ddot{f}-f''+2(\phi_k^2-1)f= 6(f_s\alpha-\phi_kf_s^2)A_s^2\,.
\end{equation}

Hence, at a quadratic order in the amplitude, the shape mode acts as a source of radiation. Assuming that the amplitude of the shape mode is given by its linear approximation, i.e., $A_s(t)=A(t)\cos(\omega_st)$, the asymptotic solution of the radiation is \cite{manton-1997}
\begin{equation}
    f(x,t)=\frac{3\pi A^2}{2\sinh(\sqrt{2}\pi)}\sqrt{\frac{3}{8}}\cos\left(2\sqrt{3}t-2\sqrt{2}x-\arctan\sqrt{2}\right)\,.
\end{equation}

Thus, we find that the radiation has twice the frequency of the shape mode.  From the previous expression, one can obtain the average energy flux away from the wobbling kink. Furthermore, since this quantity must equal the rate of change of the energy of the excited kink, energy conservation allows us to infer the decay of the shape mode's amplitude, which is given by
\begin{equation}
    A(t)=\frac{A_0}{\sqrt{0.03A_0^2t+1}}\,,
\end{equation}
where $A_0$ is the initial amplitude.

Therefore, in the scalar case, the amplitude of the shape mode exhibits a power-law decay. This behavior has been confirmed by extensive numerical simulations reported in \cite{Blanco-Pillado:2020smt} showing remarkable agreement with the analytical prediction. In the following sections, we turn to the study of the fermionic decay; however, this result will be revisited to enable a direct comparison with the fermionic case.

\section{Field theory of Dirac fermions in time-dependent kinks}
\label{sec:classicaldyn}
\subsection{General considerations}
Consider now a real scalar field interacting with a (massless) Dirac field, $\psi$, through the following Lagrangian density,
\begin{equation}
    \lag = \frac{1}{2}\partial_\mu \phi \partial^\mu \phi +V(\phi)+i\Bar{\psi}\gamma^\mu \partial_\mu \psi-g\phi\Bar{\psi}\psi\,
    \label{Lagrangian}
\end{equation}
where $\gamma^\mu$ are matrices satisfying the Clifford algebra $\{\gamma^\mu,\gamma^\nu\}=2\eta^{\mu\nu}$. In the following, we will choose the representation $\gamma^0=\sigma_1,\gamma^1=i\sigma_3$, where $\sigma_i$ are the corresponding Pauli matrices.

The potential $V(\phi)$ is chosen so that the scalar sector admits a kink solution $\phi_k(x)$. We can consider the spectrum of fermion modes on the scalar field ignoring backreaction. This is a well-justified approximation in the semi-classical limit \cite{Mussardo:2015xua,Weigel:2023fxe}.

The equation of motion for the fermion in the presence of the kink is:
\begin{equation}
    i\gamma^\mu \partial_\mu\psi-g\phi_k(x)\psi=0\,,
    \label{EOM fermion}
\end{equation}
and multiplying the above equation by $\gamma^0$ yields
\begin{equation}
    i\partial_t\psi=H_0\psi\,,\qquad {\rm where}\quad H_0=-i\sigma_2\partial_x+g\phi_k\sigma_1\,
    \label{EOM fermion 2}
\end{equation}
which is the flat spacetime Dirac equation with an effective, spacetime-dependent mass $m_{\rm eff}(x)=g\phi_k(x)$. This is analogous as if we had considered fermions in the vacuum sector but on a generally curved background \cite{Koke:2016etw}.

The operator $H_0$ can always be diagonalized by a set of eigenfunctions $\{\psi^+_n,\psi_n^-\}$:
\begin{equation}
    H_0 \psi^{\pm}_n=\pm E_n\psi^{\pm}_n\,, \qquad E_n>0\,.
    \label{eigenvalue p}
\end{equation}

Depending on the spectrum of $H_0$, we can find three different types of modes:
\begin{enumerate}
    \item \emph{Zero mode}: A mode characterized by $E_0=0$, i.e. the mode does not evolve in time.
    \item \emph{Normal modes}: A finite, discrete set of eigenfunctions $\psi_n$, with $n=1,2,\cdots N$ whose corresponding energies, $E_n$, are smaller than the mass of the fermions far away from the kink. We denote this energy threshold by $E_m$. 
    \item \emph{Scattering states}: An infinite, continuous set of eigenfunctions $\psi_k$ parametrized by a wavenumber $k\in \mathbb{R}$. The corresponding energies $E_k$ are larger than the mass threshold $E_m$.
\end{enumerate}

The zero and normal modes are only present due to the non-trivial kink background. In the case of a one-kink sector, they are typically localized around the soliton center.

A scalar product can be defined in the space of solutions $\mathcal{S}$ spanned by the eigenfunctions of $H_0$, also known as Dirac product:
\begin{equation}
    \langle \psi,\phi\rangle_D\doteq\int\Bar{\psi}(x)\gamma^0\phi (x)dx\equiv \int \psi^\dagger (x)\phi(x) dx\,.
    \label{D_prod}
\end{equation}

Since $H_0$ does not depend on time, the evolution of any fermion field configuration $\psi(x,t)$ can be trivially obtained by its expansion on $H_0$ eigenmodes:

\begin{equation}
    \psi(x,t)=a_0\psi_0(x)+\sumint dk\qty[b_k\psi_k^+(x)e^{-iE_kt}+d_k\psi_{k}^{-}(x)e^{iE_kt}]\,,
\end{equation}
where
\begin{equation}
    \sumint dk=\int dk +\sum\limits_n ~,
    \label{sumint}
\end{equation}
denotes both integration over scattering modes and sum over the discrete normal modes. All these modes are normalized under the Dirac product \footnote{For simplicity, we have collectively denoted by $\delta_{kk'}$ the Kronecker delta in case both states correspond to bound states, or the Dirac delta $\delta(k-k')$ if, on the other hand, we are dealing with scattering states. }:
\begin{equation}
    \langle \psi^{r}_k,\psi^{s}_{k'} \rangle _D=\delta_{rs}\delta_{kk'}~ ,
    \label{normalization}
\end{equation}
and satisfy a completeness relation:
\begin{equation}
    \sumint dk[\psi^+_k(x')\psi^{+\dagger}_k(x)+\psi^-_k(x')\psi^{-\dagger}_k(x)]=\delta(x'-x)~.
\end{equation}

We will be interested in the non-trivial evolution of the Dirac field under a perturbed kink. Let us now introduce a space and time-dependent scalar perturbation of the kink profile as before
\begin{equation}
    \phi(x,t)=\phi_k(x)+\varphi(x,t)~,
    \label{time dep perturbation}
\end{equation}

which is now switched on at a finite time in the past and vanishes in the asymptotic future; that is, the perturbation is active only over a finite, transient interval such that,
\begin{equation}
    \lim\limits_{t\to \pm \infty}\varphi(x,t)=0~.
    \label{on off switch}
\end{equation}

The new equation of motion for the Dirac field will be 
\begin{equation}
    i\partial_t\psi=H_D\psi\equiv H_0\psi+g\varphi\sigma_1\psi~.
    \label{EOM time dep}
\end{equation}

In the asymptotic past, a complete set of eigenfunctions $\{\psi^+_n(x),\psi_n^-(x)\}$  of the time-independent Hamiltonian can be found. Furthermore, integrating the time-dependent Dirac equation forward in time using such modes as the initial condition gives a set of solutions $\{\psi^{(\rm in)+}_n(x,t),\psi_n^{(\rm in)-}(x,t)\}$ satisfying the boundary condition:
\begin{equation}
    \psi^{(\rm in)\pm}_n(x,t)\xrightarrow[t\to -\infty]{}\psi^\pm_n(x)e^{\mp iE_n t}
\end{equation}
and the same can be done for the asymptotic future:
\begin{equation}
        \psi^{(\rm out)\pm}_n(x,t)\xrightarrow[t\to +\infty]{}\psi^\pm_n(x)e^{\mp iE_n t}~.
\end{equation}

Both the ``in'' and ``out'' sets of modes form complete, orthonormal bases of the space of solutions $\mathcal{S}$, since the Dirac product defined in \cref{D_prod} does not depend on time. However, these two sets are generally different. Thus, we have found two different sets of basis functions in terms of which we can express the general solution of the full, time-dependent Dirac equation \eqref{EOM time dep}. The physical meaning of these solutions and the connection between them will become clear once we deal with the canonical quantization of the Dirac field. 

\subsection{\texorpdfstring{$\lambda\phi^4$ model}{φ4 model}}
Let us now apply the general formalism presented above to a simple fermionic extension of the previously reviewed $\lambda\phi^4$ model, whose action is given by
\begin{equation}
    \mathcal{S}=\int d^2x\left(\frac{1}{2}\partial_\mu\phi\partial^\mu\phi - \frac{\lambda}{4}(\phi^2-\eta^2)^2+i\Bar{\psi}\gamma^\mu \partial_\mu \psi-g\phi\Bar{\psi}\psi\right).
\end{equation}

This model presents two mass scales in the vacuum \cite{rajaraman1982solitons}. Firstly, we have the mass of the scalar perturbations $m_s=\sqrt{2\lambda}\eta$. Secondly, for the fermionic sector, the Yukawa interaction generates a mass term of the form $m_f=g\eta$.

Before solving the Dirac equation, we first extend the change to dimensionless variables introduced in Section 2.1 to include the fermionic part, namely,
\begin{equation}
    \tilde{g}=\frac{2m_f}{m_s}=\sqrt{\frac{2}{\lambda}}g\,, \quad 
    \psi=\left(\frac{\lambda}{2}\eta^6\right)^{1/4}\tilde{\psi}\,.
\end{equation}

 By doing so, the previous action becomes
\begin{equation}
    S=\eta^2\int d\tilde{x}^2\left(\frac{1}{2}(\tilde{\partial}_\mu\tilde{\phi})^2-\frac{1}{2}(\tilde{\phi}^2-1)^2+i\tilde{\Bar{\psi}}\gamma^\mu\tilde{\partial}_\mu\tilde{\psi}-\tilde{g}\tilde{\phi}\tilde{\Bar{\psi}}\tilde{\psi}\right)\,.
\end{equation}

Consequently, the only parameter that enters our theory is the dimensionless coupling constant $\tilde{g}$, defined now as a scaled quotient between the masses of the fermion and scalar perturbations. From now onward, unless explicitly stated, we will be using these dimensionless quantities. tildes will be dropped for notational ease.

\subsubsection{Static solutions to the Dirac equation}
In order to find the fermion modes, we must solve the Dirac equation for our specific kink solution. Solutions of this problem for the model in consideration are well known; for completeness, the full derivation is presented in \cref{Appendix: A}. Here, we simply summarize the results as follows:

\begin{itemize}
    \item There exists a zero mode, $E_0=0$, corresponding to a single, non-degenerate eigenfunction:
    
    \begin{equation}
        \psi_0=N^{(0)}\begin{pmatrix}
            \cosh^{-g}{x}\\
0
        \end{pmatrix}\,.
    \end{equation}
  
   \item In addition to the zero mode there is a finite discrete set of bound states with energies 
   $E_n=\sqrt{n(2g-n)}$ defined for integer values of $n$ up to the largest integer strictly smaller than $g$. This bound defines the threshold separating discrete modes from scattering states. The corresponding eigenfunctions take the form 
    \begin{equation}
        \psi_n^+=N^{(n)}(e^x+e^{-x})^{n-g}\begin{pmatrix}
             F(-n,2g-n+1,g-n+1,\frac{e^{-x}}{e^x+e^{-x}})\\
    \frac{n}{E_n}F(-n+1,2g-n,g-n+1,\frac{e^{-x}}{e^x+e^{-x}})
        \end{pmatrix}\,.
        \label{bound_modes}
    \end{equation}

\item Once above the energy threshold, the scattering states form an infinite, continuous set of eigenfunctions of energies $E_k=\sqrt{k^2+g^2}$, characterized by a wavenumber $k>0$. The form of these continuous states is given by
\begin{equation}
        \psi_k^+=N^{(k)}(e^x+e^{-x})^{ik}\begin{pmatrix}
             F(-ik-g,-ik+g+1,1-ik,\frac{e^{-x}}{e^x+e^{-x}})\\
    \frac{ik+g}{E_k}F(-ik-g+1,-ik+g,1-ik,\frac{e^{-x}}{e^x+e^{-x}})
        \end{pmatrix}\,.
        \label{scatt_modes}
\end{equation}
\end{itemize}

In the above expressions, the constant $N^{j}$ in front of each mode is the normalization constant. The normalization procedure is also discussed in \cref{Appendix: A}. Moreover, the function $F$ in \cref{bound_modes,scatt_modes} denotes the Gaussian (ordinary) hypergeometric function.

\begin{figure}[hbt!]
    \centering
    \includegraphics[width=0.7\linewidth]{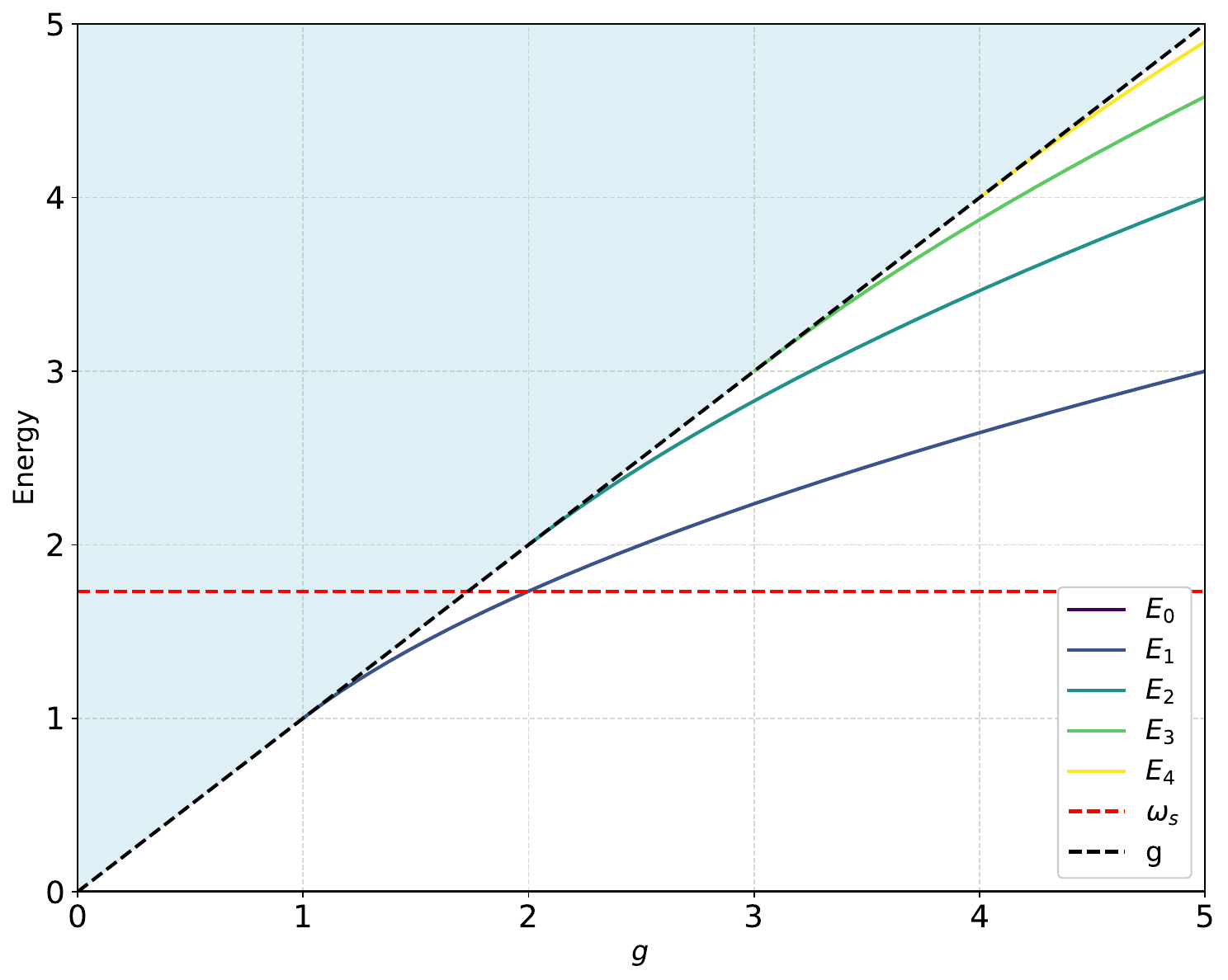}
    \caption{Energy spectrum of the time-independent Dirac equation in the kink background as a function of $g$. Independently of the value of $g$, a zero fermion mode will always be present. Whenever $g$ surpasses an integer value, a new bound fermion mode can be found. Above the mass threshold, represented by a dashed black line, scattering fermion modes exist. The energy of the shape mode is represented by a dashed red line.}
    \label{fig:energies graphs} 
\end{figure}

The spectrum of the Dirac Hamiltonian in the kink background is shown in \cref{fig:energies graphs} as a function of the Yukawa coupling. The figure illustrates the emergence of additional bound states, which branch off at specific integer threshold values of $g$.

It is worth emphasizing that the asymptotic behavior of these continuous states at spatial infinities is

\begin{equation}
u_k^+=
\begin{cases}
N_u^{(k)}\left[ \frac{ \Gamma(1 - ik) \Gamma(-ik) e^{ikx}}{\Gamma\left(-ik-g\right) \Gamma\left(-ik+g+1\right)} + \frac{ \Gamma(1 - ik) \Gamma(ik) e^{-ikx}}{\Gamma\left(g+1\right) \Gamma\left(-g\right)} \right]\,, & x \to -\infty\,, \\
N_u^{(k)} e^{ikx}\,, & x \to +\infty\,,
\end{cases}
\label{asymptotic behaviour 1}
\end{equation}
and
\begin{equation}
v_k^+=
\begin{cases}
N_v^{(k)}\left[ \frac{ \Gamma(1 - ik) \Gamma(-ik) e^{ikx}}{\Gamma\left(-ik-g+1\right) \Gamma\left(-ik+g\right)} + \frac{ \Gamma(1 - ik) \Gamma(ik) e^{-ikx}}{\Gamma\left(g\right) \Gamma\left(1-g\right)} \right]\,, & x \to -\infty\,, \\
N_v^{(k)} e^{ikx}\,, & x \to +\infty\,.
\end{cases}
\label{asymptotic behaviour 2}
\end{equation}

In both cases this behavior represents an incident wave coming from $x\to-\infty$ moving to the right, a reflected wave going back to $x\to-\infty$ and a transmitted wave moving to $x\to\infty$. Furthermore, since the gamma function diverges for any negative integer as well as 0, whenever $g$ is integer valued, the coefficient multiplying $e^{-ikx}$ vanishes in both components of the spinor, so there are no reflected states. In other words, whenever $g$ is integer valued, the potentials (\ref{potentials fermion}) are reflectionless. 

Finally, for future reference, let us remark that, since positive and negative energy spinors have the same expression up to a minus sign in the energy, they can be related via the following operation:
\begin{equation}
     \psi_{n,k}^-=\sigma_3\psi_{n,k}^+ ~.
     \label{+- relation}
\end{equation}

\subsubsection{Solutions to the time-dependent Dirac equation}

To investigate the decay of an excited kink configuration, we introduce a time-dependent perturbation of the kink background corresponding to the shape mode. In dimensionless quantities, the full expression for the shape mode is given by
\begin{equation}
    \varphi_s(x,t)=\Re(e^{-i\omega_s t})f_s(x)=\cos{(\omega_st)}\sqrt{\frac{3}{2}}\sech{x}\tanh{x}\,,
\end{equation}
where, as we described earlier, $\omega_s=\sqrt{3}$. Hence, the complete expression for the scalar field is
\begin{equation}
    \phi(x,t)=\phi_k(x)+\varphi_s(x,t)=\tanh{x}+ F(t) \cos{(\omega_st)}\sqrt{\frac{3}{2}}\sech{x}\tanh{x}\,.
\end{equation}

As outlined previously, we are interested in the scenario in which the perturbation is activated for a finite interval of time \footnote{The explicit form of the switching function $F(t)$ will be specified later in the text.}. In this setting, the kink undergoes oscillations at frequency $w_s=\sqrt{3}$ due to the influence of the shape mode, while approaching a static configuration in the asymptotic limits $t\to\pm\infty$ . It is precisely this oscillatory behavior that induces fermionic particle production. To investigate this mechanism, one must solve the time-dependent Dirac equation \eqref{EOM time dep} with $\varphi\equiv\varphi_s(x,t)$.

In order to proceed, let us consider the most general time-dependent solution for the fermion field as
an expansion of the form,
\begin{equation}
    \psi(x,t)=\sumint  dk\left[\xi_k(t)\psi^+_k(x)+\eta_k(t)\psi_k^-(x)\right]\,,
    \label{time dep expansion}
\end{equation}
where $\xi_k(t)$ and $\eta_k(t)$ are time-dependent functions that fulfill the following dynamical equations \footnote{A detailed derivation of these coupled differential equations is carried out in Appendix \ref{Appendix: B}. }
\begin{align}
    \quad i  \dot{\xi}_{k}(t) -  \xi_{k}(t) E_{k} - g \sumint  dk'\left[  \xi_{k'}(t) R_{kk'}(t) +  \eta_{k'}(t) Q_{kk'}(t) \right] = 0\label{dyn eq1} \,,\\
     \quad i  \dot{\eta}_{k}(t) +  \eta_{k}(t) E_{k} + g \sumint  dk'\left[  \xi_{k'}(t) Q_{kk'}(t) + \eta_{k'}(t) R_{kk'}(t) \right] = 0\,,\label{dyn eq2}
\end{align}
with $Q_{kk'}$ and $R_{kk'}$ defined as
 \begin{align}
    Q_{kk'}=\int dx (\psi_{k}^+(x))^\dagger \varphi_s(x,t) \sigma_1 \psi_{k'}^-(x)\,,\label{Qmatrix}\\
    R_{kk'}=\int dx (\psi_{k}^+(x))^\dagger \varphi_s(x,t) \sigma_1 \psi^+_{k'}(x)\,.\label{Rmatrix}
\end{align}

Therefore, the knowledge of the spectrum of the time-independent Dirac equation allows us to reduce the time-dependent problem to a set of (infinitely many) coupled, first-order ordinary differential equations.
We note that the spatial non-homogeneity of the background perturbation implies that, even though the equations of motion for the mode amplitudes are linear, there is a non-trivial mode mixing that couples different modes in the time evolution. Indeed, were the perturbation $\varphi_s$ just a function of time, the mode mixing matrices (\cref{Qmatrix,Rmatrix}) would become proportional to the identity and each mode would evolve independently. Such mode mixing phenomenon is reminiscent of other instances in which the background breaks spatial homogeneity, for example in the case of particle creation due to cosmological inhomogeneities (see \cite{Garani:2025qnm} for a recent review) or in finite cavities \cite{GarciaMartin-Caro:2024qpk}. 

The only remaining requirement to solve the time evolution is to specify the initial conditions for $\xi_k(t)$ and $\eta_k(t)$. 
The time-dependent solutions we are interested in are $\psi_q^{\rm{(in)}\pm}(x,t)$ and $\psi_q^{\rm{(out)}\pm}(x,t)$, which, as explained at the beginning of the section, tend to the static solutions in the asymptotic past and future, respectively. 
In particular, let us consider the $\psi_q^{\rm{(in)}+}(x,t)$ mode. As with any other time-dependent solution, we can expand it as in eq. (\ref{time dep expansion}),
\begin{equation}
    \psi_q^{\rm{(in)}+}(x,t)=\sumint  dk\left[\xi^q_k(t)\psi^+_k(x)+\eta^q_k(t)\psi_k^-(x)\right]\,.
    \label{in + decomposition}
\end{equation}

It can be seen that, in order $\psi_q^{\rm{(in)}+}(x,t)$ to become the static solution $\psi_q^+(x)$ at the asymptotic past, the time-dependent functions must obey
\begin{equation}
    \xi^q_k(t=-\infty)=\delta_{qk}\quad\text{and}\quad\eta_k^q(t=-\infty)=0\,.
    \label{ICs plus}
\end{equation}

Equivalently, the decomposition of $\psi_q^{\rm{(in)}-}(x,t)$ is
\begin{equation}
    \psi_q^{\rm{(in)}-}(x,t)=\sumint  dk\left[\xi^q_k(t)\psi^+_k(x)+\eta^q_k(t)\psi_k^-(x)\right]\,.
    \label{in - decomposition}
\end{equation}

Since this mode must tend to $\psi_q^-(x)$ at $t\to-\infty$, the initial conditions in the case of $\psi_q^{\rm{(in)}-}(x,t)$ should be
\begin{equation}
    \xi^q_k(t=-\infty)=0\quad\text{and}\quad\eta_k^q(t=-\infty)=\delta_{qk}\,.
    \label{ICs minus}
\end{equation}

It is worth noting that, for any given mode from the set of $out$ modes $\{\psi^{\rm{(out)}\pm}_q\}$, the conditions that $\xi_k^q(t)$ and $\eta_k^q(t)$ must satisfy are the same ones as (\ref{ICs plus}) and (\ref{ICs minus}), but evaluated at the asymptotic future \footnote{Just for clarification, the superindex $q$ on the time-dependent functions $\xi_k^q(t)$ and $\eta_k^q(t)$ is used to label the mode one is doing the expansion of, nothing else. Hence, whenever a system of dynamical equations has to be solved, the superindex will be the same for all the functions.}.

\section{Quantum field theory of fermions in a kink background}
\label{sec: quant_dyn}
As it is well known, fermions obey Pauli's exclusion principle, which states that two of them cannot occupy the same quantum state. The only way to take this into account in a consistent manner is to further analyze our previous model through the framework of quantum field theory. This section is divided into two parts. In the first one, we review the canonical quantization of the previous model. After the quantization is carried out, the second part addresses the fermion production phenomenon in the background of a classical excited kink.
\subsection{Canonical quantization of the Dirac Lagrangian in a non-trivial scalar background }
The effective Lagrangian density for a Dirac field with a classical scalar Yukawa source is
\begin{equation}
    \mathcal{L}=i\Bar{\psi}\gamma^\mu\partial_\mu\psi-g\phi(x,t)\Bar{\psi}\psi\,.
\end{equation}

One can associate the following canonical momenta to the fields $\psi$ and $\Bar{\psi}$
\begin{equation}
    \pi_\psi=\frac{\partial\mathcal{L}}{\partial\Dot{\psi}}=i\Bar{\psi}\gamma^0=i\psi^\dagger\,,\qquad\pi_{\Bar{\psi}}=\frac{\partial\mathcal{L}}{\partial\Dot{\Bar{\psi}}}=0\,.
\end{equation}

With the canonical momenta obtained, one can compute the Hamiltonian density of the theory by means of a Legendre transformation
\begin{equation}
     \mathcal{H}=\pi_\psi\Dot{\psi}+\pi_{\Bar{\psi}}\Dot{\Bar{\psi}}-\mathcal{L}=-i\pi_\psi H_D\psi\,,
\end{equation}
where $H_D$ is the time-dependent Dirac Hamiltonian appearing in (\ref{EOM time dep}), 
\begin{equation}
    H_D=-i\gamma^0\gamma^1\partial_1+\gamma^0g\phi(x,t)\,.
\end{equation}

Hence, the Hamiltonian of the theory can be written as
$
    H=\int dx\mathcal{H}=-i\int dx \pi_\psi H_D\psi.
   \label{Hamiltonian}
$ The canonical quantization of the theory is achieved by promoting the Dirac spinors $\psi$ and $\Bar{\psi}$ to field operators $\Hat{\psi}$ and $\Hat{\Bar{\psi}}$, as well as replacing the conjugate momenta $\pi_\psi$ and $\pi_{\Bar{\psi}}$ by their corresponding quantum operators $\Hat{\pi}_\psi$ and $\Hat{\pi}_{\Bar{\psi}}$, respectively. All these operators will satisfy equal time Canonical Anti-Commutation Relations (CARs):
\begin{equation}
    \{\hat\psi(x,t),\hat\pi_\psi(x',t)\}\equiv i\{\hat\psi(x,t),\hat\psi^\dagger(x',t)\}=i\delta(x-x')\,.
\label{CARs}
\end{equation}

On the other hand, working in the Heisenberg picture, the time evolution of the field operator $\Hat{\psi}(x,t)$ can be described via Heisenberg's equation of motion:
\begin{equation}
 \partial_t\hat\psi(x,t)=i[\hat{H}(x',t),\hat\psi(x,t)]=-   iH_D\hat{\psi}(x,t)\,.
 \label{Heis EOM}
\end{equation}

Thus, the field operator $\Hat{\psi}$ is found to satisfy the time-dependent Dirac equation (\ref{EOM time dep}) too. This comes in handy, since then $\Hat{\psi}$ can be expanded in terms of classical solutions of our Dirac equation. In particular, let us consider a general set of solutions of the Dirac equation, $\{\psi_k(x,t)\}$, that form a complete basis. In that case, one can expand the field operator and its conjugate as
\begin{equation}
    \hat{\psi}(x,t)=\sumint dk \hat{b}_k\psi_k(x,t)\quad\text{and} \quad \hat{\psi}^\dagger(x,t)=\sumint dk \hat{b}^\dagger_k\psi^\dagger_k(x,t)\,,
\end{equation}
where $\hat{b}_k$ and $\hat{b}^\dagger_k$ are the creation and annihilation operators of fermionic particles in a state $k$. The expressions for $\hat{b}_k$ and $\hat{b}^\dagger_k$ can be found by projecting $\Hat{\psi}(x,t)$ on a solution of the previous set, i.e.,
\begin{equation}
    \int dx \psi_q^\dagger(x,t)\hat{\psi}(x,t)=\sumint dk \hat{b}_k\int dx \psi_q^\dagger(x,t)\psi_k(x,t)=\sumint dk \hat{b}_k\delta_{qk}=\hat{b}_q\,,
\end{equation}
and $\hat{b}_q^\dagger$ is just the Hermitian conjugate of the above.

Using this results, equivalent CARs for the creation/annihilation operators can be constructed:
\begin{equation}
    \{\hat{b}_k,\hat{b}_q^\dagger\}=\delta_{kq}\,,\quad\{\hat{b}_k,\hat{b}_q\}=\{\hat{b}_k^\dagger,\hat{b}_q^\dagger\}=0\,.
    \label{CARs 2}
\end{equation}

It is worth mentioning that even if $\hat{b}_k$ and $\hat{b}^\dagger_k$ are the creation and annihilation operators of fermionic particles in a state $k$, in our case the intrinsic notion of particles can be ambiguous. In fact, if we were in flat QFT with free fields, the vacuum state would be uniquely defined, and since it would be the same at any instant of time, one could use the same set of creation/annihilation operators such that the notion of particles would be associated to eigenstates which would be solutions of the given equation of motion. There would not be any ambiguous notion neither of vacuum nor of particles \cite{fulgado-claudio-2023}.

However, in our case, where a non-trivial background suffers a time evolution, the notion of particles becomes ambiguous, as the vacuum is not unique (since it becomes coordinate dependent). Even if we can find a proper set of operators $\{\hat{b}_k/\hat{b}_k^\dagger\}$ associated to a set of solutions of the Dirac equation at a specific instant of time, after sufficient time has elapsed, those eigenstates will no longer be solutions of the Dirac equation. Therefore, the easiest way to deal with this problem is to restrict ourselves to consider creation and annihilation operators of particles only in asymptotic times, when the background is static.

Nevertheless, a problem still remains: even in the static scenario, the spectrum of the Hamiltonian is not bounded from below and, as a consequence, the energy of the system can have arbitrarily large negative values. In order to address this problem, the concept of antiparticle is introduced. The space of solutions of the classical Dirac equation can be divided into two subspaces containing, respectively, the positive and negative energy eigenfunctions.

Because of the division, the Fock space, $\mathcal{F}$, also gets split: $\mathcal{F}=\mathcal{F}^+\oplus\mathcal{F}^-$. Due to this, any operator acting on it can be expressed as a tensor product between operators acting in each of the split spaces, in particular,
\begin{equation}
\hat{b}_k \rightarrow 
\begin{cases} 
\hat{b}_k \otimes \bm{1} & (E_k > 0) \\ 
 \bm{1} \otimes \hat{b}_k & (E_k < 0)
\end{cases}\,,
\end{equation}
and the same goes for $\hat{b}_k^\dagger$. The antiparticle creation and annihilation operators, denoted as $\hat{d}_k^\dagger/\hat{d}_k$ respectively, are thus defined as
\begin{equation}
    \hat{d}^\dagger_k\equiv \hat{b}_k\Bigr|_{\mathcal{F}^-} \quad\text{and}\quad \hat{d}_k\equiv \hat{b}^\dagger_k\Bigr|_{\mathcal{F}^-}\,,
    \label{antipart_def}
\end{equation}
which means that creating an antiparticle is equivalent to destroying a particle in the negative energy Fock space and vice-versa, destroying an antiparticle is analogous to creating a particle in $\mathcal{F}^-$. Note that the redefinition \eqref{antipart_def} leaves the CARs untouched, but allows us to write a Hamiltonian with a spectrum that is bounded from below by normal-ordering with respect to the two sets of creation and annihilation operators:
\begin{equation}
    :\hat{H}:=\sumint dk|E_k|\qty(\hat{d}^\dagger_k \hat{d}_k+\hat{b}^\dagger_k \hat{b}_k)\,.
\end{equation}

Moreover, the field operator in the Heisenberg picture gets split into two contributions involving positive and negative energies,
\begin{equation}
    \hat{\psi}(x,t)=\sumint dk[\hat b_k \psi^+_k(x)e^{-iE_k t}+\hat d^\dagger_k \psi^-_k(x)e^{iE_k t}]\,.
\end{equation}
and the vacuum state of the theory is defined as the element of the Fock space that satisfies
$
    \hat{b}_k\ket{0}=\hat{d}_k\ket{0}=0,\,\forall k,
$
while the rest of the spectrum is constructed by the consecutive application of $\hat{b}_k^\dagger$ and $\hat{d}_k^\dagger$ on the ground state.

\subsection{Bogoliubov transformations and fermion production}

In this subsection we will explore how the quantization of the theory results in (fermion) particle production by means of the so-called Bogoliubov coefficients. For that sake, let us recall that in \cref{sec:classicaldyn} we presented two sets of time-dependent basis functions, $\{\psi^{(\rm in)\pm}_k(x,t)\}$ and $\{\psi^{(\rm out)\pm}_k(x,t)\}$. Consequently, the field operator can be decomposed in terms of these eigenstates in two possible ways:
\begin{align}
    \hat{\psi}(x,t)&=\sumint dk[\hat b_k^{(\rm in)} \psi^{(\rm in)+}_k(x,t)+\hat d^{(\rm in) \dagger}_k \psi^{(\rm in)-}_k(x,t)]\,,\\
     \hat{\psi}(x,t)&=\sumint dk[\hat b^{(\rm out)}_k \psi^{(\rm out)+}_k(x,t)+\hat d^{(\rm out) \dagger}_k \psi^{(\rm out)-}_k(x,t)]\,.
\end{align}

As a result of these decompositions, two different vacuum states can be identified: the one corresponding to the set of ingoing modes, $\ket{0;\rm in}$, and the one corresponding to the set of outgoing modes, $\ket{0;\rm out}$. The important point is that the $in$ and $out$ modes can be related as follows
\begin{align}
    \psi_k^{(\rm out)+}(x,t)=\sumint dk\qty[\alpha_{kq} \psi_q^{(\rm in)+}(x,t)+\beta_{kq}\psi_q^{(\rm in)-}(x,t)]\,,\\
    \psi_k^{(\rm out)-}(x,t)=\sumint dk\qty[\beta_{kq}^*\psi_q^{(\rm in)+}(x,t)+\alpha_{kq}^* \psi_q^{(\rm in)-}(x,t)]\,,
\end{align}
where $\alpha_{kq}$ and $\beta_{kq}$ are the so-called Bogoliubov coefficients, defined as
\begin{equation}
    \alpha_{kq}=\expval*{\psi^{(\rm in)+}_q,\psi^{(\rm out)+}_k}_D\qquad\text{and}\qquad \beta_{kq}=\expval*{\psi^{(\rm in)-}_q,\psi^{(\rm out)+}_k}_D\,.
    \label{alphabetadefs}
\end{equation}

The derivation of the Bogoliubov coefficients is shown in Appendix \ref{Appendix: C}. Just as with the $in$ and $out$ modes, one can also build a relation between the creation/annihilation operators associated to the two different representations as follows:
\begin{align}
    \hat{b}_k^{\rm (in)}=\sumint dq\qty[\alpha_{kq}\hat{b}_q^{\rm (out)}+\beta^*_{kq}\hat{d}_q^{\rm (out)\dagger}]\label{operator b in:1}\,,\\
    \hat{d}_k^{\rm (in)}=\sumint dq\qty[\beta^*_{kq}\hat{b}_q^{\rm (out)\dagger}+\alpha_{kq}\hat{d}_q^{\rm (out)}]\label{operator d in:2}\,.
\end{align}

 These are called Bogoliubov transformations. Additionally, in order the creation/annihilation operators to satisfy the canonical anti-commutation relations (\ref{CARs 2}), two closure relations must be satisfied \footnote{Both the Bogoliubov transformations and the closure relations are derived in Appendix \ref{Appendix: C}}:
 \begin{equation}
     \sumint dk(\alpha^*_{kq}\alpha_{kp}+\beta^*_{kq}\beta_{kp})=\delta_{qp},\qquad\text{and}\qquad \sumint dk (\alpha_{kq} \beta_{kp}^*+\beta_{kq}^* \alpha_{kp})=0\,.
\label{clos_rels}
 \end{equation}

A direct consequence of the Bogoliubov transformations is that the notion of vacuum is not unique for a quantized field defined on a non-trivial background. To see this, we will assume that in the asymptotic past our system is empty, i.e. the vacuum state is defined as
\begin{equation}
    \hat{b}_k^{\rm (in)}\ket{0;\rm in}=\hat{d}_k^{\rm (in)}\ket{0;\rm in}=0\,,\quad\text{where}\quad\bra{0;\rm in}\ket{0;\rm in}=1\,.
\end{equation}

If we let the system evolve in time according to the time-dependent Dirac equation, due to the time evolution under the excited kink, we expect to encounter a non-zero amount of fermion particles and antiparticles in the asymptotic future, when the wobbling stops. For that, we define the expected number of fermion particles in the asymptotic future as
\begin{equation}
    n_{b,k}=\ev{\Hat{n}_{b,k}^{\rm (out)}}{0;\rm in}=\ev{\hat{b}_k^{\rm(out)\dagger}\hat{b}_k^{\rm(out)}}{0;\rm in}\,,
\end{equation}
and in the case of antiparticles:
\begin{equation}
    n_{d,k}=\ev{\Hat{n}_{d,k}^{\rm (out)}}{0;\rm in}=\ev{\hat{d}_k^{\rm(out)\dagger}\hat{d}_k^{\rm(out)}}{0;\rm in}\,,
\end{equation}
where $\Hat{n}_{b,k}^{\rm (out)}=\hat{b}_k^{\rm(out)\dagger}\hat{b}_k^{\rm(out)}$ and $\Hat{n}_{d,k}^{\rm (out)}=\hat{d}_k^{\rm(out)\dagger}\hat{d}_k^{\rm(out)}$ are the fermion and antifermion number operators (for a given mode $k$) in the asymptotic future, respectively.

The Bogoliubov transformations allow us to relate creation/annihilation operators from different representations, so that the fermion number operator can be rewritten as
\begin{equation}
\begin{split}
    &\Hat{n}_{b,k}^{\rm (out)}=\hat{b}_k^{\rm(out)\dagger}\hat{b}_k^{\rm(out)}=\\
    &=\sumint dq\left(|\alpha_{kq}|^2\hat{b}_q^{\rm(in)\dagger}\hat{b}_q^{\rm(in)}+|\beta_{kq}|^2\hat{d}_q^{\rm(in)}\hat{d}_q^{\rm(in)\dagger}+\alpha_{kq}^*\beta_{kq}\hat{b}_q^{\rm(in)\dagger}\hat{d}_q^{\rm(in)\dagger}+\alpha_{kq}\beta_{kq}^*\hat{b}_q^{\rm(in)}\hat{d}_q^{\rm(in)}\right)\,.
\end{split}
\end{equation}

The same process can be carried out for the antiparticle case, yielding
\begin{equation}
\begin{split}
    &\Hat{n}_{d,k}^{\rm (out)}=\hat{d}_k^{\rm(out)\dagger}\hat{d}_k^{\rm(out)}=\\
    &=\sumint dq\left(|\beta_{kq}|^2\hat{b}_q^{\rm(in)}\hat{b}_q^{\rm(in)\dagger}+
    |\alpha_{kq}|^2\hat{d}_q^{\rm(in)\dagger}\hat{d}_q^{\rm(in)}+
    \beta_{kq}^*\alpha_{kq}\hat{b}_q^{\rm(in)}\hat{d}_q^{\rm(in)}+
    \alpha_{kq}^*\beta_{kq}\hat{d}_q^{\rm(in)\dagger}\hat{b}_q^{\rm(in)\dagger}\right)\,.
\end{split}
\end{equation}

As a result, the expected number of fermion and antifermion particles for a given mode $k$ in the asymptotic future is expressed as 
\begin{align}
    n_{b,k}&=\ev{\hat{b}_k^{\rm(out)\dagger}\hat{b}_k^{\rm(out)}}{0;\rm in}=\sumint dq|\beta_{kq}|^2\,,\label{probability p}\\
    n_{d,k}&=\ev{\hat{d}_k^{\rm(out)\dagger}\hat{d}_k^{\rm(out)}}{0;\rm in}=\sumint dq|\beta_{kq}|^2\label{probability ap}\,,
\end{align}
respectively. Therefore, unless the Bogoliubov coefficient $\beta_{kq}$ is null for every $q$, the vacuum state $\ket{0;\rm in}$ will contain particles in the asymptotic future, leading to the occurrence of particle creation phenomena. Besides, the fact that both quantities have the same expression is an immediate result of the system conserving the lepton number while evolving in time and that, if the values are high enough, the kink will radiate away particle-antiparticle pairs.

Thus, as we have just seen, different choices of representation (that is, $in$ and $out$ bases) will result in different notions of vacuum and, as a consequence, distinct notions of particles. Although this outcome is a common characteristic of quantum field theory formulated in curved spacetimes \cite{mukhanov-2007}, our case is a perfect example of how particle production can occur even in flat quantum field theories (with a non-trivial background) as well.

It is noteworthy that, due to the first of the closure relations (\ref{clos_rels}), nor $n_{b,k}$ neither $n_{d,k}$ can be higher than 1. Hence, rather than as expected numbers of particle/antiparticles, we can treat both quantities as probability densities of finding a fermion particle/antiparticle in a certain state labelled by $k$ in the asymptotic future, when the time-dependent perturbation has finally stopped.
 
Another key point is that the Bogoliubov coefficients can be related with the time-dependent functions $\eta^q_k$ and $\xi^q_k$ defined in \cref{sec:classicaldyn}. Indeed, from the definition of $\beta_{qk}$ (\cref{alphabetadefs}), and knowing that in the asymptotic future $\psi^{\rm(out)+}_k(x,t\to \infty)\to \psi^+_k(x)$ and that $\psi^{\rm(in)-}_q(x,t)$ can be expanded in terms of static solutions as in equation (\ref{in - decomposition}), we can evaluate the Bogoliubov coefficient $\beta_{kq}$ in the asymptotic future to write
\begin{equation}
\begin{split}
    &\beta_{kq}=\langle \psi^{\rm(in)-}_q(x,\infty), \psi^{\rm(out)+}_k(x,\infty)\rangle_D=\\
    &=\sumint_pdp\left[(\xi^q_p(\infty))^*\langle\psi_p^+(x),\psi^+_k(x)\rangle_D+(\eta_p^q(\infty))^*\langle\psi_p^-(x),\psi^+_k(x)\rangle_D\right]=(\xi_k^q(\infty))^*\,.
    \label{Bogo beta}
\end{split}
\end{equation}

Because of this, the probability densities $n_{b,k}$ and $n_{d,k}$ can be rewritten as
\begin{equation}
    n_{b,k}=n_{d,k}=\sumint dq|\xi^q_k(\infty)|^2\,.
\end{equation}

Therefore, we have encountered a way of expressing $n_{b,k}$ and $n_{d,k}$, quantities that arise from the quantization procedure of the theory, in terms of $\xi^q_k(\infty)$, functions that are obtained via classical field theory, from the evolution of the Dirac equation. Hence, the problem of obtaining the probability densities can be reduced to solving the coupled differential equations (\ref{dyn eq1}) and (\ref{dyn eq2}) with the initial conditions (\ref{ICs minus}).

As a final remark, note that one can also obtain the expression for the remaining Bogoliubov coefficient, $\alpha_{kq}$, by means of the same procedure. Evaluating $\alpha_{kq}$ at the asymptotic future yields
\begin{equation}
    \alpha_{kq}=\expval*{\psi^{(\rm in)-}_q(x,\infty),\psi^{(\rm out)+}_k(x,\infty)}_D=(\xi^q_k(\infty))^*\,.
    \label{Bogo alfa}
\end{equation}

It may seem that both Bogoliubov coefficients are related to the same time-dependent function $\xi^q_k(\infty)$ and, thus, they are equal. However, this is not the case since, from their definitions, $\beta_{kq}$ is constructed with $\psi^{\rm(in)-}_q$ while $\alpha_{kq}$ is built with $\psi^{\rm(in)+}_q$. Therefore, $\xi^q_k(\infty)$ from (\ref{Bogo beta}) satisfies the initial condition $\xi^q_k(t=-\infty)=0$ whereas $\xi^q_k(\infty)$ from (\ref{Bogo alfa}) fulfills $\xi^q_k(t=-\infty)=\delta_{qk}$.

The knowledge of both $\alpha_{kq}$ and $\beta_{kq}$ will be necessary to verify if the closure relations (\ref{clos_rels}) are being satisfied, which at the same time is going to indicate whether our numerical calculations are proceeding correctly. Our computations show that the numerical results exhibit only small deviations from these relations.

\section{Numerical results}
\label{sec: numerical_results}
Once the theoretical background has been established, we shift our focus to the details of the numerical calculations. Because the probability amplitudes of fermion particles and antiparticles are the same, in this section no distinction will be made and we will refer to them as $n_k$.
\subsection{Some preliminaries}

\subsubsection{Time dependent perturbation and switching function}

As first introduced in Section 3, the time-dependent perturbation $\varphi_s(x,t)$ must satisfy the condition $\lim_{t \to \pm \infty} \varphi_s(x,t) = 0$, ensuring that the kink remains at rest in the asymptotic past and future. To enforce this requirement, it is necessary to introduce a switching function that explicitly fulfills these boundary conditions. In the present analysis, this function is chosen to be
\begin{equation}
    F(t)=\frac{\cal{A}}{2}\left(\tanh(\frac{t+T}{s})-\tanh(\frac{t-T}{s})\right)\,,
\end{equation}
where the parameter $\cal{A}$ is the amplitude of the perturbation, $T$ specifies the times when the function is switched on and off and $s$ parametrizes how fast the switching occurs, that is, the smaller the parameter $s$, the faster the $F(t)$ transitions from 0 to its maximum value and vice-versa. Hence, the complete, time-dependent perturbation 
we are insterested in is given by,
\begin{equation}
    \varphi(x,t)=F(t)\cos{(\omega_st)}f_s(x)=A(t)\sqrt{\frac{3}{2}}\sech{x}\tanh{x}\,.
\end{equation}

\begin{figure}[htb!]
    \centering
    \begin{subfigure}{0.47\textwidth}
        \centering
        \includegraphics[width=\textwidth]{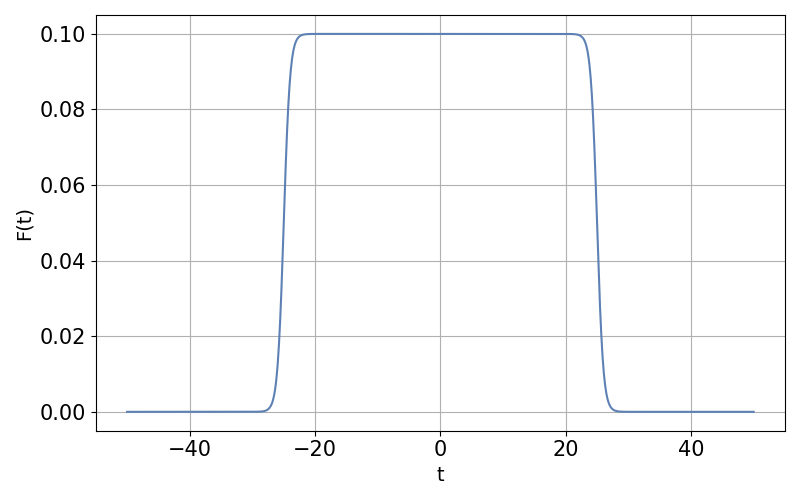}
        \label{fig:image1}

    \end{subfigure}\hfill
    \begin{subfigure}{0.47\textwidth}
        \centering
        \includegraphics[width=\textwidth]{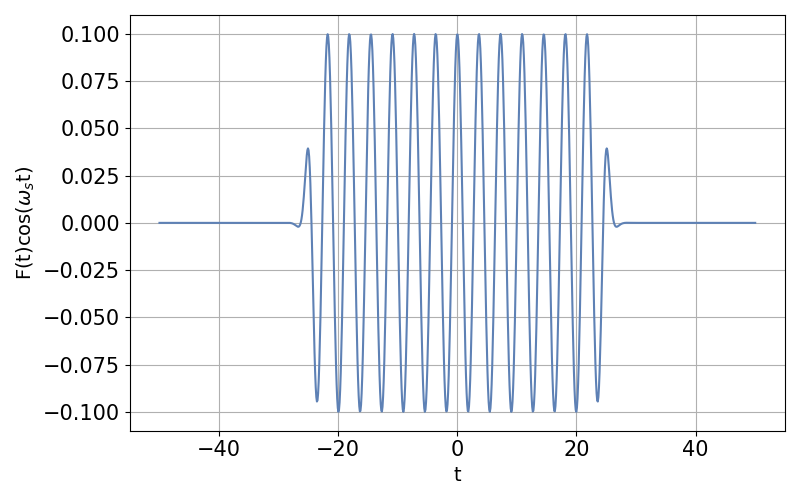}
        \label{fig:image2}
        
    \end{subfigure}
    
    \caption{Profile of the switching function $F(t)$ (left panel) and time-dependent part of the perturbation (right panel). The parameters of the switching function are taken to be ${\cal A}=0.1$, $T=25$ and $s=1$. The asymptotic times are chosen to be $\pm50$.}
    \label{fig: switching function}
\end{figure}

$A(t)$ in the last step encompasses all the time dependence of the perturbation, and thus, it will be possible to take it out from spatial integrals such as the ones for the matrices $Q_{kk'}$ and $R_{kk'}$. Both $F(t)$ and the time-dependent perturbation can be seen in Figure \ref{fig: switching function} \footnote{Unless otherwise specified, the parameters used here are the ones that we will be using from now onward.}.

\subsubsection{Discretisation of scattering states}
During the previous sections we have used the condensed notation in \cref{sumint} to express a sum over a discrete set of bound states and an integral over a continuous set of scattering states.

However, it is impossible to work with a continuous, infinite set of states in a numerical calculation and therefore we have to first discretize the spectrum by considering a set of $N$ modes separated by a constant space in $k$-space, namely, $\Delta k$.

The maximum number of discretized states considered directly affects the computational cost of the numerical calculation. For instance, the dimensions of the matrices $Q$ and $R$ in equations (\ref{dyn eq1}) and (\ref{dyn eq2}) are given by $(\lceil g \rceil\times N)^2$, and consequently, the number of integrals that need to be calculated grows like $N^2$. In fact, for integrals involving scattering states, there are not in general any parity arguments leading to their cancellation, and hence all of them must be computed. This implies that the number of differential equations increases linearly with $N$, but also the number of terms appearing in the sum on each equation does also grow like $N$. Furthermore, in order to find the complete spectrum, the number of times the system must be solved also increases linearly with $N$. The fact that all steps required to obtain numerical results increase in complexity leads to a higher computational cost and, consequently, a longer time to complete each simulation. For this reason, unless otherwise specified, we will consider in our simulations $N=60$ scattering modes, which will be uniformly distributed in the interval $k\in(0,2.5)$. A comment on the convergence of our numerical results with growing $N$ is relegated to \cref{Appendix D}.

Furthermore, the discretized version of our modes also implies that the Dirac delta appearing in the normalization condition must be changed accordingly, so that
\begin{equation}
\langle \psi_k^r,\psi_{k'}^s\rangle_D\to \frac{1}{\Delta k}\delta_{rs}\delta_{kk'}\,,
\end{equation}
whereas the integral over scattering states has to be substituted by a Riemann sum, i.e.,
\begin{equation}
    \int_{k_0}^\infty dk\to\sum_k\Delta k~.
\end{equation}

 Hence, the smaller $\Delta k$ gets, the more terms will be added to the sum, and we recover the continuum limit.

\subsection{Fermion production}

In this section, we analyze the power radiated through fermionic particle production induced by the excitation of the shape mode in the kink configuration.
We first note that while both bound states and scattering states will generally be excited due to energy transferring from the excited bosonic shape mode, only the excitation of the latter can lead to the emission of energy away from the kink, in the form of radiated fermionic particles. On the other hand, as we will show below, the maximum energy emission will take place for resonant scattering modes, i.e. for the combination of modes with energy close to the frequency of the shape mode, $\omega_s$. Since the model presents a mass gap at $\omega_*=g$, we will find that there are two clearly distinct regimes distinguished by the value of the coupling constant, namely, the resonant and the non-resonant regimes, for $g<\omega_s$  and $g>\omega_s$, respectively.

We reiterate for clarity that by numerically solving the dynamical equations (\ref{dyn eq1}) and (\ref{dyn eq2}) we can obtain the time-dependent functions $\xi_k^q(t)$ and $\eta_k^q(t)$ for any given fermion mode $\psi_q(x,t)$. As demonstrated in the previous section, these functions can also be used to calculate the probability amplitudes for detecting a fermion particle/antiparticle in the asymptotic future. Let us start analyzing the probability amplitudes evaluated at the asymptotic future for both bound and scattering fermion modes. In the following figure, these probability amplitudes are plotted for different values of the Yukawa coupling constant.
\begin{figure}[htb!]
    \centering

    \begin{subfigure}{0.48\textwidth}
        \centering
        \includegraphics[width=\textwidth]{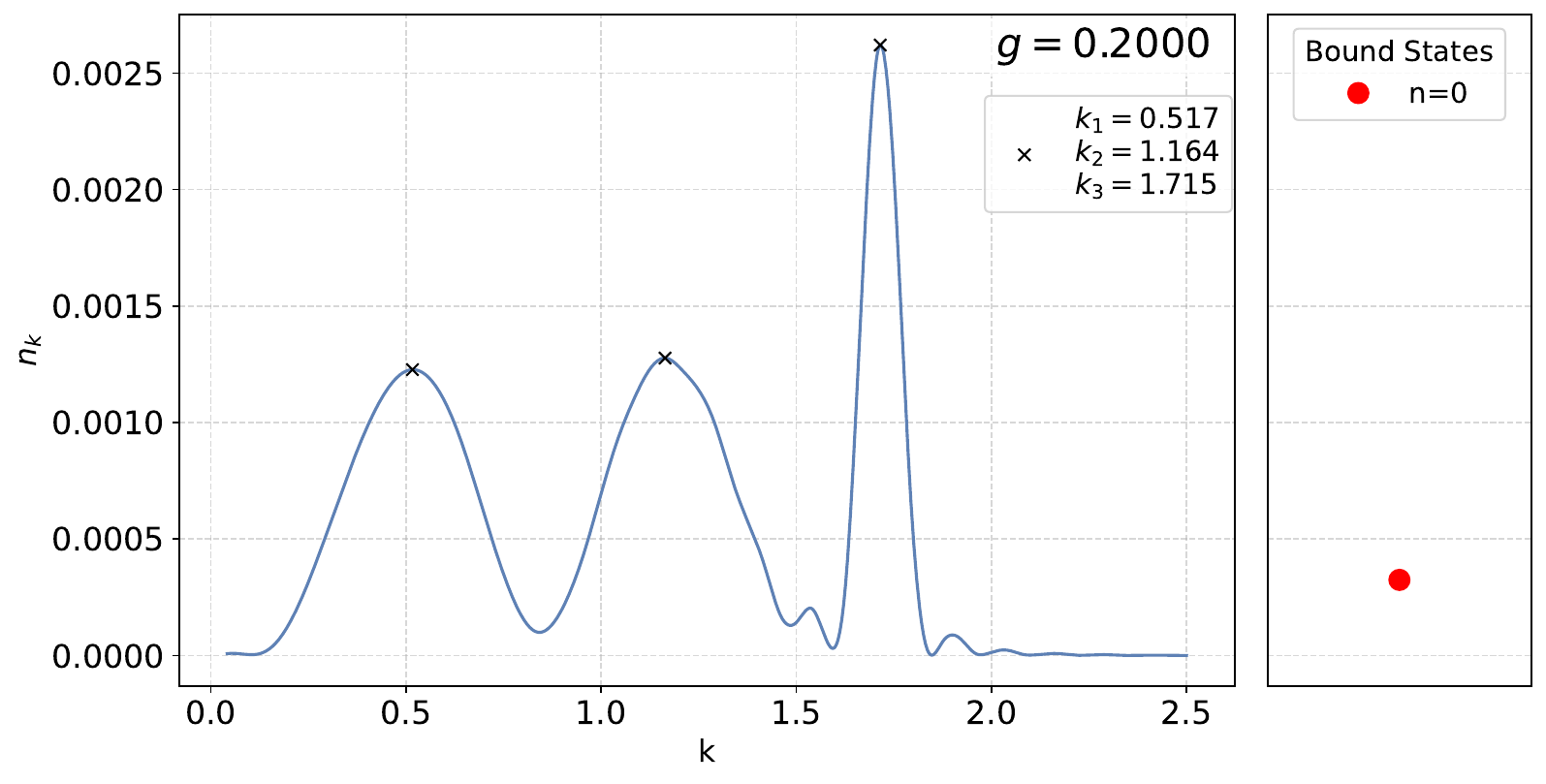}
        
    \end{subfigure}
    \hspace{0.02\textwidth}
    \begin{subfigure}{0.48\textwidth}
        \centering
        \includegraphics[width=\textwidth]{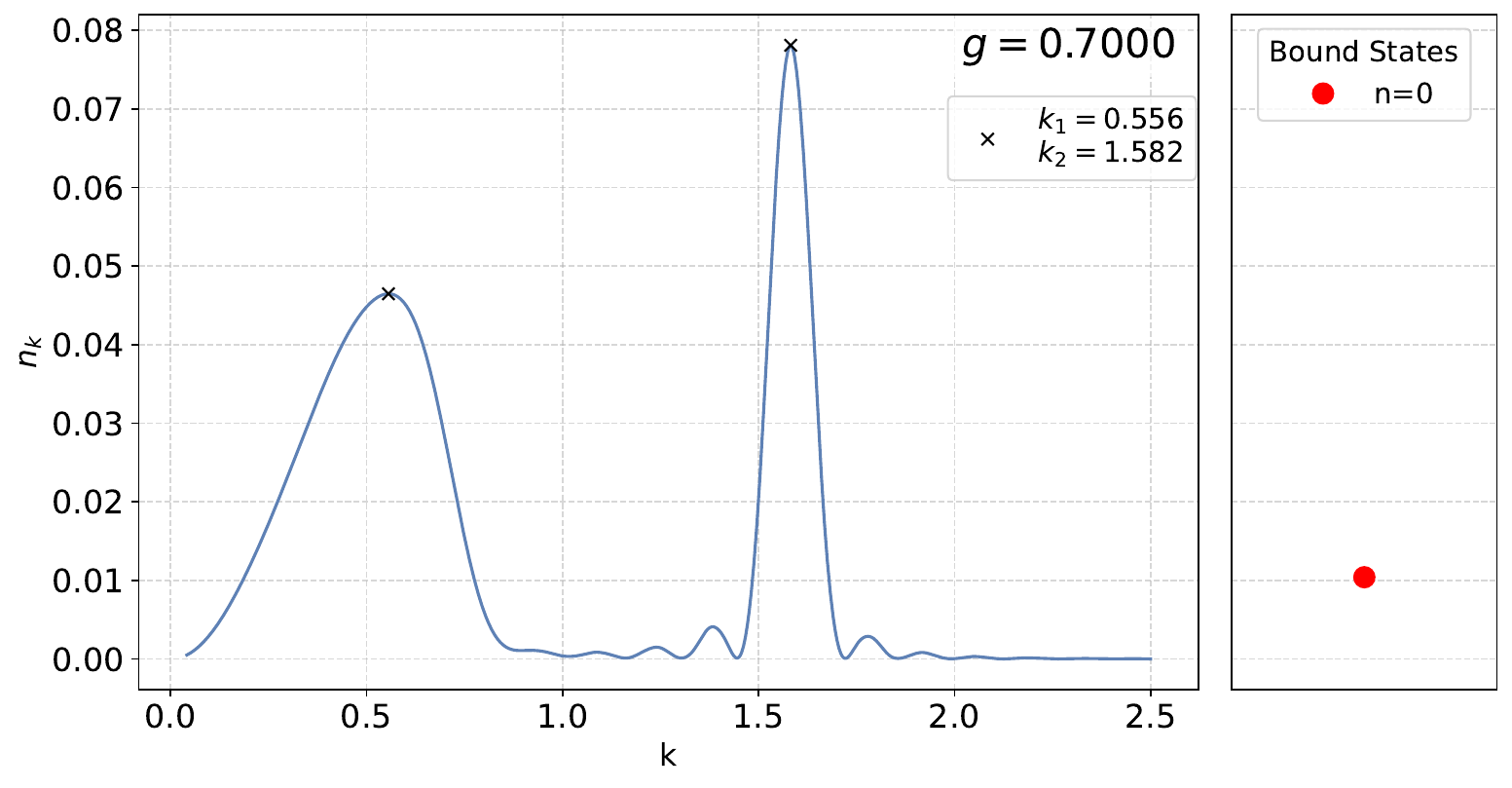}
        
    \end{subfigure}

    \begin{subfigure}{0.48\textwidth}
        \centering
        \includegraphics[width=\textwidth]{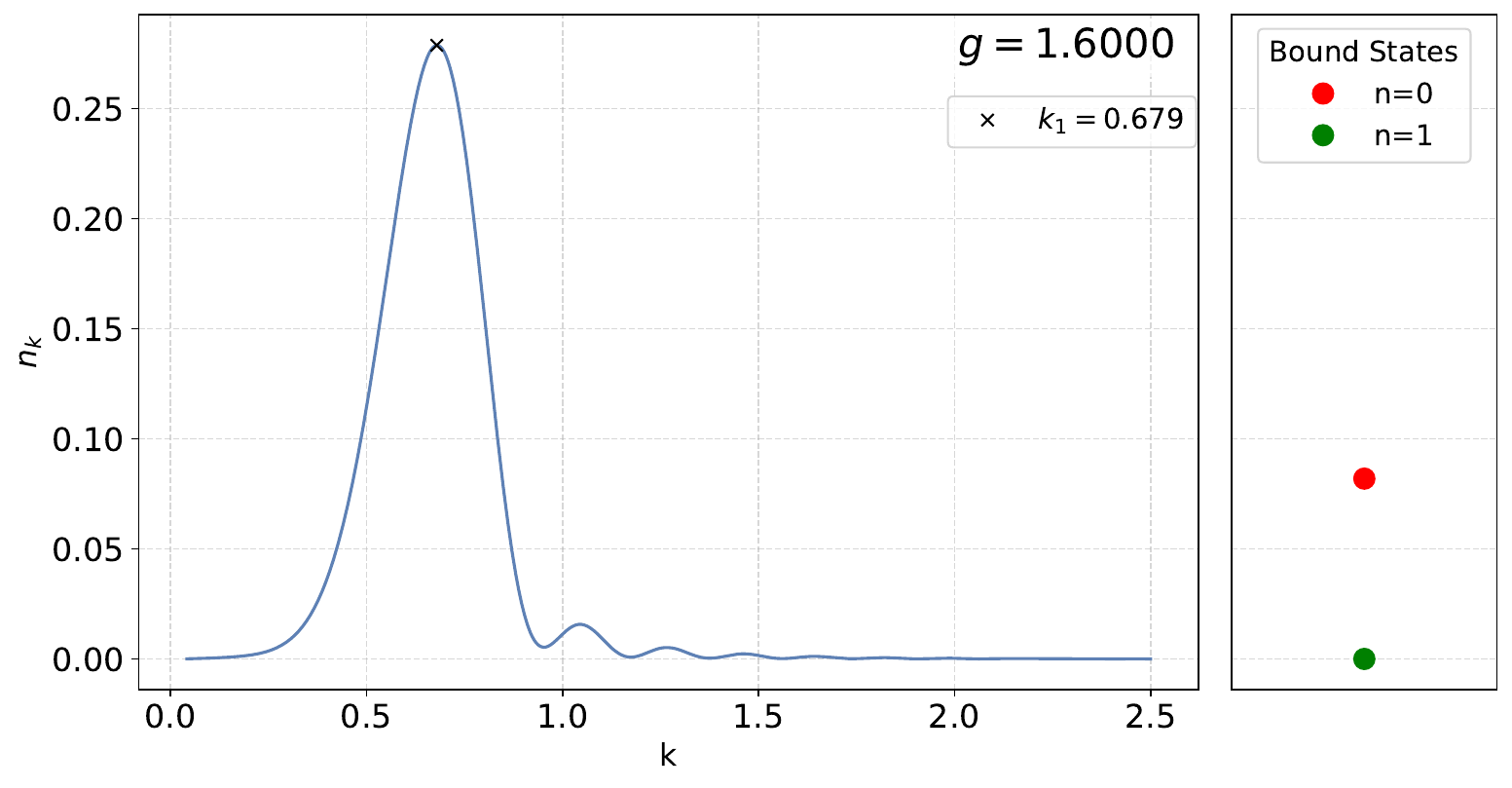}
        
    \end{subfigure}
    \hspace{0.02\textwidth}
    \begin{subfigure}{0.48\textwidth}
        \centering
        \includegraphics[width=\textwidth]{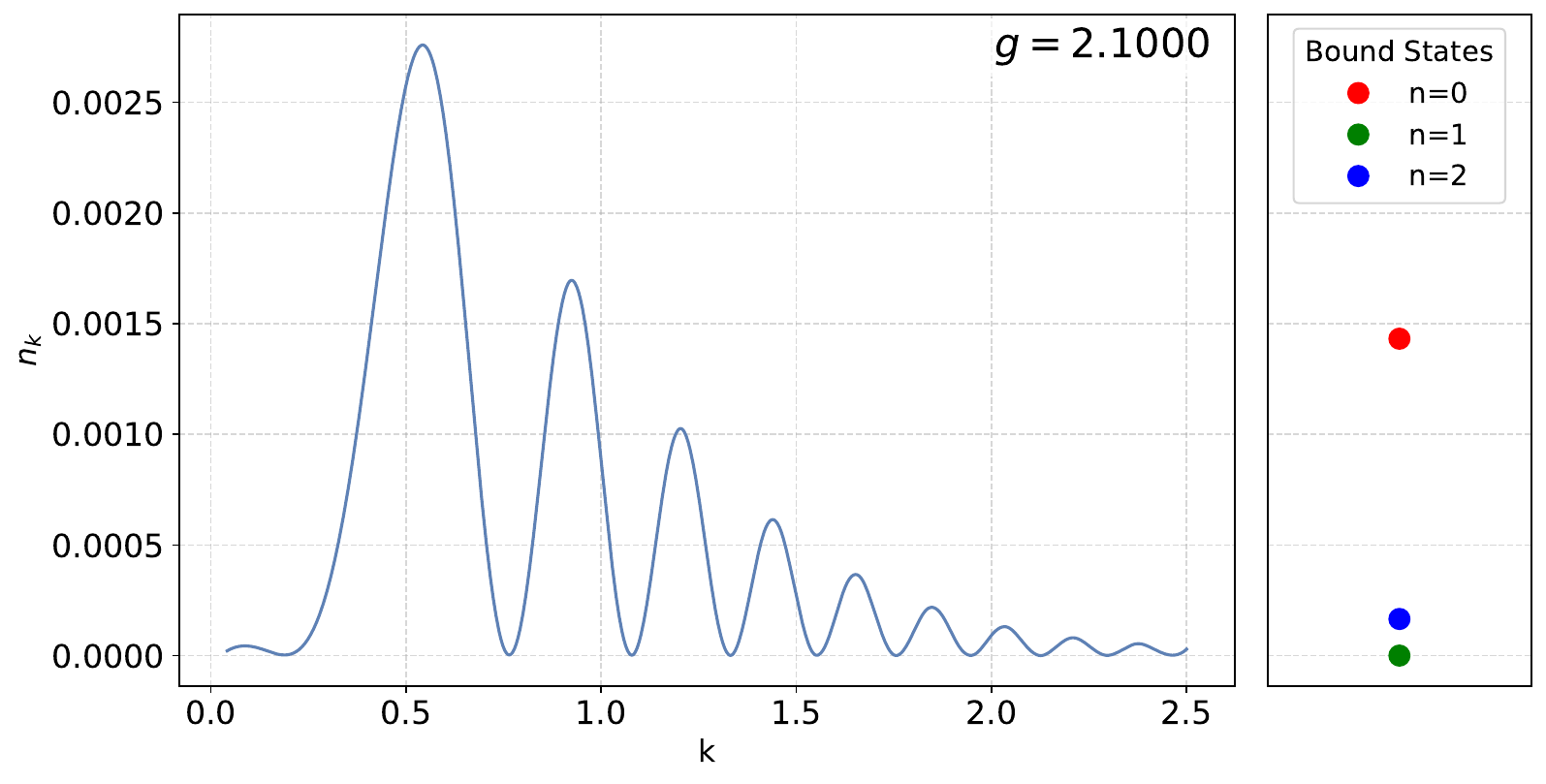}
        
    \end{subfigure}

    \caption{Probabilities of scattering states $n_k$ evaluated at the asymptotic future with respect to their wave number $k$, for increasing values of the Yukawa coupling $g$. The narrower panels to the right of each of the graphs show the discrete probabilities associated to the existing bound fermion states.}
    \label{fig: nk vs k}
\end{figure}

It can be seen that different values of $g$ yield notoriously distinct probabilities, with the only common feature being the presence of pronounced peaks around specific values of $k$. The existence of these maxima can be explained as follows: on the one hand, regarding the more prominent peaks (the absolute maxima at each plot), one can verify that each peak is centered around a scattering state $k$ whose static energy fulfills $E_k = \omega_s$, or equivalently, $k = \sqrt{\omega_s^2 - g^2}$.

This result suggests that this mode should be excited together with the
localized zero mode such that the total energy of the pair 
matches that of the shape mode: $E_k + E_0 = \omega_s$.  In fact, integrating the probability density over a region around the maximum confirms that the resulting probability is comparable to that of exciting the zero mode itself.

Moreover, following this argument, it can be concluded that this peak occurs only for $g<\omega_s$, which is consistent with the observation in Figure \ref{fig:energies graphs}, where, for higher values of $g$, the shape mode falls below the mass threshold that separates scattering and bound fermion modes. This not only explains the presence of a peak around that specific value of $k$, but also the absence of such maxima beyond $g = \omega_s$. The latter becomes evident in the bottom-right image, where the probability density, in comparison to the previous cases, has decreased in overall magnitude and is more evenly spread.

\begin{figure}[h]
    \centering
    \includegraphics[width=0.55\textwidth]{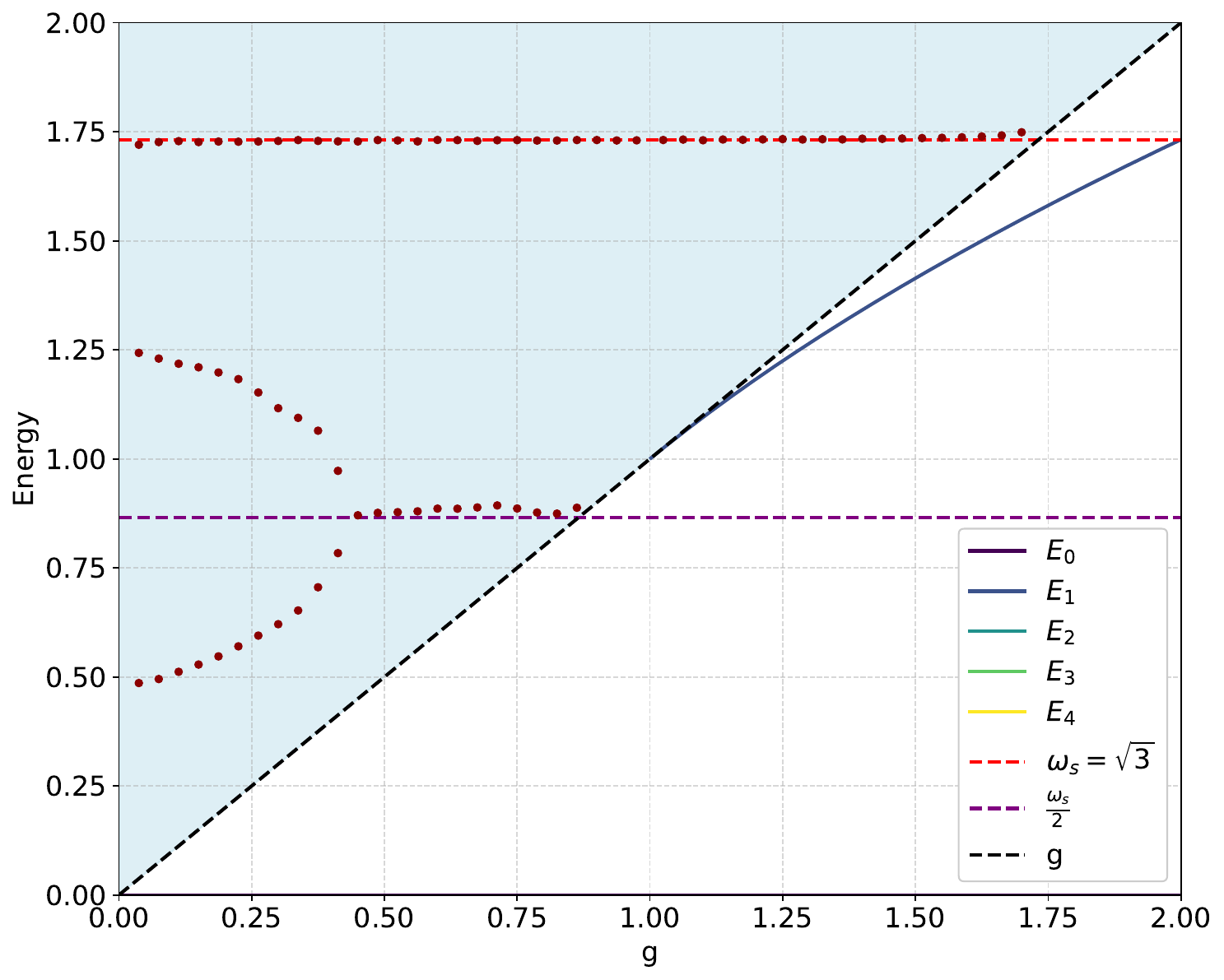} 
    \caption{Energies corresponding to the maxima in the probabilities in Figure \ref{fig: nk vs k} (dark red).}
    \label{fig:Energies_vs_g_plus_nk}
\end{figure}

On the other hand, regarding the shorter peaks in the upper two images,  a similar argument can be given for their existence, given that they are centered around scattering states that satisfy $E_{k_1}+E_{k_2}=\omega_s$ (upper left) and $2E_{k_1}=\omega_s$ (upper right). As with the previous case, they cannot persist indefinitely. Indeed, as seen in Figure $\ref{fig:Energies_vs_g_plus_nk}$, as $g$ increases, this initial `double peak' structure gradually transitions into a single maximum with half the energy of $\omega_s$, until $g=\frac{\omega_s}{2}$, where they cease to exist, for the same reason discussed in the previous paragraph. 

With respect to the remaining bound states, a particularly notable feature is that $n_1$ is found to be nearly vanishing. This can be explained by parity arguments involved in the construction of the matrices $Q_{kk'}$ (\ref{Qmatrix}) and $R_{kk'}$ (\ref{Rmatrix}). As mentioned in \cite{campos-2021}, these elements of $Q$ and $R$ vanish when the difference of the indices of two bound states is an odd number. This, in turn, simplifies the system of coupled dynamical equations (\ref{dyn eq1}) and (\ref{dyn eq2}). Indeed, it can be checked that if one solely considers bound states in the system of dynamical equations, the probability $n_1$ is directly null, since the $\xi^q_1(t)$ contributing to the probability of $n_1$ and multiplying to non-vanishing elements of $Q$ and $R$ are decoupled from the rest of the time-dependent functions.

Going back to the scattering fermion modes, we can clearly distinguish two different regimes: the resonance regime ($g<\omega_s$) and the non-resonance regime ($g>\omega_s$). 

\begin{figure}[h]
    \centering
    \resizebox{0.9\textwidth}{!}{
        \begin{tabular}{cc}
            \begin{subfigure}{0.49\textwidth}
                \centering
                \includegraphics[width=\linewidth]{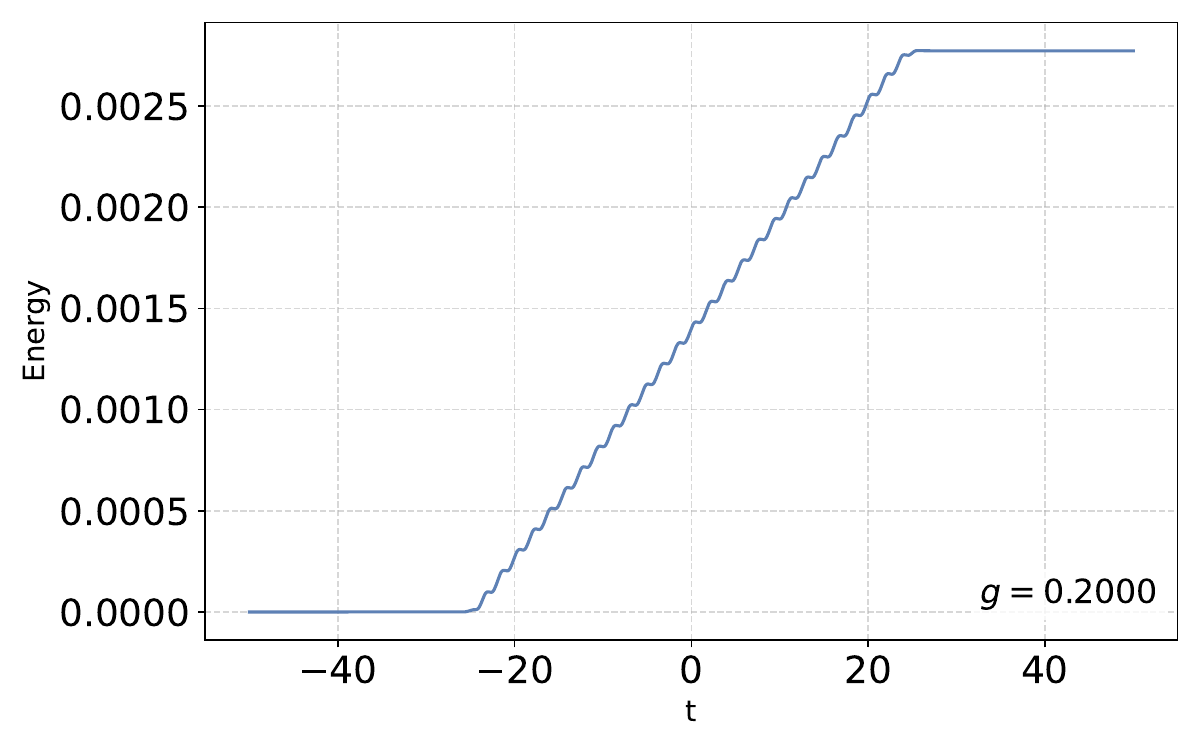}
            \end{subfigure} &
            \begin{subfigure}{0.49\textwidth}
                \centering
                \includegraphics[width=\linewidth]{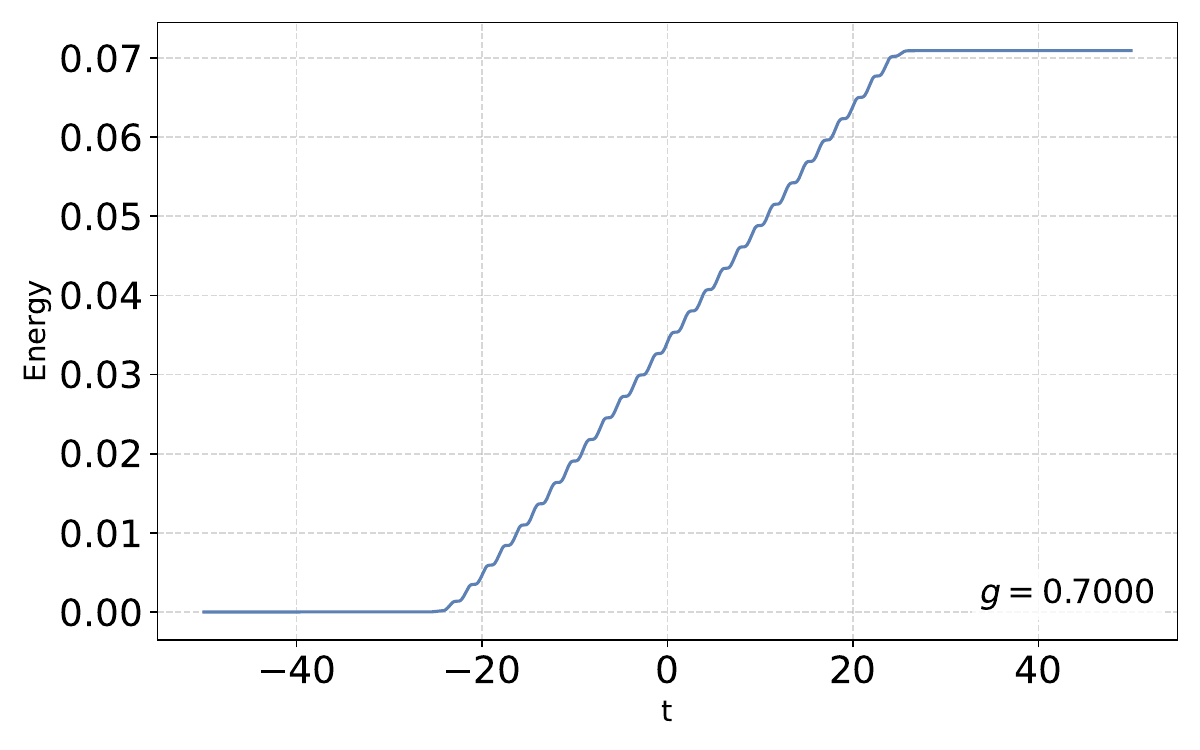}
            \end{subfigure} \\
            \begin{subfigure}{0.49\textwidth}
                \centering
                \includegraphics[width=\linewidth]{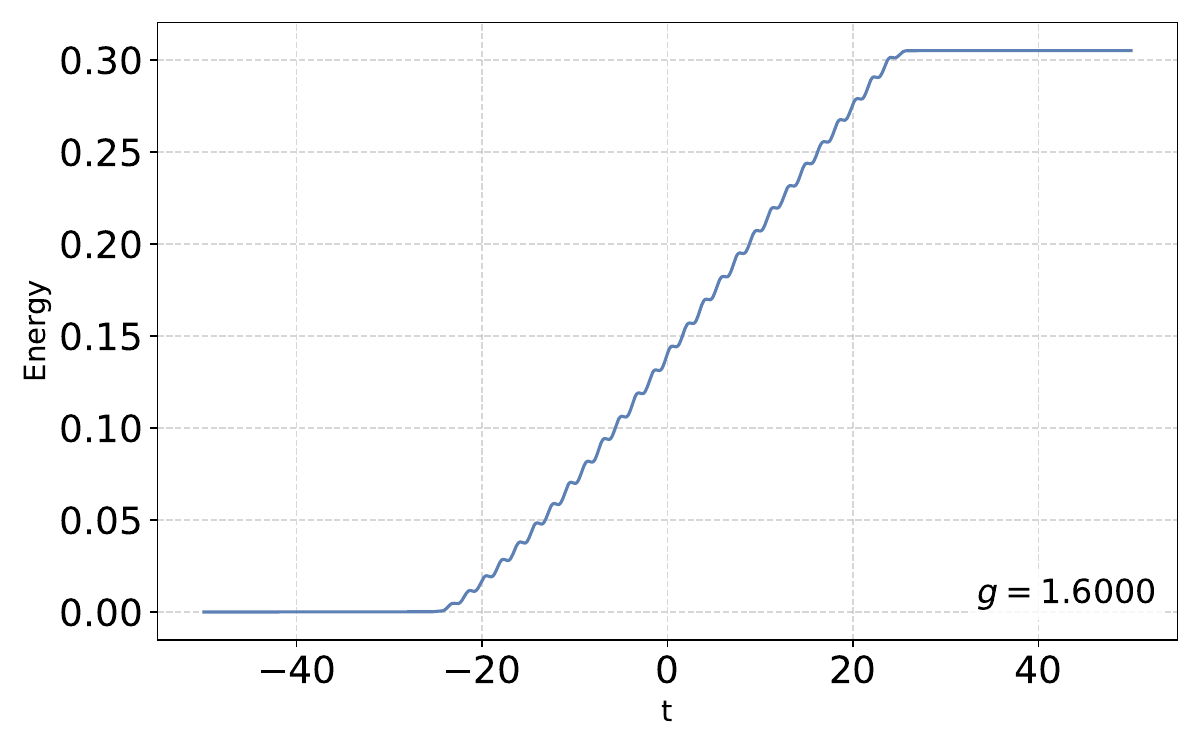}
            \end{subfigure} &
            \begin{subfigure}{0.49\textwidth}
                \centering
                \includegraphics[width=\linewidth]{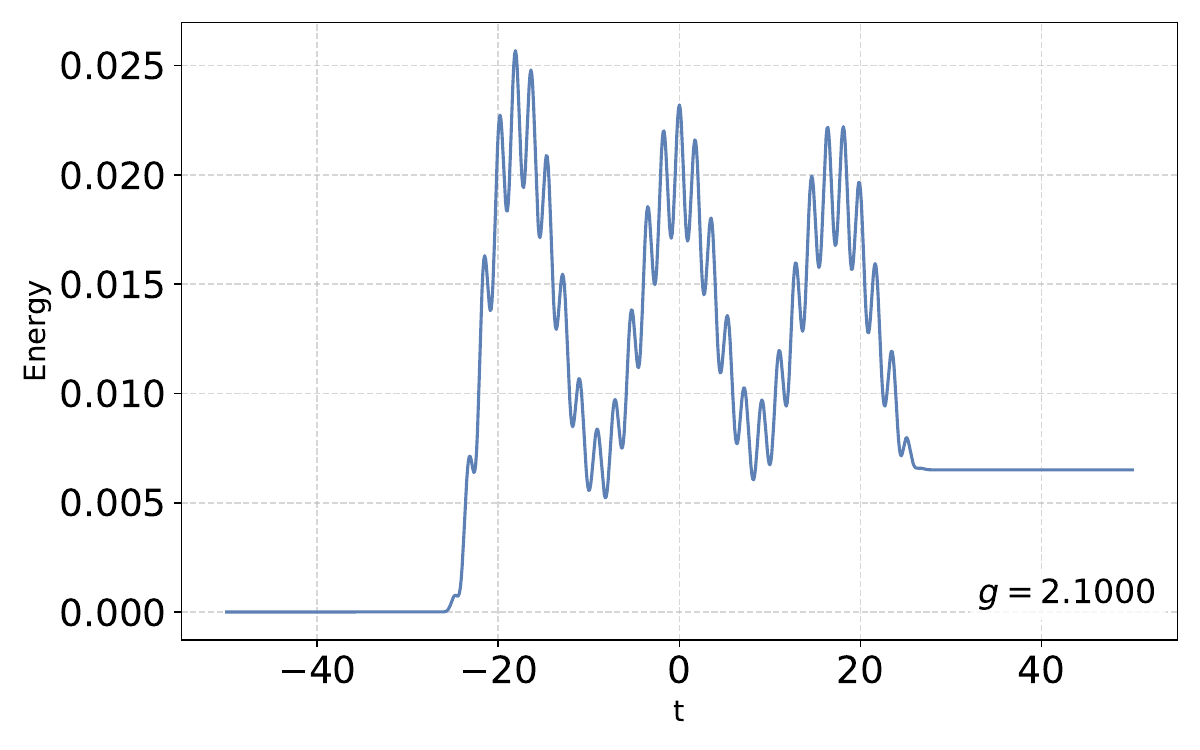}
            \end{subfigure}
        \end{tabular}
    }
    \caption{Time evolution of the total energy for different values of $g$.}
    \label{fig: TotalEnergies}
\end{figure}

By examining the resonant case, we expect significant fermion emission coming from the excited kink, which will ultimately affect the amplitude of the shape mode. We can further analyze this by noting that from the expected number of fermions in each mode $k$ one can extract the total energy transferred to the fermion field simply by 
\begin{equation}
    E(t)=\sumint dk E_k(t)=\sumint dk \omega_k n_k(t)\,,
\end{equation}
where $\omega_k$ is the static energy each mode has, either $E_n=\sqrt{n(2g-n)}$ or $E_k=\sqrt{g^2+k^2}$, and the sum accounts for both fermions and antifermions. The time evolution of the total energy for different values of $g$ is shown in Figure \ref{fig: TotalEnergies}. As mentioned before, below $g<\omega_s$, resonance will take place. This results in a total energy that evolves almost monotonically over time, apart from small oscillations associated with the shape mode's frequency. Once $g$ surpasses $\omega_s$, resonance is no longer possible. Instead of increasing continuously over time, the behavior changes: it is modulated by a lower-frequency dynamic that rapidly oscillates around a specific energy value (bottom right).

\begin{figure}[htb!]
    \centering
    \resizebox{0.95\textwidth}{!}{
        \begin{tabular}{cc}
            \begin{subfigure}{0.49\textwidth}
                \centering
                \includegraphics[width=\linewidth]{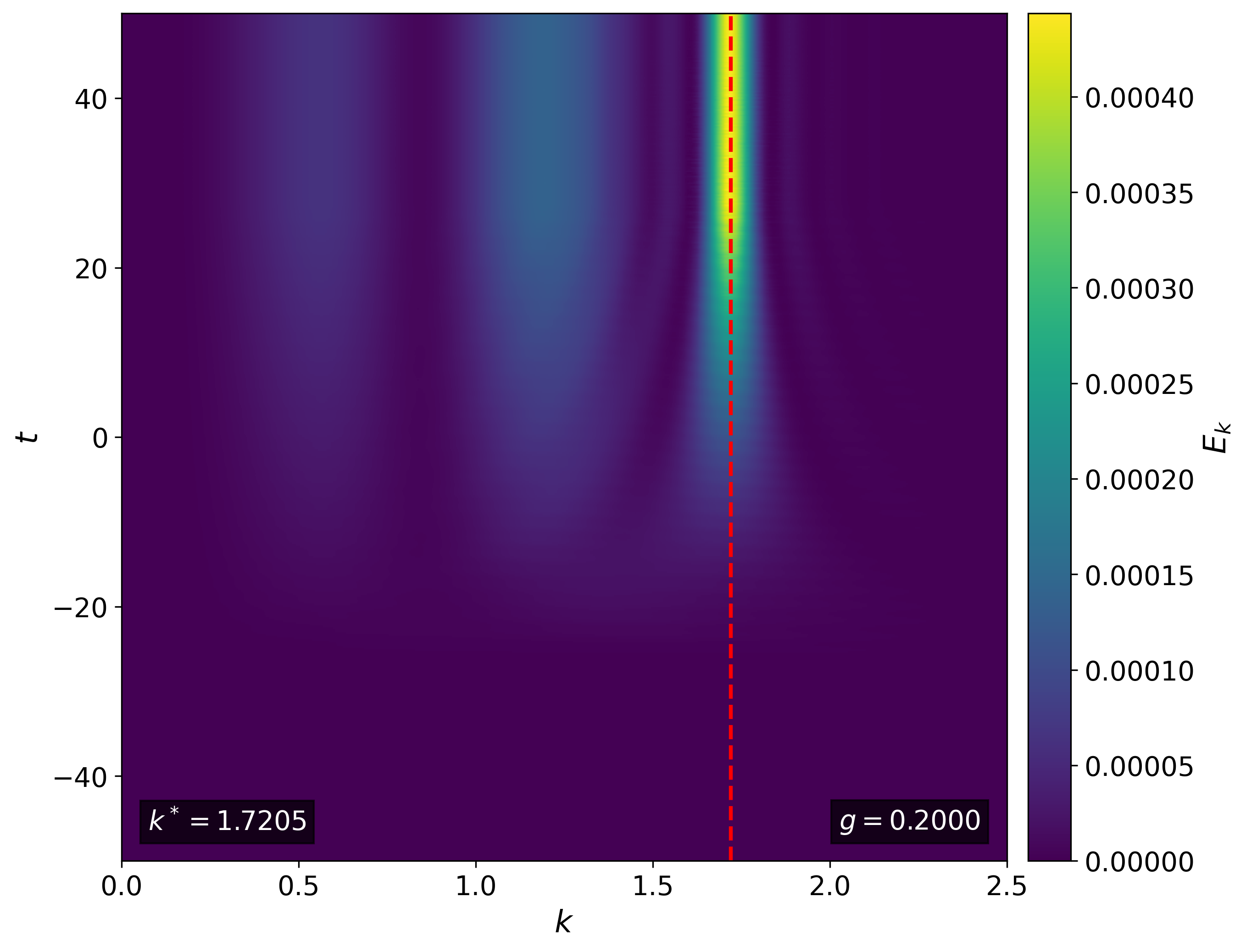}
            \end{subfigure} &
            \begin{subfigure}{0.49\textwidth}
                \centering
                \includegraphics[width=\linewidth]{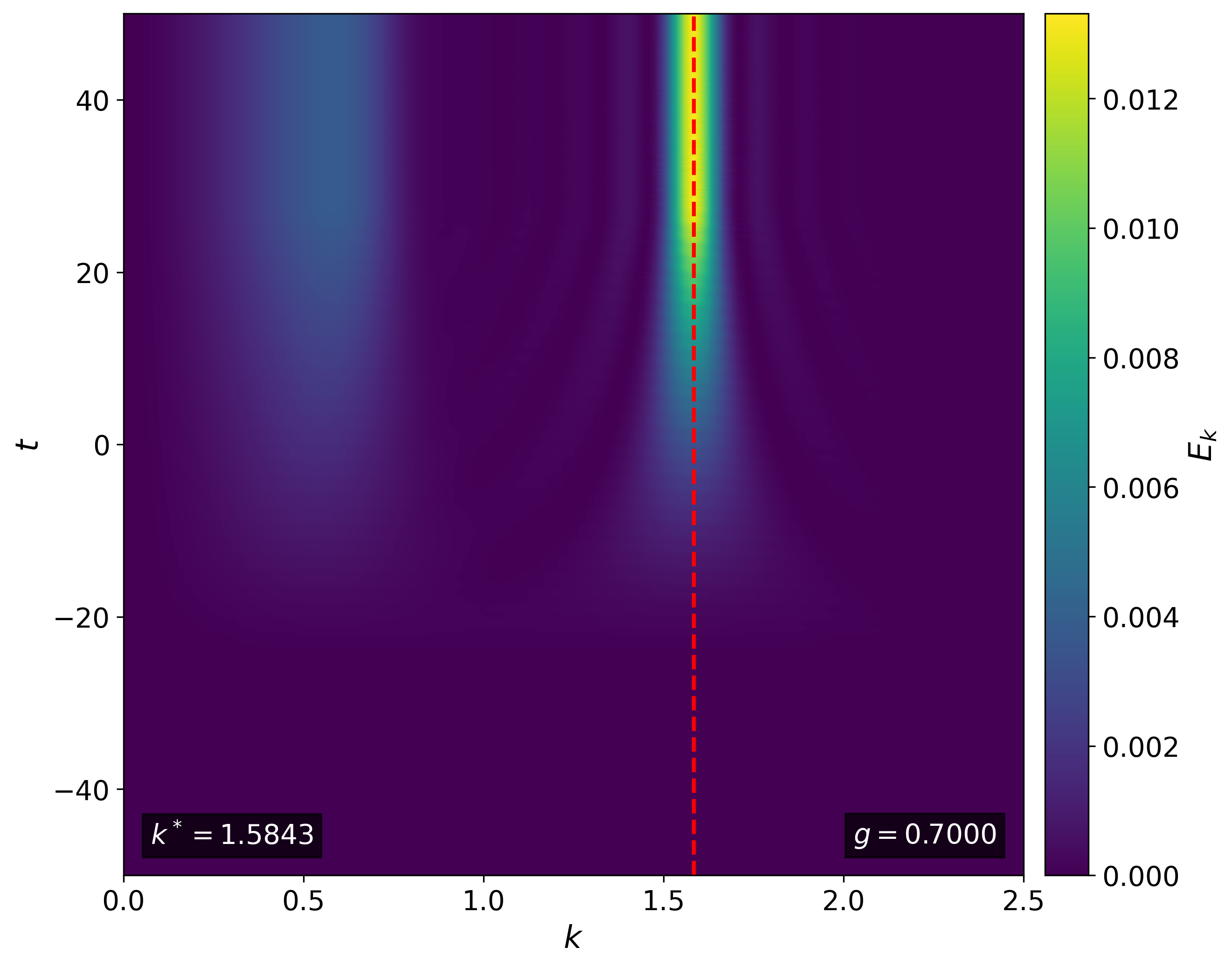}
            \end{subfigure} \\
            \begin{subfigure}{0.49\textwidth}
                \centering
                \includegraphics[width=\linewidth]{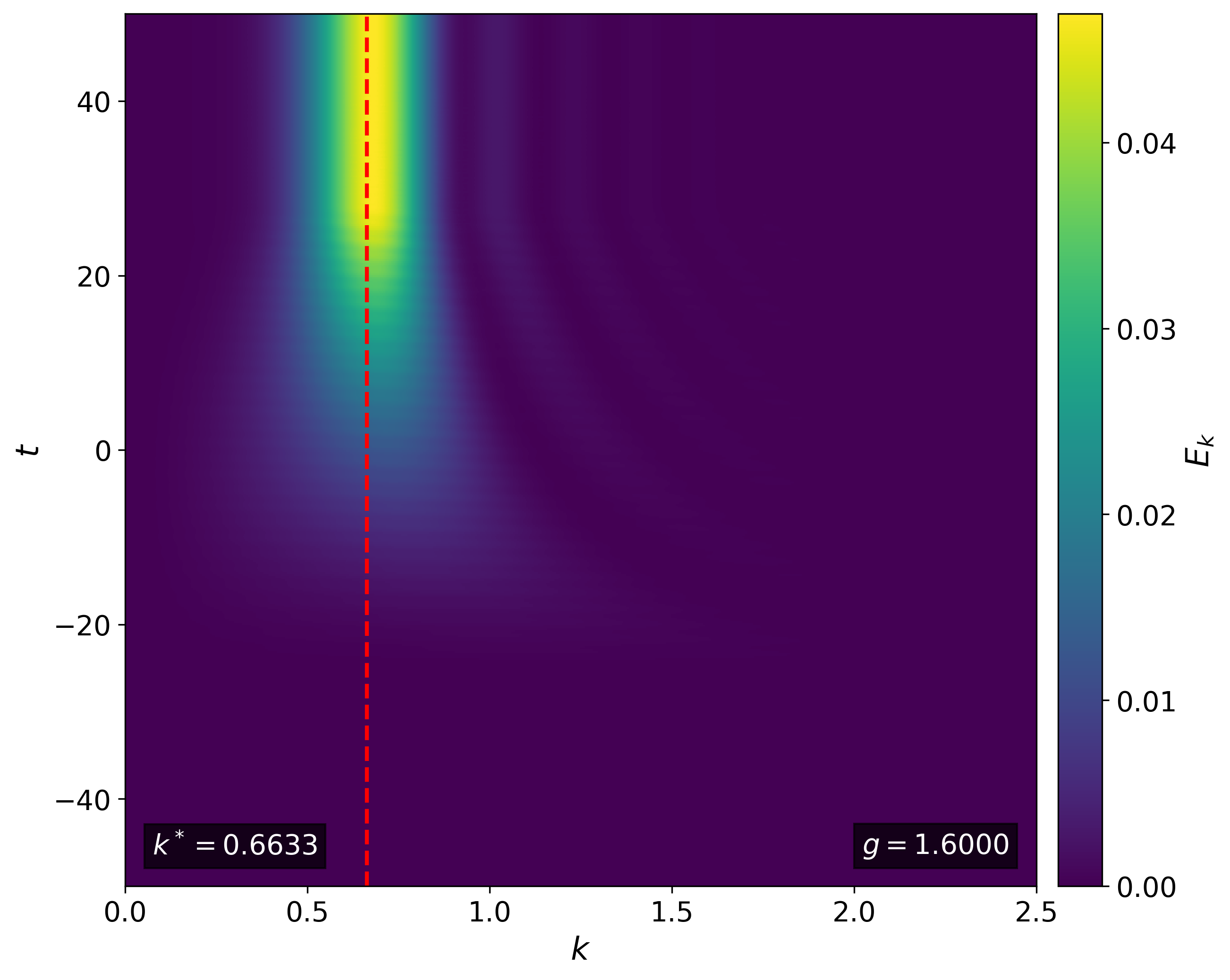}
            \end{subfigure} &
            \begin{subfigure}{0.49\textwidth}
                \centering
                \includegraphics[width=\linewidth]{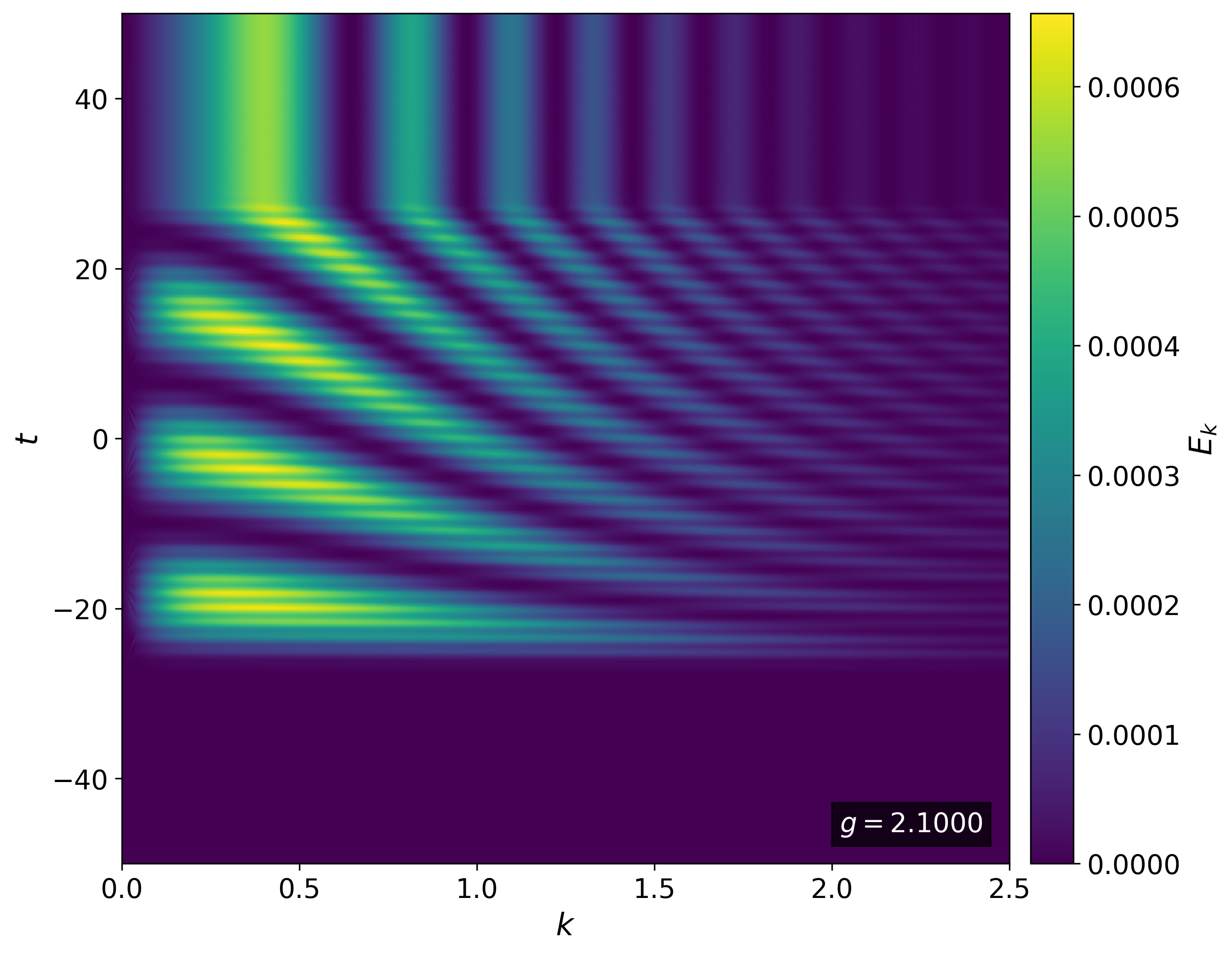}
            \end{subfigure}
        \end{tabular}
    }
    
    \caption{Time evolution of all the scattering states $k$ contributing to the total energy for different values of $g$. The dashed red line marks the scattering fermion mode $k^*=\sqrt{\omega_s^2-g^2}$.}
    \label{fig:EnergyContributions}
\end{figure}

In order to better understand this behavior, let us analyze the contribution each mode has in the previous graphs. In Figure \ref{fig:EnergyContributions}, the time evolution of each scattering mode $k$ is presented for different values of $g$. As expected, when $g<\omega_s$, only regions of scattering states around the maxima of the probabilities from Figure \ref{fig:Energies_vs_g_plus_nk} are being excited. These modes are the dominant contributors to the total energy profile in their respective cases.

For $g > \omega_s$, no specific region of states stands out as predominantly excited. Instead, the energy becomes more broadly distributed across different modes, while also decreasing by at least one order of magnitude. Additionally, note how the region centered around $k^*$ gradually broadens as the coupling constant approaches $\omega_s$ from below, signaling a smooth transition from the resonant to the non-resonant regime.

Equally relevant is the analysis of the power averaged over the time interval during which the kink's shape mode is excited, i.e.
\begin{equation}
    P(g)=\lim\limits_{T\to\infty}\frac{1}{2T}\int^{T}_{-T}\dv[]{E}{t} dt\,.
\end{equation}

\begin{figure}[htb!]
    \centering
    \includegraphics[width=0.65\textwidth]{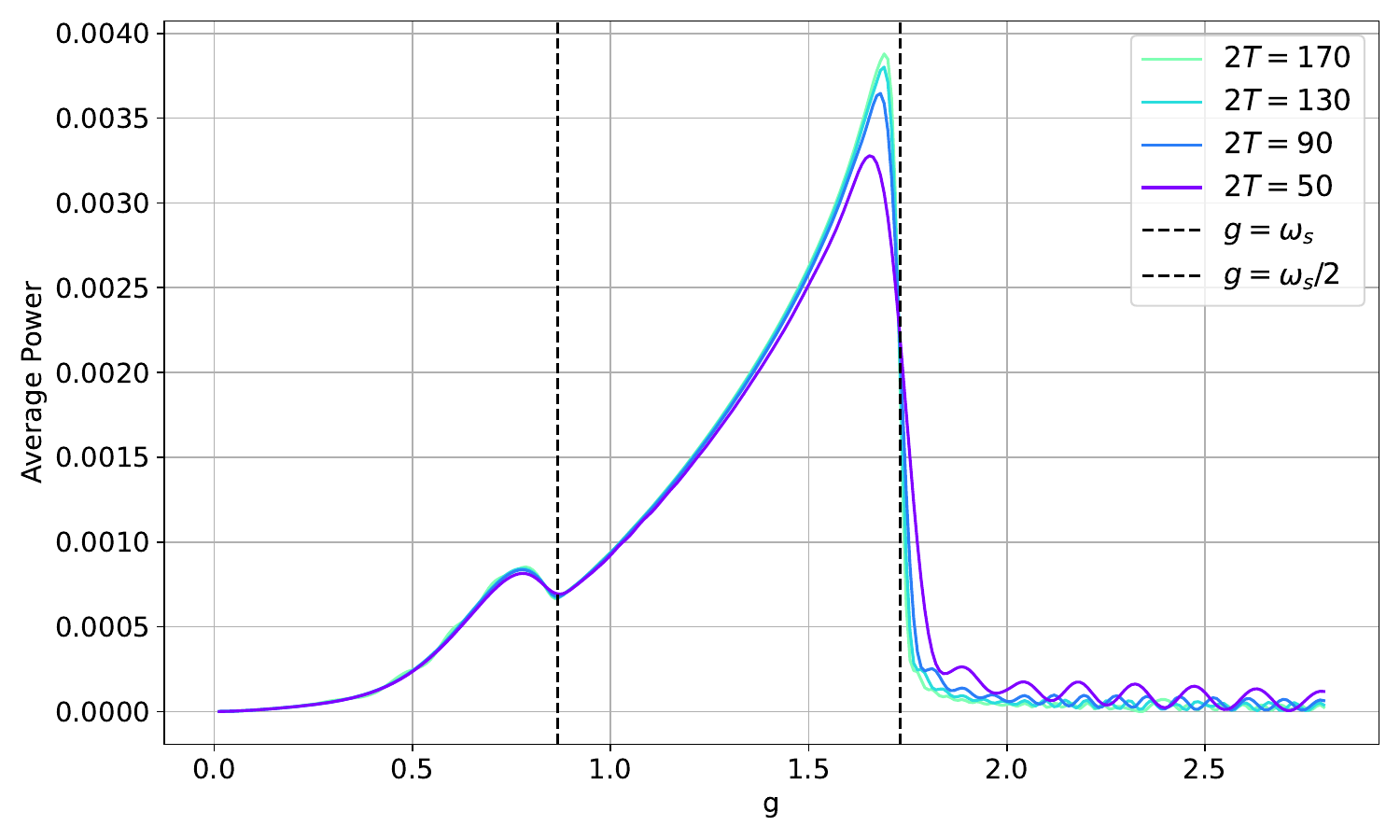} 
    \caption{Average power with respect to the Yukawa coupling constant $g$, for different times the shape mode is switched on. The vertical dashed lines represent $g=\omega_s$ and $g=\omega_s/2$. }
    \label{fig:P vs. g}
\end{figure}

This quantity provides a reliable proxy for the instantaneous power radiated by a physically excited kink. In Figure \ref{fig:P vs. g}, this average power is plotted in terms of Yukawa coupling constant. The average power has a quadratic increase (aside from the small `bump' around $g=\omega_s/2$) until it reaches its maximum before surpassing $g=\omega_s$. After reaching $g=\omega_s$, it sharply drops due to the resonating scattering states having gradually lower energy as $g$ becomes larger. From that point onward, the average power tends to zero as no resonance-like phenomena can take place for $g>\omega_s$, thus suppressing almost entirely the radiation for higher values of the coupling constant.

The presence of the `bump' can be explained by the fact that, until $g = \frac{\omega_s}{2}$, there remains a contribution from the scattering states associated with the lower maxima in the upper two images of Figure \ref{fig: nk vs k}, also observed in Figure \ref{fig:Energies_vs_g_plus_nk}. However, once the coupling constant exceeds this value, the only significant contribution arises from the scattering states $k$ whose energy satisfies $E_k = \omega_s$.

Notice as well how as the asymptotic time (hence $T$) increases, the average power gradually rises, with the location of its maximum slowly approaching the limiting value $g = \omega_s$. For larger asymptotic times, the decline in power beyond $g > \omega_s$ becomes steeper, resulting in a progressive suppression of radiation from the non-resonant regime. This behavior suggests that in the ideal limit $T \to \infty$, radiation originating from $g > \omega_s$ would be entirely suppressed, leaving only the resonance regime as the relevant contribution.

\subsection{Insights from perturbation theory}
In order to make sense of the observed resonance channels, let us try to understand the limit of small $g$, for which we can resort to standard perturbation theory for particle production from a classical source. We will assume that the fermion field is quantized as a free field and a classical source $j(x,t)=\cos{(\omega_s t)}f_s(x)$ is introduced via the interaction Hamiltonian 
\begin{equation}
    H_I=g \int j(x,t)\bar{\psi}\psi dx\,.
\end{equation}

The production of a fermion-antifermion pair with momenta $k$ and $k'$ respectively, is then given by the first nonzero term in the Dyson expansion of the evolution operator, i.e. by the matrix element
\begin{align}
   \mathcal{M}_{k\bar k'}=&\,_{in}\bra{k,\bar{k}'}ig \int\int  j(x,t) \bar{\psi}\,\psi \,dx dt \ket{0}_{in}=\notag \\=&ig\int dt\int dx (\psi^{(in)+}_k(x,t))^\dagger\sigma_1\psi^{(in)-}_{k'}(x,t) j(x,t)=\\
=&ig\sumint dq\sumint dq'\int [R_{kk'}( \xi^q_k(t)\xi^{q'*}_{k'} (t)+\eta^q_k(t)\eta^{q'*}_{k'} (t))+ Q_{kk'}( \xi^q_k(t)\eta^{q'*}_{k'} (t)+\eta^q_k(t)\xi^{q'*}_{k'} (t))]\cos(\omega_s t) dt\,.\notag
\end{align}

Now, from the dynamical equations \cref{dyn eq1,dyn eq2}, and the initial conditions \cref{ICs minus,ICs plus}, we find that in the asymptotic past and to first order in $g$,
\begin{equation}
    \xi_k^{q}\approx \delta_{qk}e^{-iE_k}\,,\quad \xi_{k'}^{q'}\approx 0\approx \eta^q_k\,,\quad\eta^{q'}_{k'}\approx \delta_{q'k'}e^{iE_{k'}}\,.
\end{equation}

Hence, we have
\begin{equation}
    \mathcal{M}_{k\bar k'}= ig Q_{kk'}\,\delta(E_k+E_{k'}-\omega_s)+\order{g^2}\,,
\end{equation}
i.e. the pair production is dominated, at the perturbative limit, by the tree-level process pictured in the following Feynman diagram, 
\begin{equation*}
\begin{tikzpicture}[baseline=(v.base)]
\begin{feynman}
  \vertex (v) at (0,0);
  \vertex[draw,circle,pattern=north east lines,minimum size=20pt] (blob) at (-2,0) {};
  \vertex (k)  at (1.0,-1.5);
  \vertex (kp) at (1.0, 1.5);

  \diagram*{
    (blob) -- [scalar] (v),
    (k)  -- [fermion, edge label=\(k\)] (v),
    (v)  -- [fermion, edge label'=\(k'\)] (kp),
  };
\end{feynman}
\end{tikzpicture}
\;\approx\; i g Q_{kk'}
\end{equation*}
in which the sum of the energies of the created pairs equals the frequency of the shape mode. 
In principle, any pair of fermions satisfying such condition can be created through this process. However, the corresponding vertex is given by the mode mixing matrix $Q_{kk'}$, and therefore, the preferred pairs of modes will be those of maximum (spatial) overlap with the source. We can identify this decay channel in the two low-energy peaks from the power spectrum at low values of $g$ in the first panel of \cref{fig:EnergyContributions}.

On the other hand, the method employed here to compute the excitation of the fermionic modes using Eqs. (\ref{dyn eq1}, \ref{dyn eq2}) admits a variety of additional terms describing state mixings. Within the framework of standard perturbative techniques, such excitations would be interpreted as arising from higher-order diagrams. However, as discussed above, the most significant results appear to be well 
captured by this simple perturbative analysis.

\subsection{Power emitted and amplitude decay: bosonic vs fermionic channels}
 
\begin{figure}[htb!]
    \centering
    \resizebox{0.95\textwidth}{!}{ 
        \begin{tabular}{cc}
            \begin{subfigure}{0.49\textwidth}
                \centering
                \includegraphics[width=\linewidth]{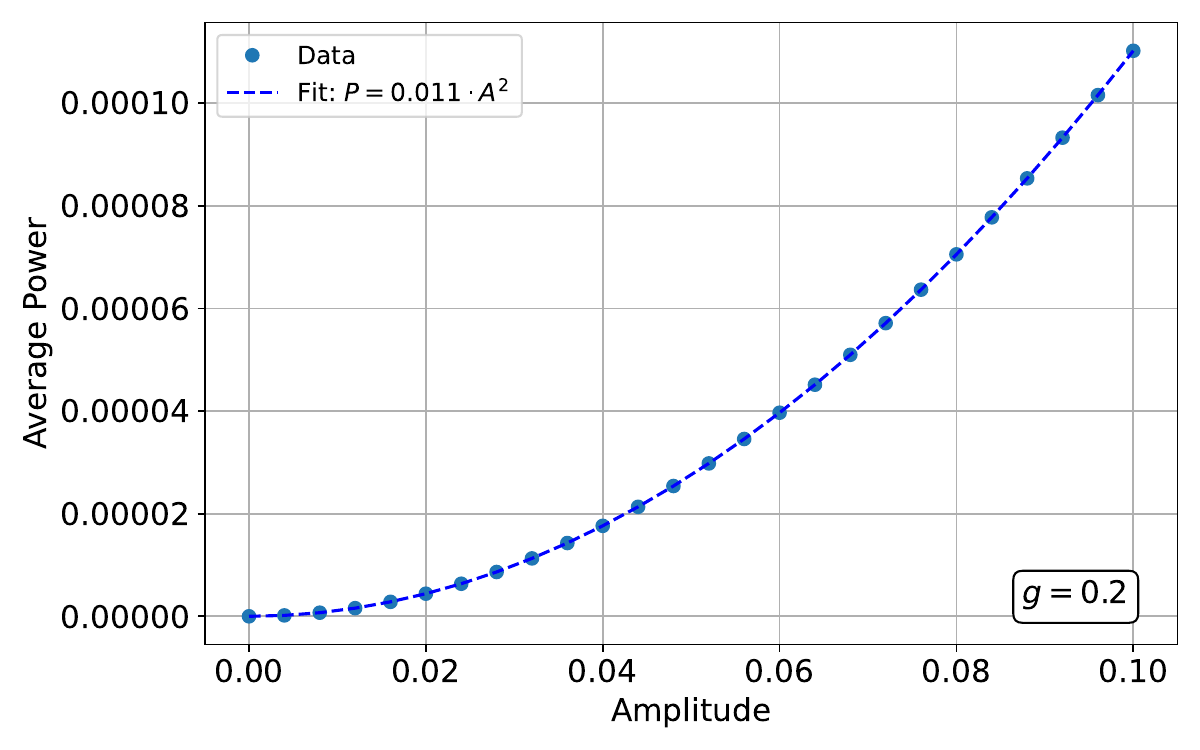}
            \end{subfigure} &
            \begin{subfigure}{0.49\textwidth}
                \centering
                \includegraphics[width=\linewidth]{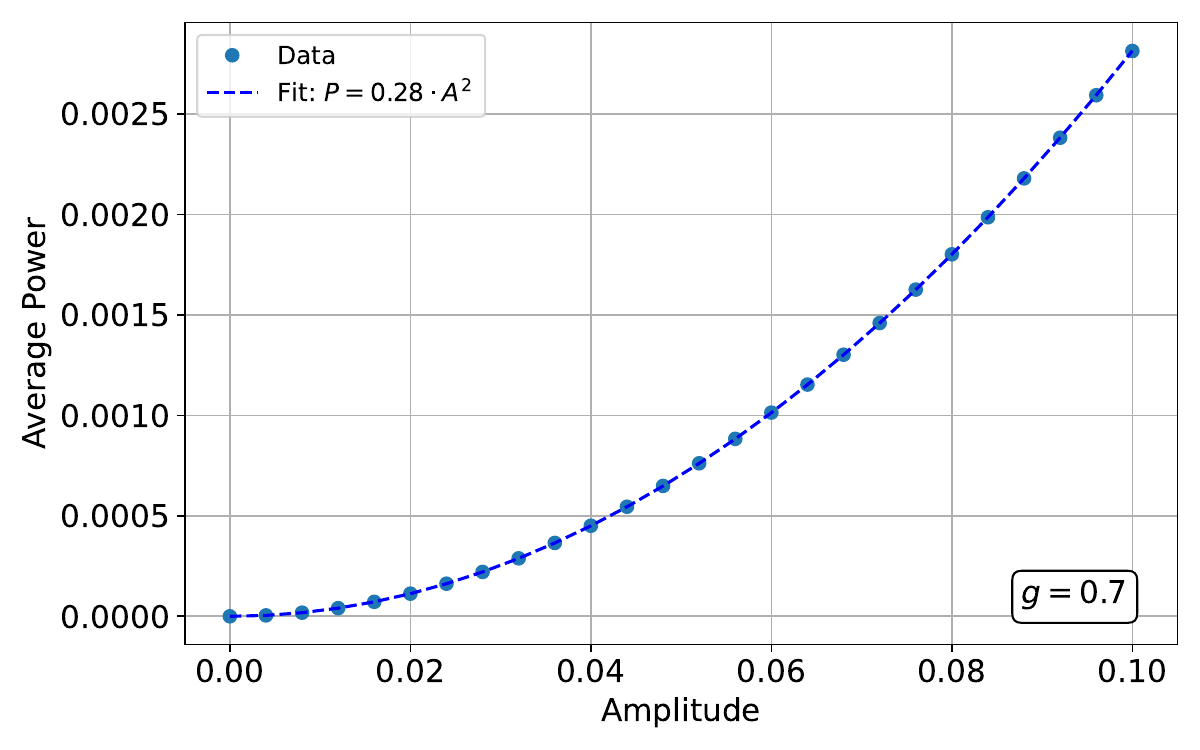}
            \end{subfigure} \\
            \begin{subfigure}{0.49\textwidth}
                \centering
                \includegraphics[width=\linewidth]{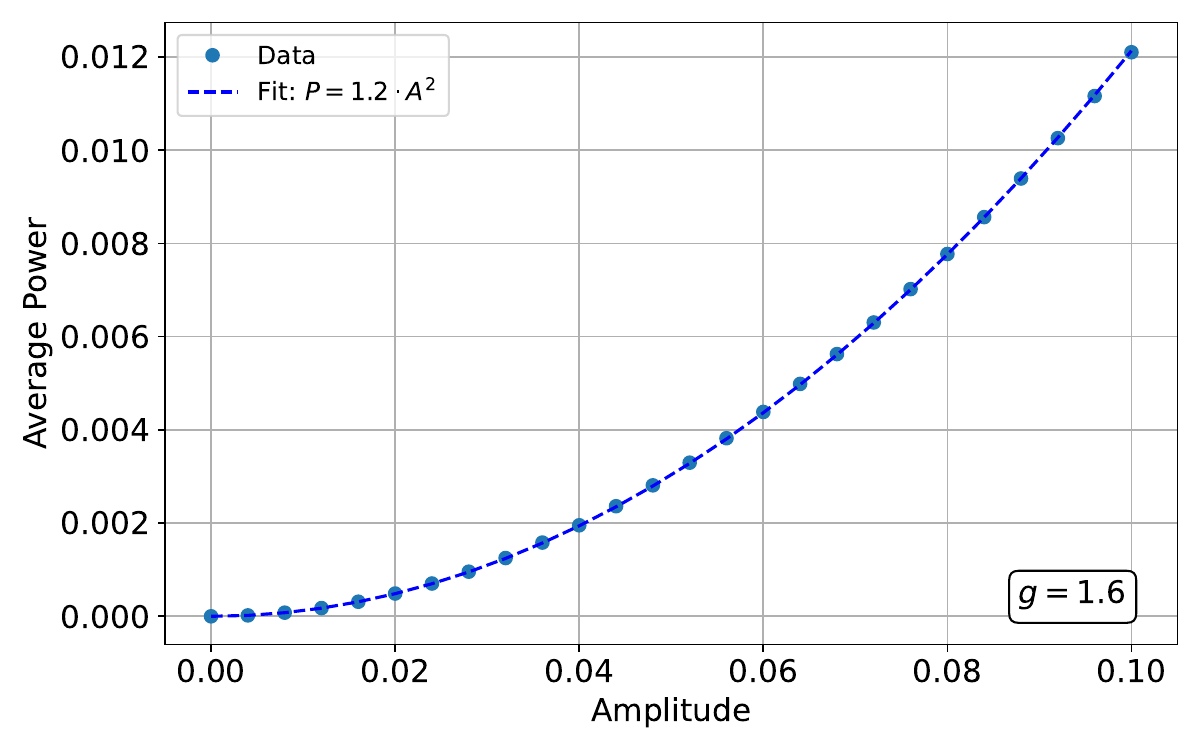}
            \end{subfigure} &
            \begin{subfigure}{0.49\textwidth}
                \centering
                \includegraphics[width=\linewidth]{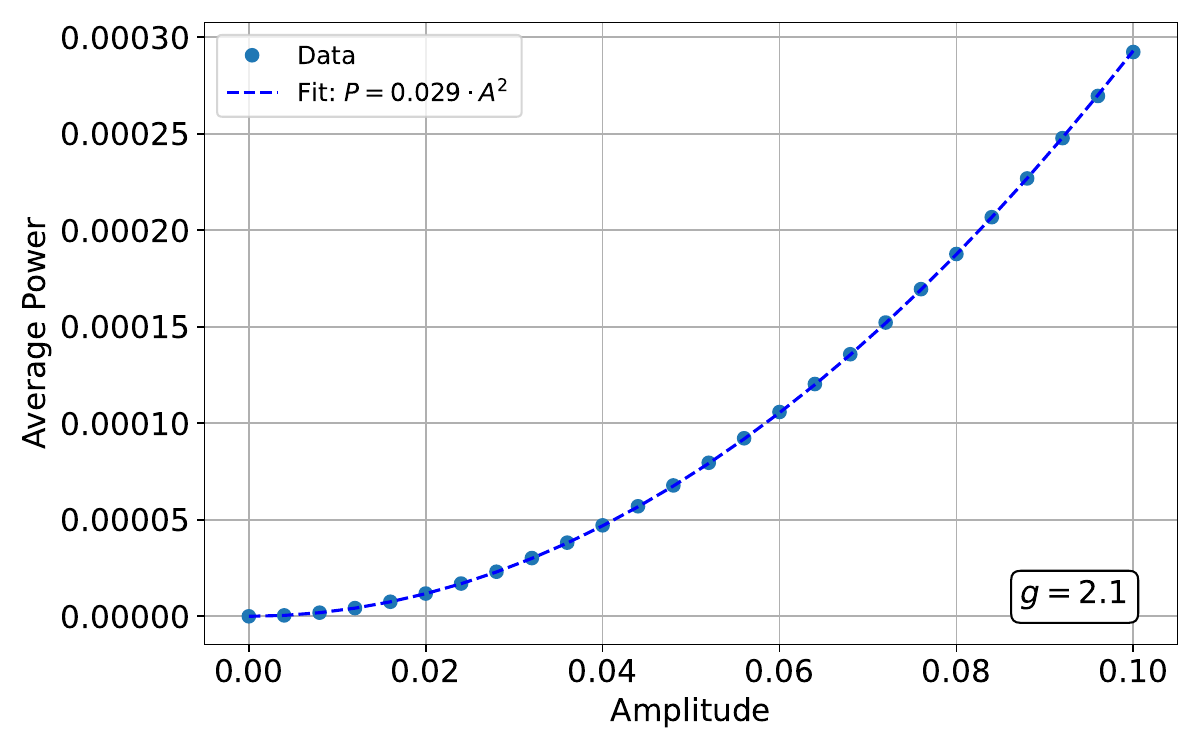}
            \end{subfigure}
        \end{tabular}
    }
    
    \caption{Average power in terms of the shape mode's amplitude.}
    \label{fig: AvPower_vs_A}
\end{figure}

The remaining question to address is whether the proposed decay mechanism can be compared to the decay in the purely scalar radiation case, if it dominates over it, or vice versa, and under which conditions it plays a relevant role. In order to answer this, one must study how the average power behaves in relation to the amplitude of the shape mode for different values of $g$. This yields the rate of energy loss through fermionic radiation from the shape mode. As shown in Figure \ref{fig: AvPower_vs_A}, the emitted average power grows quadratically with the shape mode's amplitude, provided the latter is sufficiently small.

\begin{figure}[htb!]
    \centering
    \includegraphics[width=0.55\textwidth]{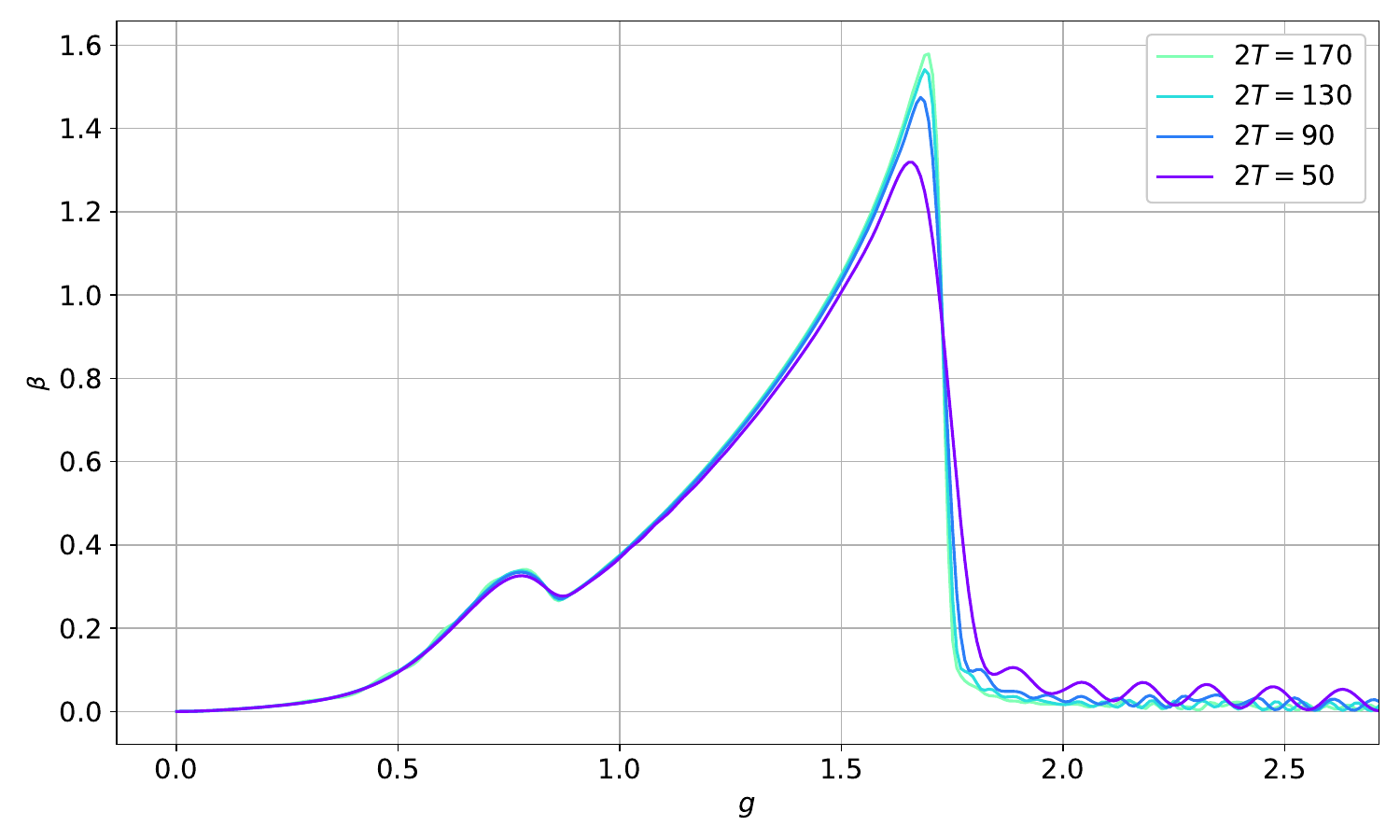}  
    \caption{$\beta$ function for different times the shape mode is switched on. }
    \label{fig:beta vs. g}
\end{figure}

This quadratic scaling suggests that, for $A_0\ll 1$, the average power can be approximated as
\begin{equation}
    P(g,A_0) \approx \beta(g) A_0^2\,,
\end{equation}
where $\beta(g)$ modulates the average power's behavior and, hence, it is no surprise that it has its same profile, as depicted in \cref{fig:beta vs. g}.

Additionally, given the energy of our scalar field configuration, an expression for the amplitude’s decay over time can be derived using energy conservation. The energy of the shape mode is given by $E=\frac{3}{2}A_0^2$, and therefore we may write the following energy balance equation
\begin{equation}
    \dv[]{E}{t}=-P(g,A_0)\implies \dv[]{A}{t}=-\frac{\beta(g)}{3}A\,.
    \label{dedt_1}
\end{equation} 

The above expression implies that the decay of the shape mode's amplitude due to fermionic emission is exponential, which differs from the well-known power law governing the amplitude's decay in the scalar case. As a first approximation, we may take the scalar channel into account as well in  the right-hand side of \eqref{dedt_1}, becoming:

\begin{equation}
       \dv[]{A}{t}=-\frac{\beta(g)}{3}A-\frac{3\alpha}{2}A^3\,,
    \label{dedt_2}
\end{equation}
where $\alpha=0.01$ in the $\lambda \phi^4$ model \cite{manton-1997}. This equation can be integrated to 
\begin{equation}
    A(t)=\sqrt{\frac{\beta}{\alpha}}\frac{A_0\sqrt{2}}{\sqrt{(9A_0^2+2\frac{\beta}{\alpha})e^{2\beta t/3}-9A_0^2}}\,.
\end{equation}

The first term of the expansion around $\beta\sim0$ recovers the decay of the amplitude in the purely scalar case, namely
\begin{equation}
    A(t)\sim \frac{A_0}{\sqrt{3\alpha A_0^2t+1}}\,.
\end{equation}

\begin{figure}[htb!]
    \centering
    \includegraphics[width=0.65\textwidth]{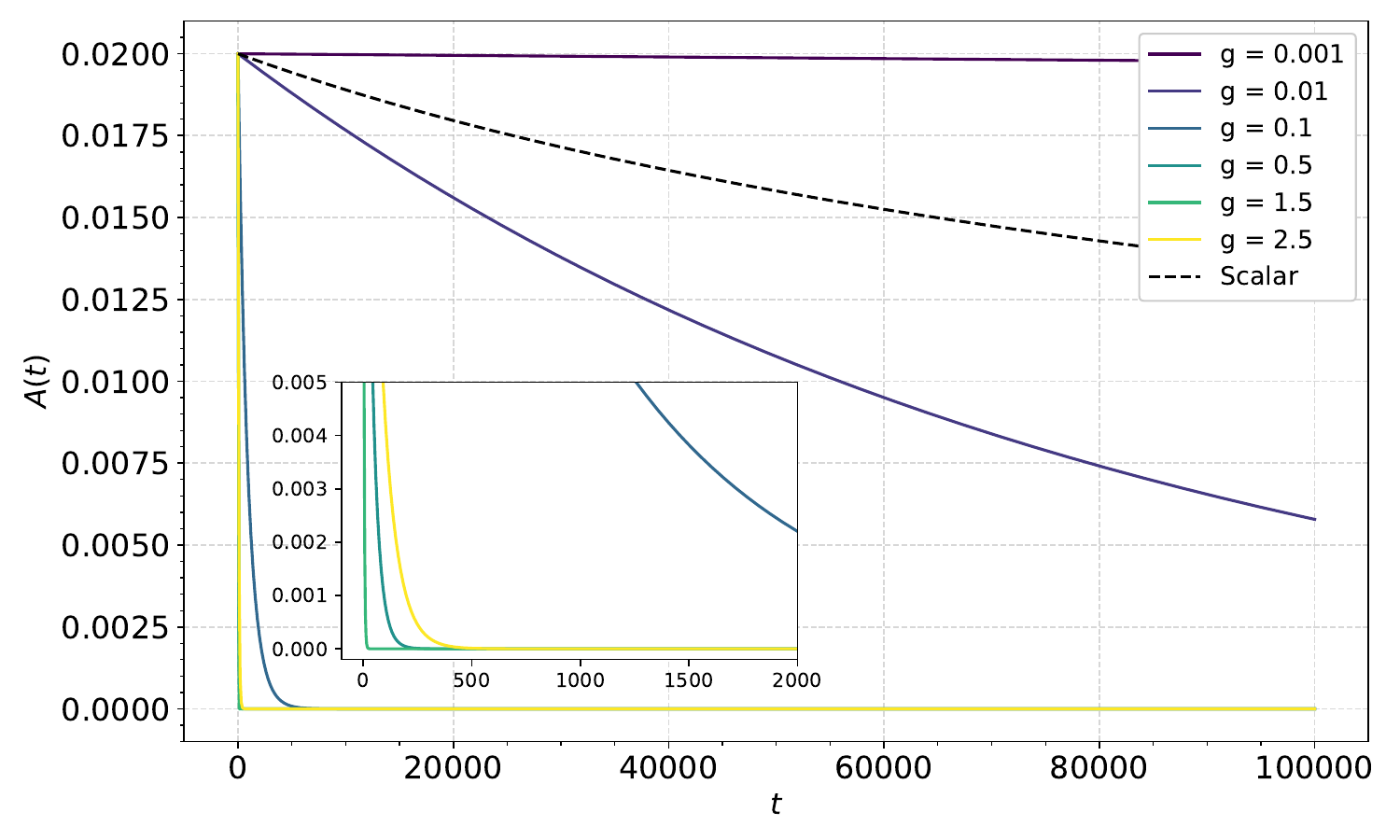}  
    \caption{Fermionic amplitude decay over time for several values of $g$ for an initial amplitude of $A_0=0.02$. The scalar decay is represented by a dashed black line. }
    \label{fig:Amplitude decay}
\end{figure}

We have therefore obtained the expected result: in the resonant regime, the fermion production dominates over scalar radiation, due to exponential decay of the amplitude in time (instead of power-like). In the non-resonant regime, the exponential suppression of emitted power implies an \emph{extremely} small exponential decay, so it may be neglected at some point with respect to the bosonic emission, particularly in the $T\to\infty$ limit. The decay of the shape mode's amplitude is plotted in Fig. \ref{fig:Amplitude decay} as a function of different values of $\beta$ and $g$. We remark that this constant is actually a function of the dimensionless Yukawa coupling $g$, hence it is fixed by the model. 

\section{Conclusions}

In this work we have analyzed the decay of the shape mode's amplitude of a kink via the emission of quanta from a fermionic quantum field, in the semi-classical approximation, namely, considering the excited kink as a classical, non-dynamical background. As opposed to previous approaches \cite{campos-2021}, we have studied the fermionic production fully non-perturbatively, by comparing the (time evolved) field modes associated to the asymptotic past and future and computing the associated Bogoliubov coefficients, as it is typically done for particle production from vacuum in other contexts such as quantum field theory in curved spacetimes. To achieve this goal, we have first reviewed the general procedure for the canonical quantization of a Dirac field in a time-dependent background and particularized it for the case of a kink with a time-dependent shape mode's amplitude. Further, we have reduced the problem of computing the Bogoliubov coefficients to solving a system of first order, coupled linear ordinary differential equations for a set of time-dependent functions which is easily solved using numerical methods.

We have found that, as long as the energy of the shape mode is larger than the mass gap of the asymptotic fermion states, not only will fermionic production happen but it will dominate over scalar emission, which in turn implies an exponential decay of the shape mode's amplitude. In other words, a small Yukawa coupling to a (massless) Dirac fermion de-stabilizes the otherwise linearly stable shape mode, accelerating its decay through the fermionic channel. This could have important consequences for the cosmological relevance of bound states of this kind in other solitonic solutions.

An interesting extension of this work would be to study the same process in a full quantum field theoretical framework, in which both the scalar and fermion fields are considered as quantum fields. The quantum decay of a scalar kink due to its own self-interaction has been recently computed in the full quantum regime in \cite{Evslin:2022wyx}, for which the result of Manton and Merabet \cite{manton-1997} is recovered when considering the classical limit. Another possible extension of this work would be to consider models with a much richer spectrum of fluctuations such as higher codimension defects, like vortices and monopoles, or models with a larger number of fields, with non-trivial features such as Feshbach resonances \cite{GarciaMartin-Caro:2025zkc}.  

\section*{Acknowledgements}

We thank Jarah Evslin for valuable discussions on this subject and related topics.
This work has been supported in part by the PID2021-123703NB-C21 grant funded by
MCIN/AEI/10.13039/501100011033/ and by ERDF;“ A way of making Europe”; the Basque
Government grant (IT-1628-22) and the Basque Foundation for Science (IKERBASQUE).
The work of AGMC was also supported by Grants No.ED481B-2025/059 and ED431B-2024/42 (Consellería de Cultura, Educación, Formación Profesional y Universidades, Xunta de Galicia).

\appendix
\section{Static solutions of the Dirac equation in the kink background}
\label{Appendix: A}

For completeness, in this appendix we detail the method to obtain solutions of the static Dirac equation in the presence of a $\lambda\phi^4$ kink. This amounts to solving (\ref{eigenvalue p}) for a two-component spinor of the form
\begin{equation}
    \psi^\pm(x)=\begin{pmatrix}
u^\pm(x)\\
v^\pm(x)
\end{pmatrix}\,.
\end{equation}

Therefore, from equation (\ref{eigenvalue p}) we get 
\begin{equation}
    \begin{pmatrix}
    0 & -\partial_x+g\phi_k(x)\\
    \partial_x+g\phi_k(x)&0
\end{pmatrix}
\begin{pmatrix}
u^\pm(x)\\
v^\pm(x)
\end{pmatrix}= \pm E\begin{pmatrix}
u^\pm(x)\\
v^\pm(x)
\end{pmatrix}\,.
\label{Dirac eq.}
\end{equation}

We shall start with the positive energy case. From the previous equation we obtain two coupled ordinary differential equations 
\begin{equation}
    \partial_xu^+=-g\phi_ku^++Ev^+\,,
    \label{coupled eq 1}
\end{equation}
\begin{equation}
    \partial_xv^+=g\phi_kv^+-Eu^+\,.
    \label{coupled eq 2}
\end{equation}

We can decouple them, in which case we arrive to two Schrödinger-like differential equations
\begin{equation}
    -\partial_x^2u^++(g^2\phi_k^2-g\partial_x\phi_k)u^+=E^2u^+\,,
    \label{decoupled 1}
\end{equation}
\begin{equation}
    -\partial_x^2v^++(g^2\phi_k^2+g\partial_x\phi_k)v^+=E^2v^+\,.
    \label{decoupled 2}
\end{equation}

Substituting the kink solution, they become
\begin{equation}
    -\partial_x^2u^++(g^2\tanh^2{x}-g\sech^2{x})u^+=E^2u^+\,,
    \label{diff eq u plus}
\end{equation}
\begin{equation}
    -\partial_x^2v^++(g^2\tanh^2{x}+g\sech^2{x})v^+=E^2v^+\,.
    \label{diff eq v plus}
\end{equation}

We can see that they are almost the same differential equations, they just differ on the potential
\begin{equation}
    U_\pm(x)=(g^2\tanh^2{x}\mp g\sech^2{x})\,.
    \label{potentials fermion}
\end{equation}

Let us focus on $u^+$ first. For any value of $g>0$ its potential has the shape of a potential well with a maximum $g^2$ at $x=\pm\infty$ and a minimum of $-g$ at $x=0$. In one dimension, this potential will have at least one bound state for any value of $g$. Moreover, since the potential becomes deeper for increasing $g$, it is logical to expect a higher number of bound states for larger values of $g$. Rearranging terms in (\ref{diff eq u plus}) we arrive at,
\begin{equation}
     -\partial_x^2u^++(g^2+g)(1-\sech^2{x})u^+=(E^2+g)u^+\,.
\end{equation}

This differential equation belongs to the same class as the Schrödinger-type equation discussed in \cref{sec: lambda_phi_4}, which yields the spectrum of perturbations of the $\lambda\phi^4$ kink, and its solution is provided in \cite{morse-1953}. Following the prescription shown there, the bounded levels are
\begin{equation}
    u_n^+=\frac{N_u^{(n)}}{(e^x+e^{-x})^{g-n_u}}F(-n_u,2g+1-n_u,g+1-n_u,\frac{e^{-x}}{e^x+e^{-x}})\,,
    \label{u bound}
\end{equation}
with energies
\begin{equation}
     E_{n_u}=\sqrt{n_u(2g-n_u)}>0\,,\quad n_u\in\mathbb{N},\quad n_u< g\,.
\end{equation}

Now we focus on $v^+$. Although the potentials are almost the same, for $0<g<1$ we have a potential barrier instead of a potential well, in which case the only possible bound state is with $E=0$. For this case we can take the initial coupled equations (\ref{coupled eq 1}), (\ref{coupled eq 2}) and see that their solutions are
\begin{equation}
    u_0^+=N^{(0)}_u\cosh^{-g}{x}\quad\text{and}\quad v_0^+=N^{(0)}_v\cosh^{g}{x}\,.
\end{equation}

However, unlike $u$, the expression for $v$ is not normalizable. Consequently, we must take $v=0$ by imposing $N_v^{(0)}=0$, which gives a single, non-degenerate solution for E$_0$=0:
\begin{equation}
    \psi^+_0=N_u^{(0)}\begin{pmatrix}
\cosh^{-g}{x}\\
0
\end{pmatrix}
\,.
\end{equation}

When $g>1$, we recover a potential well and thus more bound states are expected to be found. In this case, we can again make use of the previous prescription in order to find them. For that sake, equation (\ref{diff eq v plus}) can be rewritten as
\begin{equation}
     -\partial_x^2v^++(g^2-g)(1-\sech^2{x})v^+=(E^2-g)v^+\,,
\end{equation}
which, again, is solved in \cite{morse-1953}. In this case, the bound states are
\begin{equation}
    v_n^+=\frac{N^{(n)}_v}{(e^x+e^{-x})^{g-n_v-1}}F(-n_v,2g-n_v-1,g-n_v,\frac{e^{-x}}{e^x+e^{-x}})\,,
    \label{v bound}
\end{equation}
their energies being
\begin{equation}
    E_{n_v}=\sqrt{(n_v+1)(2g-n_v-1)}\,, \quad n_v<g-1\,.
\end{equation}

Since $u^+$ and $v^+$ are components of the same Dirac field, we expect them to have the same energy, i.e., $E_{n_u}=E_{n_v}$. This imposes the following relation
\begin{equation}
    n_u-n_v=1\,.
\end{equation}

Additionally, a relation between the normalization constants $N_u^{(n)}$, $N_v^{(n)}$ and the energy $E_n$ can be found by plugging the expressions of $u_n$ and $v_n$ back in one of the coupled differential equations (\ref{coupled eq 1}), (\ref{coupled eq 2}):
 \begin{equation}
     \frac{N_u^{(n)}}{N_v^{(n)}}=\frac{E_n}{n}\,,
     \label{Ratio Ns BoundS}
 \end{equation}
where $n\equiv n_u$ labels the nth mode. The previous relation is valid for $n\geq1$.

As far as scattering states are concerned, these ones are found for any value of energy which surpasses the mass threshold, $E_k>E_m=g$. The form of the scattering states is given in \cite{morse-1953} as well; for our case, they are expressed as
\begin{equation}
\begin{split}
     u_k^+=N^{(k)}_u(e^x+e^{-x})^{ik}F(-ik-g,-ik+g+1,1-ik,\frac{e^{-x}}{e^x+e^{-x}})\,.
     \label{scatter u}
\end{split}
\end{equation}
\begin{equation}
\begin{split}
     v_k^+=N^{(k)}_v(e^x+e^{-x})^{ik}F(-ik-g+1,-ik+g,1-ik,\frac{e^{-x}}{e^x+e^{-x}})\,.
     \label{scatter v}
\end{split}
\end{equation}

Both states have the same continuous energy $E_k=\sqrt{k^2+g^2}$, where $k>0$. Moreover, in the same way as has been done for bound states, we can obtain an analogous relation between $N^{(k)}_u$ and $N^{(k)}_v$: 
\begin{equation}
    \frac{N^{(k)}_u}{N^{(k)}_v}=\frac{E_k}{ik+g}\,.
    \label{Ratio Ns ScatteringS}
\end{equation}

The same procedure can be carried out for the negative energy case, although we can rapidly see that we arrive at the same results as the ones for positive energy. The coupled differential equations coming from (\ref{Dirac eq.}) now are
\begin{equation}
    \partial_xu^-=-g\phi_ku^--Ev^-\,,
    \label{coupled eq1 neg energy}
\end{equation}
\begin{equation}
    \partial_xv^-=g\phi_kv^-+Eu^-\,.
    \label{coupled eq2 neg energy}
\end{equation}

 After decoupling them, however, we obtain the same differential equations as (\ref{decoupled 1}) and (\ref{decoupled 2}), so the results (\ref{u bound}), (\ref{v bound}), (\ref{scatter u}) and (\ref{scatter v}) are valid for both positive and negative energies up to a change of sign in the energy.

\subsubsection{Normalization of the scattering states.}
Let us address the normalization of the solutions discussed above. The discrete part of the spectrum is given by a finite set of bound states which are square integrable on the real line and hence their normalization can be carried out without too much problem. On the other hand, the normalization of the scattering states can be harder to deal with as these are non-normalizable functions in the strict sense, and have to be normalized to the Dirac delta \eqref{normalization}.  Nevertheless, a clever approach can be made: we know these scattering states tend asymptotically to plane waves, as shown in \cref{asymptotic behaviour 1,asymptotic behaviour 2}. Consequently, rather than focusing our attention on the normalization of the whole solutions, we can try to normalize the asymptotic solutions. Each of the spinor components is a solution to its own Schrödinger equation that can be treated as a usual scattering problem from quantum mechanics. One can check that, as it is normally done in this kind of problems \cite{landau2013quantum}, the transmission and reflection coefficients, defined as
\begin{align}
    R_u &= \abs{\frac{C_{\rm{Reflected}}}{C_{\rm{Incident}}}}^2=\abs{\frac{  \Gamma(ik)\Gamma\left(-ik-g\right) \Gamma\left(-ik+g+1\right) }{ \Gamma(-ik)\Gamma\left(g+1\right) \Gamma\left(-g\right)}}^2\,,\\
    T_u &=\abs{\frac{C_{\rm{Transmitted}}}{C_{\rm{Incident}}}}^2 =\abs{\frac{\Gamma\left(-ik-g\right) \Gamma\left(-ik+g+1\right)}{ \Gamma(1 - ik) \Gamma(-ik)}}^2\,,
\end{align}

and

\begin{align}
    R_v &=\abs{\frac{\Gamma(ik)\Gamma\left(-ik+g\right) \Gamma\left(-ik-g+1\right)}{\Gamma(-ik)\Gamma\left(1-g\right) \Gamma\left(g\right)}}^2\,,\\
    T_v &=\abs{\frac{\Gamma\left(-ik+g\right) \Gamma\left(-ik-g+1\right)}{ \Gamma(1 - ik) \Gamma(-ik)}}^2\,,
\end{align}

do indeed fulfill the condition $T+R=1$. Hence, without loss of generality, the normalization constants can be chosen so that the incident wave in each of the components has unit amplitude. Therefore
\begin{align}
    N_u^{(k)}=\frac{\Gamma(-ik-g)\Gamma(-ik+g+1)}{\Gamma(1-ik)\Gamma(-ik)}\,,\\
    N_v^{(k)}=\frac{\Gamma(-ik+g)\Gamma(-ik-g+1)}{\Gamma(1-ik)\Gamma(-ik)}\,.
\end{align}

On top of that, since we are working with a two-component spinor, an additional factor of $1/\sqrt{2}$ must be imposed on the normalization constants, as well as a factor of $1/\sqrt{2\pi}$, coming from the normalization of plane waves to a Dirac delta, that is,
\begin{equation}
  \frac{1}{2\pi}\int_{-\infty}^{\infty}e^{-i(k-k')x} dx=\delta(k-k')\,.
  \label{delta_norm}
\end{equation}.

\section{\texorpdfstring{Dynamical equations for $\xi_k(t)$ and $\eta_k(t)$}{Dynamical equations for xi(t) and eta(t)}}
\label{Appendix: B}

The solutions of the time-dependent Dirac equation \eqref{EOM time dep} can be represented in the form given by \eqref{time dep expansion}, where  $\xi_k(t)$ and $\eta_k(t)$ denote the corresponding time-dependent components. These functions will fulfill their corresponding dynamical equations, which can be obtained by firstly substituting (\ref{time dep expansion}) in (\ref{EOM time dep}):
\begin{equation}
\begin{split}
     i\sumint dk[\Dot{\xi}_k(t)\psi^+_k(x)+\Dot{\eta}_k(t)\psi_k^-(x)]-\sumint dk[\xi_k(t)E_k\psi^+_k(x)-\eta_k(t)E_k\psi_k^-(x)]-\\
    -g\sumint dk[\varphi\sigma_1\xi_k(t)\psi^+_k(x)+\eta_k(t)\varphi\sigma_1\psi_k^-(x)]=0\,.
\label{eq.B1}
\end{split}
\end{equation}

At this point, one can project both sides onto $(\psi_{k'}^\pm(x))^\dagger$. Let us start first with $(\psi_{k'}^+(x))^\dagger$, which yields
    \begin{equation}
    \begin{split}
         i  \dot{\xi}_{k'}(t) -  \xi_{k'}(t) E_{k'} & -\\
        -g \sumint dk &\left[  \xi_k(t) \int dx (\psi_{k'}^+(x))^\dagger \varphi \sigma_1 \psi^+_k(x) +  \eta_k(t) \int dx (\psi_{k'}^+(x))^\dagger \varphi \sigma_1 \psi_k^-(x) \right] = 0\,.
    \end{split}
    \end{equation}
    
    After interchanging $k\leftrightarrow k'$ one arrives at
    \begin{equation}
        \quad i  \dot{\xi}_{k}(t) -  \xi_{k}(t) E_{k} - g \sumint dk' \left[  \xi_{k'}(t) R_{kk'} +  \eta_{k'}(t) Q_{kk'} \right] = 0\,,
    \end{equation}
    where
    \begin{align}
        Q_{kk'}=\int dx (\psi_{k}^+(x))^\dagger \varphi \sigma_1 \psi_{k'}^-(x)\,,\\
        R_{kk'}=\int dx (\psi_{k}^+(x))^\dagger \varphi \sigma_1 \psi^+_{k'}(x)\,.
    \end{align}
    
On the other hand, if we project onto $(\psi_{k'}^-(x))^\dagger$, we get
   \begin{equation}
    \begin{split}
        i  \dot{\eta}_{k'}(t) +  \eta_{k'}(t) E_{k'} & \\
        -g \sumint dk &\left( \xi_k(t) \int dx \, (\psi_{k'}^-(x))^\dagger \, \varphi \, \sigma_1 \, \psi^+_k(x) +  \eta_k(t) \int dx \, (\psi_{k'}^-(x))^\dagger \, \varphi \, \sigma_1 \, \psi_k^-(x) \right) = 0\,.
    \end{split}
\end{equation}

Again, after interchanging $k\leftrightarrow k'$ one arrives at
    \begin{equation}
        \quad i  \dot{\eta}_{k}(t) +  \eta_{k}(t) E_{k} - g \sumint dk' \left[  \xi_{k'}(t) Q'_{kk'} +  \eta_{k'}(t) S_{kk'} \right] = 0\,,
    \end{equation}
    where, taking into account that $\psi^-(x)=\sigma^3\psi^+(x)$,
    \begin{align}
        Q'_{kk'}=\int dx (\psi_{k}^-(x))^\dagger \varphi \sigma_1 \psi_{k'}^+(x)=-Q_{kk'}\,,\\
        S_{kk'}=\int dx (\psi_{k}^-(x))^\dagger \varphi \sigma_1 \psi^-_{k'}(x)=-R_{kk'}\,.
    \end{align}

\section{Bogoliubov coefficients, transformations and closure relations}
\label{Appendix: C}
Since both $\psi^{(in)\pm}_k$ and $\psi^{(out)\pm}_k$ form complete sets of solutions of the time-dependent Dirac equation (\ref{EOM time dep}), the $out$ modes can be expressed as an expansion in terms of the $in$ modes (and vice-versa) as follows
\begin{align}
    \psi^{(out)+}_k=\sumint dq\left[\alpha_{kq}\psi^{(in)+}_q+\beta_{kq}\psi^{(in)-}_q\right]\,,\\
    \psi^{(out)-}_k=\sumint dq\left[\gamma_{kq}\psi^{(in)+}_q+\rho_{kq}\psi^{(in)-}_q\right]\,.
\end{align}

In order to get the Bogoliubov coefficients, one can multiply both sides of these equations by $(\psi^{(in)\pm}_j)^\dagger$ and integrate over all space:
\begin{itemize}
    \item $\underline{\alpha_{kq}}:$
    \begin{equation}
    \begin{split}
        \int dx (\psi^{(in)+}_j)^\dagger\psi^{(out)+}_k=\int dx (\psi^{(in)+}_j)^\dagger\sumint dq\left[\alpha_{kq}\psi^{(in)+}_q+\beta_{kq}\psi^{(in)-}_q\right]\Leftrightarrow\\
        \Leftrightarrow \langle \psi^{(in)+}_j, \psi^{(out)+}_k\rangle_D=\langle\psi^{(in)+}_j, \sumint dq\left[\alpha_{kq}\psi^{(in)+}_q+\beta_{kq}\psi^{(in)-}_q\right]\rangle_D&=\\
        =\sumint dq\left[\alpha_{kq}\langle \psi^{(in)+}_j, \psi^{(in)+}_q\rangle_D+\beta_{kq}\langle \psi^{(in)+}_j, \psi^{(in)-}_q\rangle_D\right]=\sumint dq\alpha_{kq}\delta_{jq}=\alpha_{kj}\,.
     \end{split}
    \end{equation}

    In the last row the normalization condition for Dirac fields has been used. Thus,
    \begin{equation}
        \alpha_{kq}=\langle \psi^{(in)+}_q, \psi^{(out)+}_k\rangle_D\,.
    \end{equation}
\end{itemize}

The rest of the Bogoliubov coefficients follow by applying the same procedure:
\begin{align}
    \beta_{kq}=\langle \psi^{(in)-}_q, \psi^{(out)+}_k\rangle_D\,,\\
    \gamma_{kq}=\langle \psi^{(in)+}_q, \psi^{(out)-}_k\rangle_D\,,\\
    \rho_{kq}=\langle \psi^{(in)-}_q, \psi^{(out)-}_k\rangle_D\,.
\end{align}

 Furthermore, we can make use of the charge conjugation operation that relates particles and antiparticles, $(\psi^-(x))^c=\sigma_3\psi^+(x)^*$ \cite{charmchi-2014}, to see that only 2 of the 4 Bogoliubov coefficients are truly independent from each other. \\

Starting off with $\alpha_{kq}$ and $\rho_{kq}$:
\begin{equation}
    \rho^*_{kq}=\int (u^{(in)-}_q (u^{(out)-}_k)^*+v^{(in)-}_q (v^{(out)-}_k))^*dx\,,
\end{equation}
using the charge conjugation operation
\begin{equation}
    (\psi^-(x))^c=\sigma_3\psi^+(x)^*\leftrightarrow\begin{pmatrix}
    (u^-(x))^c\\
    (v^-(x))^c
\end{pmatrix}=\begin{pmatrix}
    (u^+(x))^*\\
    -(v^+(x))^*
\end{pmatrix}\,.
\end{equation}

Thus
\begin{equation}
    \rho^*_{kq}=\int ((u^{(in)+}_q)^* (u^{(out)+}_k)+(v^{(in)+}_q)^* v^{(out)+}_k)dx=\alpha_{kq}\,.
\end{equation}

The same can be done for $\beta_{kq}$ and $\gamma_{kq}$:
\begin{equation}
    \gamma^*_{kq}=\int (u^{(in)+}_q(u^{(out)-}_k)^*+ v^{(in)+}_q(v^{(out)-}_k)^*)dx=\int ((u^{(in)-}_q)^*u^{(out)+}_k+ (v^{(in)-}_q)^*v^{(out)+}_k)dx=\beta_{kq}\,.
\end{equation}

Consequently, we can restrict ourselves to work just with $\alpha_{kq}$ and $\beta_{kq}$.\\

By the same procedure, using the expansion of the field operator $\Hat{\psi}(x,t)$ in terms of the $in$ modes as a starting point, one can express the creation and annihilation operators of one Fock space as an expansion in terms of the operators of the other space. Firstly,
\begin{equation}
\begin{split}
    \hat{\psi}(x,t)&=\sumint dk[\hat b_k^{(\rm in)} \psi^{(\rm in)+}_k(x,t)+\hat d^{(\rm in) \dagger}_k \psi^{(\rm in)-}_k(x,t)]\Leftrightarrow\\
    \Leftrightarrow \langle\psi_q^{(in)+},\hat{\psi}\rangle_D &=\sumint dk[\hat b_k^{(\rm in)}\langle\psi_q^{(in)+},\psi^{(\rm in)+}_k\rangle_D+\hat d^{(\rm in) \dagger}_k \langle\psi_q^{(in)+},\psi^{(\rm in)-}_k\rangle_D]=\sumint dk\hat b_k^{(\rm in)}\delta_{qk}\,.
\end{split}
\end{equation}

Hence,
\begin{equation}
    \hat b_q^{(\rm in)}=\langle\psi_q^{(in)+},\hat{\psi}\rangle_D\,.
\end{equation}

Now, one can substitute $\Hat{\psi}(x,t)$ by its expansion in terms of $out$ modes. Consequently,
\begin{equation}
\begin{split}
     \Hat{b}_k^{(\rm in)}=\langle\psi_k^{(in)+},\hat{\psi}\rangle_D&=\sumint dq[\hat b^{(\rm out)}_q \langle\psi_k^{(in)+},\psi^{(\rm out)+}_q\rangle_D+\hat d^{(\rm out) \dagger}_q \langle\psi_k^{(in)+},\psi^{(\rm out)-}_q\rangle_D]\Leftrightarrow\\
     \Leftrightarrow \hat b_k^{(\rm in)}&=\sumint dq[\hat b^{(\rm out)}_q \alpha_{qk}+\hat d^{(\rm out) \dagger}_q \gamma_{qk}]=\sumint dq[\hat b^{(\rm out)}_q \alpha_{qk}+\hat d^{(\rm out) \dagger}_q \beta_{qk}^*]\,.
\end{split}
\end{equation}

The same process can be repeated in order to get $\hat{d}_k^{\rm (in)}$:
\begin{equation}
    \hat{d}_k^{\rm (in)}=\sumint dq[\hat b^{(\rm out)\dagger}_q\beta^*_{qk}+\hat d^{(\rm out)}_q \alpha_{qk}]\,.
\end{equation}

As far as the closure relations go, one just needs to develop the anti-commutation relations of the creation/annihilation operators to see that
\begin{equation}
\begin{split}
    \{\Hat{b}_k^{(\rm in)}, \Hat{b}_{k'}^{(\rm in)\dagger}\} &= \sumint dq \sumint dq' \left( \{\hat{b}^{(\rm out)}_q \alpha_{qk} + \hat{d}^{(\rm out)\dagger}_q \beta_{qk}^*, \hat{b}^{(\rm out)\dagger}_{q'} \alpha_{q'k'}^* + \hat{d}^{(\rm out)}_{q'} \beta_{q'k'}\} \right) \\
    &= \sumint dq (\alpha_{qk} \alpha_{qk'}^*+\beta_{qk}^* \beta_{qk'})\,.
\end{split}
\end{equation}

Thus, in order $\{\Hat{b}_k^{(\rm in)}, \Hat{b}_{k'}^{(\rm in)\dagger}\}=\delta_{kk'}$ to be satisfied, the Bogoliubov coefficients must fulfill the following closure relation:
\begin{equation}
    \sumint dq (\alpha_{qk} \alpha_{qk'}^*+\beta_{qk}^* \beta_{qk'})=\delta_{kk'}\,.
\end{equation}

Equivalently, the remaining closure relation can be found via the same procedure. Taking the following anti-commutation relation:
\begin{equation}
    \begin{split}
    \{\Hat{b}_k^{(\rm in)}, \Hat{d}_{k'}^{(\rm in)}\}=\sumint dq (\alpha_{qk} \beta_{qk'}^*+\beta_{qk}^* \alpha_{qk'})\,,
    \end{split}
\end{equation}
and since
\begin{equation}
     \{\Hat{b}_k^{(\rm in)}, \Hat{d}_{k'}^{(\rm in)}\}=0\,,
\end{equation}
we conclude,
\begin{equation}
    \sumint dq (\alpha_{qk} \beta_{qk'}^*+\beta_{qk}^* \alpha_{qk'})=0\,.
\end{equation}

\begin{section}{Numerical checks}
\label{Appendix D}

Throughout the numerical part of the work, we have replaced the integral over scattering states with a Riemann sum, which inevitably introduces a finite spacing between modes or, equivalently, limits the number of scattering modes that contribute to the dynamics of the system. Here we address the convergence of the results depending on the number of such states included. 

In particular, we want to focus on the problem arising with the time evolution of the system depending on the resolution in the $k$ space. This issue has been already commented on \cite{Navarro-Obregon:2023}, although in the context of the collective coordinate approach for the scalar decay of the shape mode. As explained there, the discretisation of the scattering states imposes a maximum timescale, of the order of $1/\Delta k$, for which the results are trustable. Beyond that timescale, the behavior of the system deviates from that of the ``full-spectrum" (continuum) case.

We can see this behavior in the average power emitted with respect to the coupling constant, illustrated in \cref{fig: App P vs g}. This does not seem to be an issue for asymptotic times used during the work (left picture), but may become a real problem for larger magnitude timescales. Thus, one should be careful with the number of scattering modes chosen in the latter scenario.

\begin{figure}[htbp]
    \centering
    \begin{subfigure}[b]{0.49\textwidth}
        \centering
        \includegraphics[width=\textwidth]{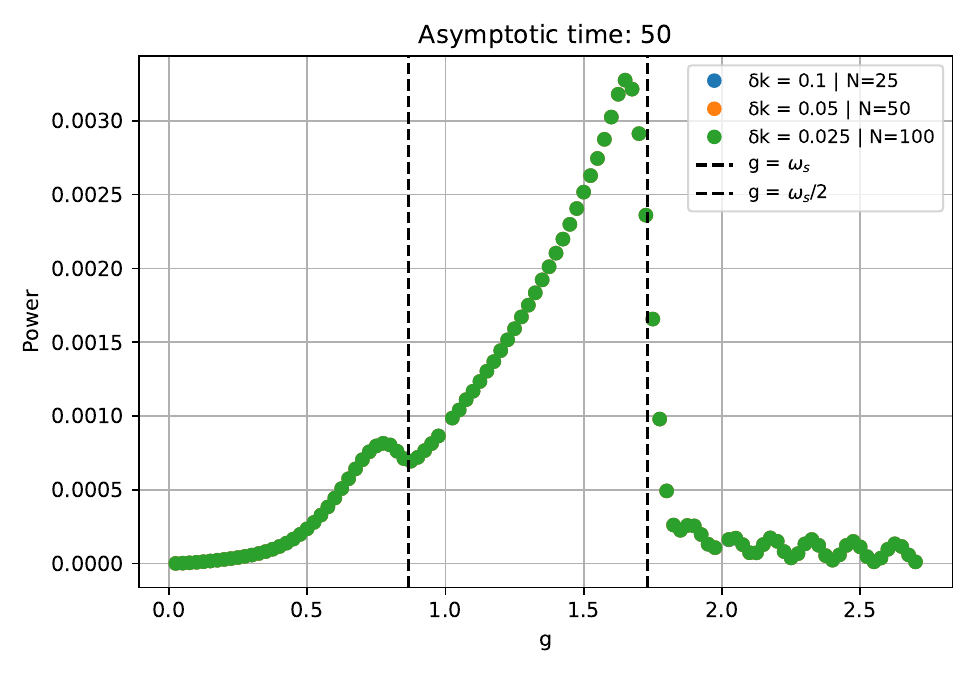}
    
        \label{fig:sub1}
    \end{subfigure}
    \hfill
    \begin{subfigure}[b]{0.49\textwidth}
        \centering
        \includegraphics[width=\textwidth]{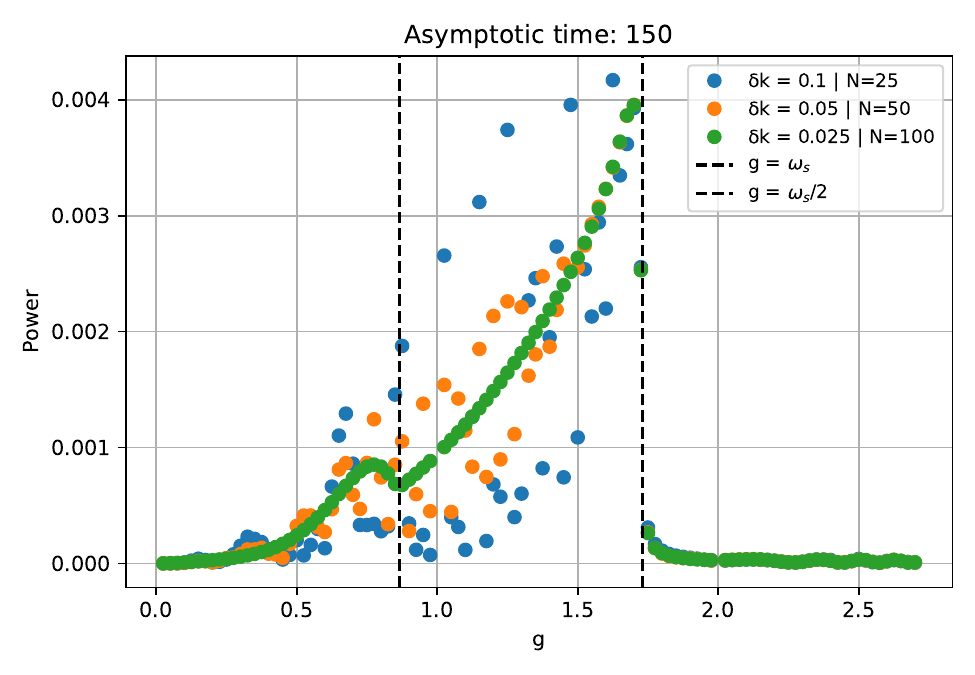}

        \label{fig:sub2}
    \end{subfigure}
    \caption{Average power as a function of the coupling constant,  for different asymptotic times and numbers of scattering states. To clarify, we note that in the left picture, points corresponding to different $N$ values lie on top of each other.}
    \label{fig: App P vs g}
\end{figure}

The underlying reason is straightforward. Since we maintain the range of the $k$ space fixed to $k\in(0,2.5)$, the lowest non-zero value of $k$ becomes gradually smaller as the spectral resolution increases. In order to ``sample" its corresponding frequency, higher times will be needed. Hence, for a small number of scattering states, the system deviates from the full-spectrum behavior for large asymptotic times, not due to a numerical error $per$ $se$, but because the relevant frequencies that should be sampled are absent. In other words, the system does not know how it is supposed to evolve at late times because the modes that govern that behavior are not present in the discretisation.

Nevertheless, provided a good equilibrium between the number of scattering modes and the asymptotic times, this issue should not become problematic. Therefore, for the timescales used in most part of this work, $t\in(-50,50)$, a discretisation with $N=60$ is sufficient. However, if the asymptotic times were to be extended, a larger $N$ should be required to ensure reliable results.

\end{section}


\begin{thebibliography}{38}%
\makeatletter
\providecommand \@ifxundefined [1]{%
 \@ifx{#1\undefined}
}%
\providecommand \@ifnum [1]{%
 \ifnum #1\expandafter \@firstoftwo
 \else \expandafter \@secondoftwo
 \fi
}%
\providecommand \@ifx [1]{%
 \ifx #1\expandafter \@firstoftwo
 \else \expandafter \@secondoftwo
 \fi
}%
\providecommand \natexlab [1]{#1}%
\providecommand \enquote  [1]{``#1''}%
\providecommand \bibnamefont  [1]{#1}%
\providecommand \bibfnamefont [1]{#1}%
\providecommand \citenamefont [1]{#1}%
\providecommand \href@noop [0]{\@secondoftwo}%
\providecommand \href [0]{\begingroup \@sanitize@url \@href}%
\providecommand \@href[1]{\@@startlink{#1}\@@href}%
\providecommand \@@href[1]{\endgroup#1\@@endlink}%
\providecommand \@sanitize@url [0]{\catcode `\\12\catcode `\$12\catcode `\&12\catcode `\#12\catcode `\^12\catcode `\_12\catcode `\%12\relax}%
\providecommand \@@startlink[1]{}%
\providecommand \@@endlink[0]{}%
\providecommand \url  [0]{\begingroup\@sanitize@url \@url }%
\providecommand \@url [1]{\endgroup\@href {#1}{\urlprefix }}%
\providecommand \urlprefix  [0]{URL }%
\providecommand \Eprint [0]{\href }%
\providecommand \doibase [0]{http://dx.doi.org/}%
\providecommand \selectlanguage [0]{\@gobble}%
\providecommand \bibinfo  [0]{\@secondoftwo}%
\providecommand \bibfield  [0]{\@secondoftwo}%
\providecommand \translation [1]{[#1]}%
\providecommand \BibitemOpen [0]{}%
\providecommand \bibitemStop [0]{}%
\providecommand \bibitemNoStop [0]{.\EOS\space}%
\providecommand \EOS [0]{\spacefactor3000\relax}%
\providecommand \BibitemShut  [1]{\csname bibitem#1\endcsname}%
\let\auto@bib@innerbib\@empty
\bibitem [{\citenamefont {Campos}\ and\ \citenamefont {Verdaguer}(1992)}]{Campos:1991ff}%
  \BibitemOpen
  \bibfield  {author} {\bibinfo {author} {\bibfnamefont {A.}~\bibnamefont {Campos}}\ and\ \bibinfo {author} {\bibfnamefont {E.}~\bibnamefont {Verdaguer}},\ }\href {\doibase 10.1103/PhysRevD.45.4428} {\bibfield  {journal} {\bibinfo  {journal} {Phys. Rev. D}\ }\textbf {\bibinfo {volume} {45}},\ \bibinfo {pages} {4428} (\bibinfo {year} {1992})}\BibitemShut {NoStop}%
\bibitem [{\citenamefont {Greene}\ and\ \citenamefont {Kofman}(1999)}]{Greene:1998nh}%
  \BibitemOpen
  \bibfield  {author} {\bibinfo {author} {\bibfnamefont {P.~B.}\ \bibnamefont {Greene}}\ and\ \bibinfo {author} {\bibfnamefont {L.}~\bibnamefont {Kofman}},\ }\href {\doibase 10.1016/S0370-2693(99)00020-9} {\bibfield  {journal} {\bibinfo  {journal} {Phys. Lett. B}\ }\textbf {\bibinfo {volume} {448}},\ \bibinfo {pages} {6} (\bibinfo {year} {1999})},\ \Eprint {http://arxiv.org/abs/hep-ph/9807339} {arXiv:hep-ph/9807339} \BibitemShut {NoStop}%
\bibitem [{\citenamefont {Parker}(1971)}]{Parker:1971pt}%
  \BibitemOpen
  \bibfield  {author} {\bibinfo {author} {\bibfnamefont {L.}~\bibnamefont {Parker}},\ }\href {\doibase 10.1103/PhysRevD.3.346} {\bibfield  {journal} {\bibinfo  {journal} {Phys. Rev. D}\ }\textbf {\bibinfo {volume} {3}},\ \bibinfo {pages} {346} (\bibinfo {year} {1971})},\ \bibinfo {note} {[Erratum: Phys.Rev.D 3, 2546--2546 (1971)]}\BibitemShut {NoStop}%
\bibitem [{\citenamefont {Baacke}\ \emph {et~al.}(1998)\citenamefont {Baacke}, \citenamefont {Heitmann},\ and\ \citenamefont {Patzold}}]{Baacke:1998di}%
  \BibitemOpen
  \bibfield  {author} {\bibinfo {author} {\bibfnamefont {J.}~\bibnamefont {Baacke}}, \bibinfo {author} {\bibfnamefont {K.}~\bibnamefont {Heitmann}}, \ and\ \bibinfo {author} {\bibfnamefont {C.}~\bibnamefont {Patzold}},\ }\href {\doibase 10.1103/PhysRevD.58.125013} {\bibfield  {journal} {\bibinfo  {journal} {Phys. Rev. D}\ }\textbf {\bibinfo {volume} {58}},\ \bibinfo {pages} {125013} (\bibinfo {year} {1998})},\ \Eprint {http://arxiv.org/abs/hep-ph/9806205} {arXiv:hep-ph/9806205} \BibitemShut {NoStop}%
\bibitem [{\citenamefont {Fuentes}\ \emph {et~al.}(2010)\citenamefont {Fuentes}, \citenamefont {Mann}, \citenamefont {Martin-Martinez},\ and\ \citenamefont {Moradi}}]{Fuentes:2010dt}%
  \BibitemOpen
  \bibfield  {author} {\bibinfo {author} {\bibfnamefont {I.}~\bibnamefont {Fuentes}}, \bibinfo {author} {\bibfnamefont {R.~B.}\ \bibnamefont {Mann}}, \bibinfo {author} {\bibfnamefont {E.}~\bibnamefont {Martin-Martinez}}, \ and\ \bibinfo {author} {\bibfnamefont {S.}~\bibnamefont {Moradi}},\ }\href {\doibase 10.1103/PhysRevD.82.045030} {\bibfield  {journal} {\bibinfo  {journal} {Phys. Rev. D}\ }\textbf {\bibinfo {volume} {82}},\ \bibinfo {pages} {045030} (\bibinfo {year} {2010})},\ \Eprint {http://arxiv.org/abs/1007.1569} {arXiv:1007.1569 [quant-ph]} \BibitemShut {NoStop}%
\bibitem [{\citenamefont {Vilenkin}\ and\ \citenamefont {Shellard}(2000)}]{Vilenkin_2000}%
  \BibitemOpen
  \bibfield  {author} {\bibinfo {author} {\bibfnamefont {A.}~\bibnamefont {Vilenkin}}\ and\ \bibinfo {author} {\bibfnamefont {E.~P.~S.}\ \bibnamefont {Shellard}},\ }\href@noop {} {\emph {\bibinfo {title} {{Cosmic Strings and Other Topological Defects}}}}\ (\bibinfo  {publisher} {Cambridge University Press},\ \bibinfo {year} {2000})\BibitemShut {NoStop}%
\bibitem [{\citenamefont {Rubakov}(2002)}]{Rubakov:2002fi}%
  \BibitemOpen
  \bibfield  {author} {\bibinfo {author} {\bibfnamefont {V.~A.}\ \bibnamefont {Rubakov}},\ }\href@noop {} {\emph {\bibinfo {title} {{Classical theory of gauge fields}}}}\ (\bibinfo  {publisher} {Princeton University Press},\ \bibinfo {address} {Princeton, New Jersey},\ \bibinfo {year} {2002})\BibitemShut {NoStop}%
\bibitem [{\citenamefont {Weinberg}(2012)}]{Weinberg:2012pjx}%
  \BibitemOpen
  \bibfield  {author} {\bibinfo {author} {\bibfnamefont {E.~J.}\ \bibnamefont {Weinberg}},\ }\href {\doibase 10.1017/CBO9781139017787} {\emph {\bibinfo {title} {{Classical solutions in quantum field theory}: {Solitons and Instantons in High Energy Physics}}}},\ Cambridge Monographs on Mathematical Physics\ (\bibinfo  {publisher} {Cambridge University Press},\ \bibinfo {year} {2012})\BibitemShut {NoStop}%
\bibitem [{\citenamefont {Cohen}\ \emph {et~al.}(1986)\citenamefont {Cohen}, \citenamefont {Coleman}, \citenamefont {Georgi},\ and\ \citenamefont {Manohar}}]{Cohen:1986ct}%
  \BibitemOpen
  \bibfield  {author} {\bibinfo {author} {\bibfnamefont {A.~G.}\ \bibnamefont {Cohen}}, \bibinfo {author} {\bibfnamefont {S.~R.}\ \bibnamefont {Coleman}}, \bibinfo {author} {\bibfnamefont {H.}~\bibnamefont {Georgi}}, \ and\ \bibinfo {author} {\bibfnamefont {A.}~\bibnamefont {Manohar}},\ }\href {\doibase 10.1016/0550-3213(86)90004-0} {\bibfield  {journal} {\bibinfo  {journal} {Nucl. Phys. B}\ }\textbf {\bibinfo {volume} {272}},\ \bibinfo {pages} {301} (\bibinfo {year} {1986})}\BibitemShut {NoStop}%
\bibitem [{\citenamefont {Clark}(2006)}]{Clark:2005zc}%
  \BibitemOpen
  \bibfield  {author} {\bibinfo {author} {\bibfnamefont {S.~S.}\ \bibnamefont {Clark}},\ }\href {\doibase 10.1016/j.nuclphysb.2006.08.019} {\bibfield  {journal} {\bibinfo  {journal} {Nucl. Phys. B}\ }\textbf {\bibinfo {volume} {756}},\ \bibinfo {pages} {38} (\bibinfo {year} {2006})},\ \Eprint {http://arxiv.org/abs/hep-ph/0510078} {arXiv:hep-ph/0510078} \BibitemShut {NoStop}%
\bibitem [{\citenamefont {Hertzberg}(2010)}]{Hertzberg:2010yz}%
  \BibitemOpen
  \bibfield  {author} {\bibinfo {author} {\bibfnamefont {M.~P.}\ \bibnamefont {Hertzberg}},\ }\href {\doibase 10.1103/PhysRevD.82.045022} {\bibfield  {journal} {\bibinfo  {journal} {Phys. Rev. D}\ }\textbf {\bibinfo {volume} {82}},\ \bibinfo {pages} {045022} (\bibinfo {year} {2010})},\ \Eprint {http://arxiv.org/abs/1003.3459} {arXiv:1003.3459 [hep-th]} \BibitemShut {NoStop}%
\bibitem [{\citenamefont {Saffin}(2017)}]{Saffin:2016kof}%
  \BibitemOpen
  \bibfield  {author} {\bibinfo {author} {\bibfnamefont {P.~M.}\ \bibnamefont {Saffin}},\ }\href {\doibase 10.1007/JHEP07(2017)126} {\bibfield  {journal} {\bibinfo  {journal} {JHEP}\ }\textbf {\bibinfo {volume} {07}},\ \bibinfo {pages} {126} (\bibinfo {year} {2017})},\ \Eprint {http://arxiv.org/abs/1612.02014} {arXiv:1612.02014 [hep-th]} \BibitemShut {NoStop}%
\bibitem [{\citenamefont {Evslin}\ \emph {et~al.}(2023)\citenamefont {Evslin}, \citenamefont {Roma{\'n}czukiewicz},\ and\ \citenamefont {Wereszczy{\'n}ski}}]{Evslin:2023qbv}%
  \BibitemOpen
  \bibfield  {author} {\bibinfo {author} {\bibfnamefont {J.}~\bibnamefont {Evslin}}, \bibinfo {author} {\bibfnamefont {T.}~\bibnamefont {Roma{\'n}czukiewicz}}, \ and\ \bibinfo {author} {\bibfnamefont {A.}~\bibnamefont {Wereszczy{\'n}ski}},\ }\href {\doibase 10.1007/JHEP08(2023)182} {\bibfield  {journal} {\bibinfo  {journal} {JHEP}\ }\textbf {\bibinfo {volume} {08}},\ \bibinfo {pages} {182} (\bibinfo {year} {2023})},\ \Eprint {http://arxiv.org/abs/2305.18056} {arXiv:2305.18056 [hep-th]} \BibitemShut {NoStop}%
\bibitem [{\citenamefont {Oll{\'e}}\ \emph {et~al.}(2019)\citenamefont {Oll{\'e}}, \citenamefont {Pujolas}, \citenamefont {Vachaspati},\ and\ \citenamefont {Zahariade}}]{Olle:2019skb}%
  \BibitemOpen
  \bibfield  {author} {\bibinfo {author} {\bibfnamefont {J.}~\bibnamefont {Oll{\'e}}}, \bibinfo {author} {\bibfnamefont {O.}~\bibnamefont {Pujolas}}, \bibinfo {author} {\bibfnamefont {T.}~\bibnamefont {Vachaspati}}, \ and\ \bibinfo {author} {\bibfnamefont {G.}~\bibnamefont {Zahariade}},\ }\href {\doibase 10.1103/PhysRevD.100.045011} {\bibfield  {journal} {\bibinfo  {journal} {Phys. Rev. D}\ }\textbf {\bibinfo {volume} {100}},\ \bibinfo {pages} {045011} (\bibinfo {year} {2019})},\ \Eprint {http://arxiv.org/abs/1904.12962} {arXiv:1904.12962 [hep-th]} \BibitemShut {NoStop}%
\bibitem [{\citenamefont {Mukhopadhyay}\ \emph {et~al.}(2022)\citenamefont {Mukhopadhyay}, \citenamefont {Sfakianakis}, \citenamefont {Vachaspati},\ and\ \citenamefont {Zahariade}}]{Mukhopadhyay:2021wmu}%
  \BibitemOpen
  \bibfield  {author} {\bibinfo {author} {\bibfnamefont {M.}~\bibnamefont {Mukhopadhyay}}, \bibinfo {author} {\bibfnamefont {E.~I.}\ \bibnamefont {Sfakianakis}}, \bibinfo {author} {\bibfnamefont {T.}~\bibnamefont {Vachaspati}}, \ and\ \bibinfo {author} {\bibfnamefont {G.}~\bibnamefont {Zahariade}},\ }\href {\doibase 10.1007/JHEP04(2022)118} {\bibfield  {journal} {\bibinfo  {journal} {JHEP}\ }\textbf {\bibinfo {volume} {04}},\ \bibinfo {pages} {118} (\bibinfo {year} {2022})},\ \Eprint {http://arxiv.org/abs/2110.08277} {arXiv:2110.08277 [hep-th]} \BibitemShut {NoStop}%
\bibitem [{\citenamefont {Rout}\ and\ \citenamefont {Altschul}(2024)}]{Rout:2024kbz}%
  \BibitemOpen
  \bibfield  {author} {\bibinfo {author} {\bibfnamefont {A.}~\bibnamefont {Rout}}\ and\ \bibinfo {author} {\bibfnamefont {B.}~\bibnamefont {Altschul}},\ }\href {\doibase 10.3390/sym16121571} {\bibfield  {journal} {\bibinfo  {journal} {Symmetry}\ }\textbf {\bibinfo {volume} {16}},\ \bibinfo {pages} {1571} (\bibinfo {year} {2024})},\ \Eprint {http://arxiv.org/abs/2410.09273} {arXiv:2410.09273 [hep-th]} \BibitemShut {NoStop}%
\bibitem [{\citenamefont {Contri}\ \emph {et~al.}(2025)\citenamefont {Contri}, \citenamefont {Dvali},\ and\ \citenamefont {Sakhelashvili}}]{Contri:2025eod}%
  \BibitemOpen
  \bibfield  {author} {\bibinfo {author} {\bibfnamefont {G.}~\bibnamefont {Contri}}, \bibinfo {author} {\bibfnamefont {G.}~\bibnamefont {Dvali}}, \ and\ \bibinfo {author} {\bibfnamefont {O.}~\bibnamefont {Sakhelashvili}},\ }\href@noop {} {\  (\bibinfo {year} {2025})},\ \Eprint {http://arxiv.org/abs/2509.08049} {arXiv:2509.08049 [hep-th]} \BibitemShut {NoStop}%
\bibitem [{\citenamefont {Jackiw}\ and\ \citenamefont {Rebbi}(1976)}]{jackiw-1976}%
  \BibitemOpen
  \bibfield  {author} {\bibinfo {author} {\bibfnamefont {R.}~\bibnamefont {Jackiw}}\ and\ \bibinfo {author} {\bibfnamefont {C.}~\bibnamefont {Rebbi}},\ }\href {\doibase 10.1103/physrevd.13.3398} {\bibfield  {journal} {\bibinfo  {journal} {Phys. Rev. D}\ }\textbf {\bibinfo {volume} {13}},\ \bibinfo {pages} {3398} (\bibinfo {year} {1976})}\BibitemShut {NoStop}%
\bibitem [{\citenamefont {Angelakis}\ and\ \citenamefont {Noh}(2014)}]{Angelakis:2013dia}%
  \BibitemOpen
  \bibfield  {author} {\bibinfo {author} {\bibfnamefont {D.~G.}\ \bibnamefont {Angelakis}}\ and\ \bibinfo {author} {\bibfnamefont {C.}~\bibnamefont {Noh}},\ }\href {\doibase 10.1038/srep06110} {\bibfield  {journal} {\bibinfo  {journal} {Sci. Rep.}\ }\textbf {\bibinfo {volume} {4}},\ \bibinfo {pages} {6110} (\bibinfo {year} {2014})},\ \Eprint {http://arxiv.org/abs/1306.2179} {arXiv:1306.2179 [quant-ph]} \BibitemShut {NoStop}%
\bibitem [{\citenamefont {{Gupta}}\ \emph {et~al.}(2024)\citenamefont {{Gupta}}, \citenamefont {{Srinivasu}}, \citenamefont {{Kumar}}, \citenamefont {{Tiwari}}, \citenamefont {{Pal}}, \citenamefont {{Wanare}},\ and\ \citenamefont {{Ramakrishna}}}]{2024Kumar}%
  \BibitemOpen
  \bibfield  {author} {\bibinfo {author} {\bibfnamefont {N.~K.}\ \bibnamefont {{Gupta}}}, \bibinfo {author} {\bibfnamefont {S.}~\bibnamefont {{Srinivasu}}}, \bibinfo {author} {\bibfnamefont {M.}~\bibnamefont {{Kumar}}}, \bibinfo {author} {\bibfnamefont {A.~K.}\ \bibnamefont {{Tiwari}}}, \bibinfo {author} {\bibfnamefont {S.~S.}\ \bibnamefont {{Pal}}}, \bibinfo {author} {\bibfnamefont {H.}~\bibnamefont {{Wanare}}}, \ and\ \bibinfo {author} {\bibfnamefont {S.~A.}\ \bibnamefont {{Ramakrishna}}},\ }\href {\doibase 10.1063/5.0175495} {\bibfield  {journal} {\bibinfo  {journal} {Applied Physics Letters}\ }\textbf {\bibinfo {volume} {124}},\ \bibinfo {eid} {091104} (\bibinfo {year} {2024})}\BibitemShut {NoStop}%
\bibitem [{\citenamefont {Jana}\ \emph {et~al.}(2019)\citenamefont {Jana}, \citenamefont {Saha},\ and\ \citenamefont {Das}}]{jana2019jackiw}%
  \BibitemOpen
  \bibfield  {author} {\bibinfo {author} {\bibfnamefont {S.}~\bibnamefont {Jana}}, \bibinfo {author} {\bibfnamefont {A.}~\bibnamefont {Saha}}, \ and\ \bibinfo {author} {\bibfnamefont {S.}~\bibnamefont {Das}},\ }\href@noop {} {\bibfield  {journal} {\bibinfo  {journal} {Physical Review B}\ }\textbf {\bibinfo {volume} {100}},\ \bibinfo {pages} {085428} (\bibinfo {year} {2019})}\BibitemShut {NoStop}%
\bibitem [{\citenamefont {Morse}\ and\ \citenamefont {Feshbach}(1953)}]{morse-1953}%
  \BibitemOpen
  \bibfield  {author} {\bibinfo {author} {\bibfnamefont {P.~M.}\ \bibnamefont {Morse}}\ and\ \bibinfo {author} {\bibfnamefont {H.}~\bibnamefont {Feshbach}},\ }\href@noop {} {\emph {\bibinfo {title} {{Methods of Theoretical Physics}}}}\ (\bibinfo  {publisher} {McGraw-Hill Science, Engineering \& Mathematics},\ \bibinfo {year} {1953})\BibitemShut {NoStop}%
\bibitem [{\citenamefont {Manton}\ and\ \citenamefont {Merabet}(1997)}]{manton-1997}%
  \BibitemOpen
  \bibfield  {author} {\bibinfo {author} {\bibfnamefont {N.~S.}\ \bibnamefont {Manton}}\ and\ \bibinfo {author} {\bibfnamefont {H.}~\bibnamefont {Merabet}},\ }\href {\doibase 10.1088/0951-7715/10/1/002} {\bibfield  {journal} {\bibinfo  {journal} {Nonlinearity}\ }\textbf {\bibinfo {volume} {10}},\ \bibinfo {pages} {3} (\bibinfo {year} {1997})}\BibitemShut {NoStop}%
\bibitem [{\citenamefont {Blanco-Pillado}\ \emph {et~al.}(2021)\citenamefont {Blanco-Pillado}, \citenamefont {Jim{\'e}nez-Aguilar},\ and\ \citenamefont {Urrestilla}}]{Blanco-Pillado:2020smt}%
  \BibitemOpen
  \bibfield  {author} {\bibinfo {author} {\bibfnamefont {J.~J.}\ \bibnamefont {Blanco-Pillado}}, \bibinfo {author} {\bibfnamefont {D.}~\bibnamefont {Jim{\'e}nez-Aguilar}}, \ and\ \bibinfo {author} {\bibfnamefont {J.}~\bibnamefont {Urrestilla}},\ }\href {\doibase 10.1088/1475-7516/2021/01/027} {\bibfield  {journal} {\bibinfo  {journal} {JCAP}\ }\textbf {\bibinfo {volume} {01}},\ \bibinfo {pages} {027} (\bibinfo {year} {2021})},\ \Eprint {http://arxiv.org/abs/2006.13255} {arXiv:2006.13255 [hep-th]} \BibitemShut {NoStop}%
\bibitem [{\citenamefont {Navarro-Obreg\'on}\ \emph {et~al.}(2023)\citenamefont {Navarro-Obreg\'on}, \citenamefont {Nieto},\ and\ \citenamefont {Queiruga}}]{Navarro-Obregon:2023}%
  \BibitemOpen
  \bibfield  {author} {\bibinfo {author} {\bibfnamefont {S.}~\bibnamefont {Navarro-Obreg\'on}}, \bibinfo {author} {\bibfnamefont {L.~M.}\ \bibnamefont {Nieto}}, \ and\ \bibinfo {author} {\bibfnamefont {J.~M.}\ \bibnamefont {Queiruga}},\ }\href {\doibase 10.1103/PhysRevE.108.044216} {\bibfield  {journal} {\bibinfo  {journal} {Phys. Rev. E}\ }\textbf {\bibinfo {volume} {108}},\ \bibinfo {pages} {044216} (\bibinfo {year} {2023})}\BibitemShut {NoStop}%
\bibitem [{\citenamefont {Mussardo}(2015)}]{Mussardo:2015xua}%
  \BibitemOpen
  \bibfield  {author} {\bibinfo {author} {\bibfnamefont {G.}~\bibnamefont {Mussardo}},\ }\href {\doibase 10.1088/1742-5468/2015/12/P12003} {\bibfield  {journal} {\bibinfo  {journal} {J. Stat. Mech.}\ }\textbf {\bibinfo {volume} {1512}},\ \bibinfo {pages} {P12003} (\bibinfo {year} {2015})},\ \Eprint {http://arxiv.org/abs/1508.05975} {arXiv:1508.05975 [cond-mat.stat-mech]} \BibitemShut {NoStop}%
\bibitem [{\citenamefont {Weigel}\ and\ \citenamefont {Saadatmand}(2024)}]{Weigel:2023fxe}%
  \BibitemOpen
  \bibfield  {author} {\bibinfo {author} {\bibfnamefont {H.}~\bibnamefont {Weigel}}\ and\ \bibinfo {author} {\bibfnamefont {D.}~\bibnamefont {Saadatmand}},\ }\href {\doibase 10.3390/universe10010013} {\bibfield  {journal} {\bibinfo  {journal} {Universe}\ }\textbf {\bibinfo {volume} {10}},\ \bibinfo {pages} {13} (\bibinfo {year} {2024})},\ \Eprint {http://arxiv.org/abs/2311.14437} {arXiv:2311.14437 [hep-th]} \BibitemShut {NoStop}%
\bibitem [{\citenamefont {Koke}\ \emph {et~al.}(2016)\citenamefont {Koke}, \citenamefont {Noh},\ and\ \citenamefont {Angelakis}}]{Koke:2016etw}%
  \BibitemOpen
  \bibfield  {author} {\bibinfo {author} {\bibfnamefont {C.}~\bibnamefont {Koke}}, \bibinfo {author} {\bibfnamefont {C.}~\bibnamefont {Noh}}, \ and\ \bibinfo {author} {\bibfnamefont {D.~G.}\ \bibnamefont {Angelakis}},\ }\href {\doibase 10.1016/j.aop.2016.08.013} {\bibfield  {journal} {\bibinfo  {journal} {Annals Phys.}\ }\textbf {\bibinfo {volume} {374}},\ \bibinfo {pages} {162} (\bibinfo {year} {2016})},\ \Eprint {http://arxiv.org/abs/1607.04821} {arXiv:1607.04821 [quant-ph]} \BibitemShut {NoStop}%
\bibitem [{\citenamefont {Rajaraman}(1982)}]{rajaraman1982solitons}%
  \BibitemOpen
  \bibfield  {author} {\bibinfo {author} {\bibfnamefont {R.}~\bibnamefont {Rajaraman}},\ }\href {https://books.google.es/books?id=1XucQgAACAAJ} {\emph {\bibinfo {title} {Solitons and Instantons: An Introduction to Solitons and Instantons in Quantum Field Theory}}},\ North-Holland personal library\ (\bibinfo  {publisher} {North-Holland Publishing Company},\ \bibinfo {year} {1982})\BibitemShut {NoStop}%
\bibitem [{\citenamefont {Garani}\ \emph {et~al.}(2025)\citenamefont {Garani}, \citenamefont {Redi},\ and\ \citenamefont {Tesi}}]{Garani:2025qnm}%
  \BibitemOpen
  \bibfield  {author} {\bibinfo {author} {\bibfnamefont {R.}~\bibnamefont {Garani}}, \bibinfo {author} {\bibfnamefont {M.}~\bibnamefont {Redi}}, \ and\ \bibinfo {author} {\bibfnamefont {A.}~\bibnamefont {Tesi}},\ }\href@noop {} {\  (\bibinfo {year} {2025})},\ \Eprint {http://arxiv.org/abs/2502.12249} {arXiv:2502.12249 [hep-th]} \BibitemShut {NoStop}%
\bibitem [{\citenamefont {García Martín-Caro}\ \emph {et~al.}(2024)\citenamefont {García Martín-Caro}, \citenamefont {García-Moreno}, \citenamefont {Olmedo},\ and\ \citenamefont {Sánchez~Velázquez}}]{GarciaMartin-Caro:2024qpk}%
  \BibitemOpen
  \bibfield  {author} {\bibinfo {author} {\bibfnamefont {A.}~\bibnamefont {García Martín-Caro}}, \bibinfo {author} {\bibfnamefont {G.}~\bibnamefont {García-Moreno}}, \bibinfo {author} {\bibfnamefont {J.}~\bibnamefont {Olmedo}}, \ and\ \bibinfo {author} {\bibfnamefont {J.~M.}\ \bibnamefont {Sánchez~Velázquez}},\ }\href {\doibase 10.1103/PhysRevD.110.025007} {\bibfield  {journal} {\bibinfo  {journal} {Phys. Rev. D}\ }\textbf {\bibinfo {volume} {110}},\ \bibinfo {pages} {025007} (\bibinfo {year} {2024})},\ \Eprint {http://arxiv.org/abs/2404.06166} {arXiv:2404.06166 [quant-ph]} \BibitemShut {NoStop}%
\bibitem [{\citenamefont {Fulgado-Claudio}\ \emph {et~al.}(2023)\citenamefont {Fulgado-Claudio}, \citenamefont {Velázquez},\ and\ \citenamefont {Bermudez}}]{fulgado-claudio-2023}%
  \BibitemOpen
  \bibfield  {author} {\bibinfo {author} {\bibfnamefont {C.}~\bibnamefont {Fulgado-Claudio}}, \bibinfo {author} {\bibfnamefont {J.~M.~S.}\ \bibnamefont {Velázquez}}, \ and\ \bibinfo {author} {\bibfnamefont {A.}~\bibnamefont {Bermudez}},\ }\href {\doibase 10.22331/q-2023-06-21-1042} {\bibfield  {journal} {\bibinfo  {journal} {Quantum}\ }\textbf {\bibinfo {volume} {7}},\ \bibinfo {pages} {1042} (\bibinfo {year} {2023})}\BibitemShut {NoStop}%
\bibitem [{\citenamefont {Mukhanov}\ and\ \citenamefont {Winitzki}(2007)}]{mukhanov-2007}%
  \BibitemOpen
  \bibfield  {author} {\bibinfo {author} {\bibfnamefont {V.}~\bibnamefont {Mukhanov}}\ and\ \bibinfo {author} {\bibfnamefont {S.}~\bibnamefont {Winitzki}},\ }\href@noop {} {\emph {\bibinfo {title} {{Introduction to Quantum Effects in Gravity}}}}\ (\bibinfo  {publisher} {Cambridge University Press},\ \bibinfo {year} {2007})\BibitemShut {NoStop}%
\bibitem [{\citenamefont {Campos}\ and\ \citenamefont {Mohammadi}(2021)}]{campos-2021}%
  \BibitemOpen
  \bibfield  {author} {\bibinfo {author} {\bibfnamefont {J.~G.~F.}\ \bibnamefont {Campos}}\ and\ \bibinfo {author} {\bibfnamefont {A.}~\bibnamefont {Mohammadi}},\ }\href {\doibase 10.1007/jhep09(2021)103} {\bibfield  {journal} {\bibinfo  {journal} {JHEP}\ }\textbf {\bibinfo {volume} {2021}} (\bibinfo {year} {2021}),\ 10.1007/jhep09(2021)103}\BibitemShut {NoStop}%
\bibitem [{\citenamefont {Evslin}\ and\ \citenamefont {Garc\'\i{}a Mart\'\i{}n-Caro}(2022)}]{Evslin:2022wyx}%
  \BibitemOpen
  \bibfield  {author} {\bibinfo {author} {\bibfnamefont {J.}~\bibnamefont {Evslin}}\ and\ \bibinfo {author} {\bibfnamefont {A.}~\bibnamefont {Garc\'\i{}a Mart\'\i{}n-Caro}},\ }\href {\doibase 10.1007/JHEP12(2022)111} {\bibfield  {journal} {\bibinfo  {journal} {JHEP}\ }\textbf {\bibinfo {volume} {12}},\ \bibinfo {pages} {111} (\bibinfo {year} {2022})},\ \Eprint {http://arxiv.org/abs/2210.13791} {arXiv:2210.13791 [hep-th]} \BibitemShut {NoStop}%
\bibitem [{\citenamefont {Garc\'\i{}a Mart\'\i{}n-Caro}\ \emph {et~al.}(2025)\citenamefont {Garc\'\i{}a Mart\'\i{}n-Caro}, \citenamefont {Queiruga},\ and\ \citenamefont {Wereszczynski}}]{GarciaMartin-Caro:2025zkc}%
  \BibitemOpen
  \bibfield  {author} {\bibinfo {author} {\bibfnamefont {A.}~\bibnamefont {Garc\'\i{}a Mart\'\i{}n-Caro}}, \bibinfo {author} {\bibfnamefont {J.}~\bibnamefont {Queiruga}}, \ and\ \bibinfo {author} {\bibfnamefont {A.}~\bibnamefont {Wereszczynski}},\ }\href@noop {} {\  (\bibinfo {year} {2025})},\ \Eprint {http://arxiv.org/abs/2501.02589} {arXiv:2501.02589 [hep-th]} \BibitemShut {NoStop}%
\bibitem [{\citenamefont {Landau}\ and\ \citenamefont {Lifshitz}(2013)}]{landau2013quantum}%
  \BibitemOpen
  \bibfield  {author} {\bibinfo {author} {\bibfnamefont {L.}~\bibnamefont {Landau}}\ and\ \bibinfo {author} {\bibfnamefont {E.}~\bibnamefont {Lifshitz}},\ }\href@noop {} {\emph {\bibinfo {title} {Quantum Mechanics: Non-Relativistic Theory}}}\ (\bibinfo  {publisher} {Elsevier Science},\ \bibinfo {year} {2013})\BibitemShut {NoStop}%
\bibitem [{\citenamefont {Charmchi}\ and\ \citenamefont {Gousheh}(2014)}]{charmchi-2014}%
  \BibitemOpen
  \bibfield  {author} {\bibinfo {author} {\bibfnamefont {F.}~\bibnamefont {Charmchi}}\ and\ \bibinfo {author} {\bibfnamefont {S.~S.}\ \bibnamefont {Gousheh}},\ }\href {\doibase 10.1103/physrevd.89.025002} {\bibfield  {journal} {\bibinfo  {journal} {Phys. Rev. D}\ }\textbf {\bibinfo {volume} {89}} (\bibinfo {year} {2014}),\ 10.1103/physrevd.89.025002}\BibitemShut {NoStop}%
\end{thebibliography}
\end{document}